\documentclass[
preprint,
superscriptaddress,aps,
prd,a4paper
preprintnumbers,amsmath,amssymb,nofootinbib,11pt
]{revtex4-1}

\usepackage{placeins}
\usepackage{amsmath}
\usepackage{amssymb}
\usepackage{amsfonts}
\usepackage{dsfont}
\usepackage{graphicx}
\graphicspath{{./figs/}}
\usepackage[usenames,dvipsnames,svgnames,table]{xcolor}
\usepackage{xfrac}
\usepackage{comment}
\usepackage{pifont}
\usepackage{physics}
\usepackage{hyperref}
\usepackage{bm}
\usepackage{enumitem}
\usepackage[normalem]{ulem}
\usepackage{fontawesome}
\usepackage{float}
\usepackage{subfigure}
\usepackage{multirow}

\usepackage{booktabs}
\usepackage{siunitx}
\usepackage[normalem]{ulem}

\definecolor{rossoferrari}{HTML}{D9073D}
\definecolor{mediumblue}{HTML}{0000CD}
\definecolor{forestgreen}{HTML}{228B22}
\definecolor{desy_blue}{HTML}{009EE2}
\definecolor{desy_orange}{HTML}{FD8800}
\definecolor{peera_green}{HTML}{008B8B}
\definecolor{peera_orange}{HTML}{B22222}
\definecolor{light_pink}{rgb}{1,0.4,0.4}
\definecolor{light_blue}{rgb}{0.284602,0.317763,0.963947}
\definecolor{peera_col}{RGB}{240, 94, 28}
\definecolor{blue_col}{RGB}{0,92,175}
\definecolor{red_col}{RGB}{203,64,66}
\usepackage{hyperref}
\hypersetup{linktocpage,
    colorlinks=true,
    linkcolor=red_col,
    filecolor=Mahogany,
    urlcolor=blue_col,
    citecolor=blue_col,
}

\begin{document}



\title{Reviving Motivated Inflationary Potentials with $K$-inflation in the light of ACT}

\author{Milad Solbi}
\email{miladsolbi@gmail.com}
\affiliation{Khon Kaen Particle Physics and Cosmology Theory Group (KKPaCT),
Department of Physics, Faculty of Science, Khon Kaen University,
123 Mitraphap Rd., Khon Kaen, 40002, Thailand}

\author{Daris Samart}
\email{darisa@kku.ac.th : corresponding author}
\affiliation{Khon Kaen Particle Physics and Cosmology Theory Group (KKPaCT),
Department of Physics, Faculty of Science, Khon Kaen University,
123 Mitraphap Rd., Khon Kaen, 40002, Thailand}

\author{Peera Simakachorn}
\email{peera.sima@gmail.com}
\affiliation{Khon Kaen Particle Physics and Cosmology Theory Group (KKPaCT),
Department of Physics, Faculty of Science, Khon Kaen University,
123 Mitraphap Rd., Khon Kaen, 40002, Thailand}


\date{\today}


\begin{abstract}
\noindent
\textcolor{black}{
The recent data release from the Atacama Cosmology Telescope (ACT) favors a higher scalar spectral index $n_s$, placing well-motivated inflationary models such as $\alpha$-attractor T-models and natural Inflation in tension with observations. We propose a K-inflation framework with a field-dependent non-canonical kinetic term $G(\phi)$ to reconcile these models with the latest joint Planck-ACT-LB-BK18 constraints. Our analysis incorporates a careful calculation of the reheating equation of state parameter $w_{\rm re}$, avoiding standard power-law approximations, and examines consistency with the Swampland Distance and de Sitter Conjectures. We demonstrate that the additional friction induced by the non-minimal kinetic coupling successfully shifts the predictions of both the T-model and natural Inflation into the favored observational regions. For the $\alpha$-attractor T-model ($n=2$), this is achieved with $\beta\sim\mathcal{O}(10)$ over a wide range of $\alpha$, while compliance with the Swampland criteria favors $\alpha\gtrsim\mathcal{O}(10^{-3})$; this scenario yields a matter-like reheating phase ($w_{\rm re}\approx 0$) and a red-tilted gravitational-wave background (GWB) that remains undetectable at near-future observatories. Importantly, the natural Inflation potentials with $n=4$ ($n=5$) are consistent with the CMB results for $\alpha\lesssim 7$ ($\alpha\lesssim 8$) and $\beta\lesssim -1$, producing a stiff reheating epoch with asymptotic $w_{\rm re}\approx 3/5$ ($w_{\rm re}\approx 2/3$) and a blue-tilted GWB scaling as $\Omega_{\rm GW}\propto f^{4/7}$ ($\Omega_{\rm GW}\propto f^{2/3}$). This signal is potentially detectable by future observatories such as LISA, Cosmic Explorer, the Einstein Telescope, DECIGO, and BBO, while remaining consistent with Big Bang Nucleosynthesis and $\Delta N_{\rm eff}$ bounds. Furthermore, the Swampland distance conjecture is satisfied only for $\alpha\lesssim 5$, indicating that GW observations combined with the Swampland criteria can hint at the class of UV completions underlying inflation.}
\end{abstract}
\maketitle
{\small \tableofcontents}
\hrulefill
\newpage
\section{Introduction}
Cosmic inflation is a cornerstone of the standard cosmological model and was initially proposed to solve the fundamental problems of the Hot Big Bang theory, such as the horizon, flatness, and primordial monopole problems \cite{Guth:1981, Linde:1982, Starobinsky:1980}. Beyond addressing these issues, the inflationary paradigm also provides a compelling mechanism for generating the primordial density perturbations that can seed the large-scale structures of the Universe \cite{Mukhanov:1981, Hawking:1982, Guth:1982} and explain the Cosmic-Microwave-Background (CMB) anisotropies. This scenario has gained strong support over the past decades, from a series of observations, including WMAP \cite{Hinshaw:2013}, Planck \cite{Planck:2020}, Atacama Cosmology Telescope (ACT) \cite{Louis:2025, Calabrese:2025}, and the Dark Energy Spectroscopic Instrument (DESI) \cite{DESI:2024mwx}, and the BICEP/Keck array \cite{BICEP:2021xfz}.

The recent joint analysis incorporating the ACT data release 6 (DR6) with Planck, DESI BAO, and BICEP/Keck results provides updated constraints on cosmological parameters of the cosmological concordance model or the so-called ``$\Lambda$CDM" model \cite{Calabrese:2025}. 
Notably, the new analysis favors a larger spectral index for the primordial scalar power spectrum, $n_s = 0.9743 \pm 0.0034$ \cite{Calabrese:2025,BICEP:2021xfz}, creating tensions with many well-motivated inflationary models that had previously agreed with the joint analysis based on the Planck data release.
For instance, the standard $\alpha$-attractor models \cite{Kallosh:2013a, Kallosh:2013b} now locate near the $2\sigma$ boundary of the joint observations' best-fitted region, while the better agreement can be achieved by considering a modified setup, e.g., a non-instantaneous reheating period \cite{Calabrese:2025}.
As well as other inflationary models, their viable parameter spaces are also re-evaluated in light of the latest data \cite{Parvizi:2025sed,Heidarian:2025,Kallosh:2025, Aoki:2025, Dioguardi:2025, Salvio:2025, Gao:2025, Drees:2025, Liu:2025, Gialamas:2025,Pallis:2025vxo,Aldabergenov:2025kcv,Ahmed:2025sfm,Zhu:2025twm,Yuennan:2025kde,Zharov:2025zjg,Addazi:2025agg,Kumar:2025apf,Alexandre:2025ixz}.

Another avenue to modify the predictions of each inflationary model is to incorporate a non-canonical kinetic term for the inflaton field.
In this work, we focus on the $K$-inflation framework \cite{Armendariz-Picon:1999, Garriga:1999, Barenboim:2007, Lin:2020, Solbi:2021a, Solbi:2021b}, in which the non-canonical kinetic terms can arise from low-energy effective string theories \cite{Barenboim:2007}.
We will see that the non-canonical kinetic term introduces additional friction or driving terms that can shift the predictions for $n_s$ and the tensor-to-scalar ratio $r$ back into the favored region of the new joint data; this allows us to revive models that would otherwise be disfavored.
In particular, we apply the $K$-inflation framework to the $\alpha$-attractor T-models, which were previously in perfect agreement with the Planck data \cite{Planck:2020} but are now in tension with the recent ACT DR6 release \cite{Calabrese:2025}.
We also study the efficacy of the $K$-inflation when applying it to the natural-inflation potential \cite{Freese:2014nla,Kitabayashi:2023vfe,Zhang:2018wbn} of the quartic $(n=4)$ and quintic $(n=5)$ orders, which was already largely ruled out by Planck 2018 data for power indices $n \ge 1$ \cite{Planck:2020}.

In addition to being consistent with new ACT results, we need to ensure that such inflationary models also satisfy other phenomenological and observational constraints.
After inflation ends, the inflaton oscillates around the potential minimum, which is fixed by the inflationary setup and determines how the universe expands during reheating, before the inflaton transfers its energy to the primordial thermal plasma. 
However, the success of Big Bang Nucleosynthesis theory places a bound on the expansion history around MeV scales \cite{Kawasaki:1999na,Kawasaki:2000en, Hannestad:2004px, deSalas:2015glj} and can therefore be used to constrain overall expansion history and inflationary models.
In this work, we rigorously compute the equation of state $w_{\rm re}$ during reheating by solving the full inflaton evolution; see \cite{Iacconi:2025} for a similar approach.

Interestingly, the non-trivial expansion history also leaves a characteristic imprint on the gravitational-wave  background (GWB) from inflation \cite{Boyle:2005se,Watanabe:2006qe,Boyle:2007zx,Jinno:2012xb,Saikawa:2018rcs,Figueroa:2019paj,Allahverdi:2020bys,Gouttenoire:2021jhk,Gouttenoire:2021wzu, Co:2021lkc,Simakachorn:2022yjy,Duval:2024jsg,Konings:2024zvz,LIGOScientific:2025kry}, i.e., the enhanced and suppressed gravitational-wave (GW) spectrum when $w_{\rm re} > 1/3$ and $w_{\rm re} < 1/3$, respectively.
This GW signature could be searched and constrained by current and future GW observatories, such as LIGO-Virgo-KAGRA network \cite{LIGOScientific:2014qfs,LIGOScientific:2019vic}, LISA \cite{LISA:2017pwj,LISACosmologyWorkingGroup:2022jok,LISA:2024hlh}, Einstein Telescope (ET) \cite{Hild:2010id,Punturo:2010zz,ET:2025xjr}, Cosmic Explorer (CE) \cite{LIGOScientific:2016wof}, BBO \cite{Harry:2006, Crowder:2005, Corbin:2006, Yagi:2011}, DECIGO \cite{Seto:2001, Kawamura:2006, Kawamura:2011}, and the Pulsar Timing Arrays (including SKA) \cite{ NANOGrav:2023gor,NANOGrav:2023hvm,EPTA:2023fyk,Reardon:2023gzh,SKAOPulsarScienceWorkingGroup:2025oyu}.
As we shall see, some parameter spaces of the $K$-inflation models that fit the CMB results well could be tested using GW observations.

The inflation models are subjected to the ``Swampland'' criteria, which emerges as a powerful theoretical framework for distinguishing effective field theories that can be embedded in a UV-complete theory of quantum gravity (the Landscape) from those that cannot (the Swampland) \cite{Ooguri:2006in,Vafa:2005ui,Das:2018hqy,Garg:2018reu,Ooguri:2018wrx,Kehagias:2018uem}. 
Within the parameter space that fits the recent CMB results and might also accommodate an observable GW signal, we will examine whether such inflationary models require a particular class of UV completions by confronting them with the Swampland Distance and the de Sitter conjectures.


The outline of the paper is as follows. In Sect.~\ref{sec2}, we review the $K$-inflation framework and its background and perturbation equations. Section~\ref{sec3} presents the theoretical, phenomenological, and observational limits applied to our framework, including the Swampland criteria, bounds from the reheating era, and a review of the GW background from primordial inflation. We then apply our framework to the $\alpha$-attractor T-model in Sect.~\ref{sec:alpha_attractor} and natural inflation in Sect.~\ref{sec:axion}. Our conclusions are presented in Sect.~\ref{sec6}.
App.~\ref{app:power_law_validity} contains further details on our calculation of tensor power spectrum and GWB, and the alternative setup of natural inflation is discussed in App.~\ref{app:natural_inflation_minus}.

\section{$K$-inflation}
\label{sec2}

We adopt the $K$-inflation action where the kinetic term of a real scalar field is non-canonically coupled via a \emph{coupling function} $G(\phi)$ \cite{Kobayashi:2010cm,Lin:2020}
\begin{equation}
    \mathcal{S} = \int d^4x \sqrt{-g} \left[ \frac{M_{\rm Pl}^2 \mathcal{R}}{2} + (1 - 2G(\phi))X - V(\phi) \right],
    \label{eq:action}
\end{equation}
where $X = -\frac{1}{2}g^{\mu\nu}\partial_\mu\phi\partial_\nu\phi$, {$g_{\mu\nu}$ is the spacetime metric, $\mathcal{R}$ is the Ricci scalar, and $M_{\rm Pl}$ is the reduced Planck mass.
For $G(\phi) \rightarrow 0$, the action reduces to the standard inflationary model.
Assuming a homogeneous, isotropic and expanding universe, i.e., Friedmann-Lema\^{i}tre-Robertson-Walker metric, the Friedmann equations are given by \cite{Kobayashi:2010cm,Lin:2020}
\begin{align}
   3 M_{\rm Pl}^2 H^2 &= \frac{1}{2}\dot{\phi}^2\left(1-2G(\phi)\right) + V(\phi), \label{eq:friedmann1} \\
    M_{\rm Pl}^2(2\dot{H} + 3H^2) &= -\frac{1}{2}\dot{\phi}^2\left(1-2G(\phi)\right) + V(\phi),
    \label{eq:friedmann2}
\end{align}
with $H$ being the Hubble parameter.
The equation of motion for the scalar field is given by
\begin{equation}
    \ddot{\phi} + 3H\dot{\phi} + \frac{V'(\phi) - \dot{\phi}^2 G'(\phi)}{1 - 2G(\phi)} = 0,
    \label{eq:klein_gordon}
\end{equation}
where the prime and dot stand for derivatives with respect to the scalar field $\phi$ and cosmic time, respectively.

Under the slow-roll approximation (i.e., $-{\dot{H}}/{H^2}, ~ {\ddot{\phi}}/{(H\dot{\phi})} \ll 1$),
Eqs.~\eqref{eq:friedmann1} and \eqref{eq:klein_gordon} are simplified to be 
\begin{align}
    & 3 M_{\rm Pl}^2 H^2 \simeq V(\phi) \, , \label{eq:slow_roll_Friedmann1}\\
    & 3H\dot{\phi}(1-2G(\phi)) + V'(\phi) \simeq 0.
    \label{eq:slow_roll_Friedmann2}
\end{align}
In addition, during slow-roll inflation, the scalar and tensor power spectra, which describe the evolution of the corresponding perturbations, are respectively given by \cite{Kobayashi:2010cm,Lin:2020}
\begin{align}\label{eq:ps-pt}
\mathcal{P}_{s} = \frac{H^4}{4\pi^2 \dot{\phi}^2 \left( 1-2G(\phi) \right)}, ~ 
\mathcal{P}_{t} = \frac{2 H^2}{\pi^2 M_{\rm Pl}^2}.
\end{align}
The scalar (tensor) spectral index $n_s$ and tensor-to-scalar ratio $r$ in the slow-roll regime get modified by the coupling function $G$ to \cite{Lin:2020}
\begin{align}
   n_s - 1 & \simeq \frac{M_{\rm Pl}^2}{1-2G(\phi)} \left[ 2 \left(\frac{V''}{V}\right) - 3 \left(\frac{V'}{V}\right)^2 + \frac{2G'(\phi)}{1-2G(\phi)}\left|\frac{V'}{V}\right| \right] \, ,
    \label{eq:ns}\\
    n_t & \simeq -\frac{M_{\rm Pl}^2(V'/V)^2}{1-2G(\phi)} 
    \label{eq:nt}\, , \\
   r & \simeq \frac{8M_{\rm Pl}^2(V'/V)^2}{1-2G(\phi)} \, .
    \label{eq:r}
\end{align}
Note that from Eqs. (\ref{eq:nt}) and (\ref{eq:r}), the consistency relation holds as $r \simeq - 8n_t$.
We calculate $n_s$ and $r$ relevant for CMB scales by using Eqs.~\eqref{eq:ns} and \eqref{eq:r}, which are solved together with Eqs.~\eqref{eq:slow_roll_Friedmann1} and \eqref{eq:slow_roll_Friedmann2}.
Note that we solve them with the initial condition of $\phi$ set to be Note that we solve them by integrating the boundary condition at the end of inflation ($\epsilon \equiv -\dot{H}/H^2 = 1$)  backwards to determine the field value at CMB horizon crossing.

The key feature of $K$-inflation is that an appropriate form for $G(\phi)$ can help improve the consistency of inflationary models with recent joint observational data \cite{Solbi:2021a,Solbi:2021b,Lin:2020}.
In this work, we will apply the $K$-inflation framework to some of the inflationary models, namely, the $\alpha$-attractor T model and the natural inflation.
But before discussing specific details of these models, we note that inflation must end and be followed by the reheating phase.
The next section will review the implications of reheating, which can be used as model-independent constraints on generic inflationary setups.

\section{Theoretical, phenomenological, and observational limits}\label{sec3}
While the $K$-inflation model can improve the inflationary explanation of the CMB observation, there are some limits on the parameter space of the $K$-inflation beyond which some other bounds get violated.
In this section, we discuss some bounds arising from theoretical motivations (i.e., Swampland criteria) and phenomenological constraints (i.e., reheating), as well as observational prospects (i.e., GWB).
These bounds will be model-independent and will be applied in the next sections to specific inflationary setups. 

\subsection{Swampland Criteria}
\label{sec:swampland}

The Swampland program, motivated by string theories, outlines the boundary between effective field theories that can be consistently embedded into a UV-complete theory of quantum gravity (the \emph{Landscape}) and those that cannot (the \emph{Swampland}) \cite{Vafa:2005ui}. Two central conjectures within this framework are the Swampland distance conjecture and the de Sitter conjecture. 
However, any theory violating such conjectures does not mean it is excluded, but it just resides in another class of theory rather than being string-theory motivated.

\textit{Swampland Distance Conjecture (SDC).}---The range of scalar field excursions during inflation, $\Delta \phi$, in an effective theory is conjectured to be bounded by the Planck mass \cite{Ooguri:2006in}
\begin{align}\label{sdc-conj}
|\Delta \phi| \leq \mathcal{O}(M_{\rm Pl}) 
\end{align}
This conjecture implies that trans-Planckian field excursions, often required in large-field inflationary models, may be incompatible with quantum gravity.

\textit{de Sitter Conjecture (dSC).}---The dSC imposes constraints on the shape of the scalar potential $V(\phi)$ by requiring that at least one of the following conditions is satisfied \cite{Kehagias:2018uem}
\begin{align}
M_{\text{Pl}} {|V'|}/{V} \geq c_2 \quad \text{or} \quad M_{\text{Pl}}^2 {V''}/{V} \leq -c_2,
\label{eq:ds_conj}
\end{align}
where typically $c_2 \simeq \mathcal{O}(0.1-1)$.
For the standard slow-roll inflation whose tensor-to-scalar ratio is given by $r = 16\epsilon_V$ with the slow-roll parameter $\epsilon_V \equiv (|V'|/V)^2/2$, the first condition leads to $r \gtrsim 8 c_s^2 \sim 0.08$, conflicting with the current CMB bound $r \lesssim 0.032$ \cite{Planck:2020}.
I.e., the standard slow-roll inflation models consistent with the current cosmological observations would reside within the Swampland.
This conclusion is not always true for $K$-inflation models, as the inflaton's dynamics get modified. 

Apart from analyzing whether the $K$-inflation model fits the CMB data, we also chart which regions of the $ K$-inflation parameter space are consistent with either the String Landscape or the Swampland.
This result would help guide us in constructing such $K$-inflation models in the future.
Furthermore, some regions of its parameter space are bounded by phenomenological constraints or are supported by promising GW observations, as we will discuss next.



\subsection{Reheating}
\label{sec:reheating}

As inflaton ends its slow-rolling phase, it begins oscillating around the minimum of its potential \cite{Kofman:1994}. The inflaton then decays via its coupling to Standard Model particles, reheating the universe, and establishing the radiation-dominated era. We will not specify the form of Standard-Model coupling, but will characterize the reheating phase in a model-independent way by 
\begin{itemize}
    \item the reheating temperature $T_{\rm re}$, defined when the reheating phase is completed and the radiation-domination era starts, i.e., from the radiation bath $\rho_{\rm rad}(T_{\rm re}) = \rho_{\rm re} \equiv \pi^2 g_*(T_{\rm re})T_{\rm re}^4/30$,
    \item the duration, parametrized by the number of e-folds $N_{\rm re} = \log(a_{\rm re}/a_{\rm end})$ where $a_{\rm end}$ is the scale factor at the end of inflation and $a_{\rm re} = a(T_{\rm re})$.
\end{itemize}

\subsubsection{Equation of state during the reheating phase}
The cosmological evolution during the reheating phase also depends on how the inflaton energy density is diluted by cosmic expansion. This evolution can be captured by the effective equation of state parameter\footnote{$w_{\rm re} \equiv P/\rho$ where $P$ and $\rho$ are pressure and energy density of the inflaton field as a perfect fluid.} $w_{\rm re}$, i.e., $\rho(a) = \rho_i \exp\left[-3 \int_{\ln a_i}^{\ln a} 3(1+w_{\rm re}(a)) d \ln a \right]$ which reduces to $\rho \propto a^{-3(1+w_{\rm re})}$ for a constant $w_{\rm re}$.
For a scalar field with a canonical kinetic term oscillating inside a power-law potential $V(\phi) \propto \phi^k$, its time-averaged equation of state parameter is $\left<w_{\rm re}\right> = (k-2)/(k+2)$ \cite{Mukaida:2012qn}.
However, in the $K$-inflation framework, the equation of state of the inflaton does not necessarily follow this simple relation due to its non-canonical kinetic term.  
Therefore, we calculate the time-averaged equation of state by
\begin{align}
    \langle w_{\rm re}(t) \rangle = T^{-1} \int_{t}^{t+T} w(t') dt' \, ,
    \label{eq:w_integral}
\end{align}
where $T$ is the period of the field oscillation, and we solved numerically Eq.~\eqref{eq:klein_gordon} for the inflaton dynamics after inflation and the instantaneous equation-of-state parameter
\begin{align}
    w(t) = \frac{\frac{1}{2}\dot{\phi}^2(t) \left[1 - 2G(\phi(t))\right] - V(\phi(t))}{\frac{1}{2}\dot{\phi}^2(t) \left[1 - 2G(\phi(t))\right] + V(\phi(t))} \, .
    \label{eq:w_re_instant}
\end{align}
%
Although $K$-inflation generally allows for deviations from reheating dynamics after standard slow-roll inflation, the $K$-inflation dynamics reduce to 
its canonical limit at later times during reheating.
We will see, from the numerical results in our examples, that $\left<w_{\rm re}\right>$ (using Eq.~\eqref{eq:w_integral}) evolves to an asymptotic value
$\left<w_{\rm re}\right> = (k-2)/(k+2)$, as if the inflaton with canonical kinetic term oscillates in a potential $V(\phi)\propto 
\phi^k$. 

For simplicity in our calculation, we will approximate that $\left<w_{\rm re}\right>$ reaches its final value right after the inflation's end and will use this $\langle w_{\rm re} \rangle$ to consistently cast  constraints on the parameter space of each inflationary model.\footnote{We have checked that, for our example models, the constraints on e-folding number could change by a factor of $\lesssim 1$, when using the precise $\left<w_{\rm re}\right>$ from Eq.~\eqref{eq:w_integral}.}
Hereafter, we shall use $w_{\rm re}$ to denote $\langle w_{\rm re}(t) \rangle$ for brevity.



\subsubsection{Bounds on number of inflationary efolds}
Consider the cosmological evolution starting when the perturbation of comoving scale $k = a_k H_k$ (with $a_k$ and $H_k$ being the scale factor and Hubble parameter at that time) exited the horizon during inflation. This mode left the horizon $N_k \equiv \ln (a_{\rm end}/a_k)$ e-folds before inflation ends and reenters the horizon at the CMB scale, i.e., $k = 0.05 \, {\rm Mpc^{-1}}$.
By extracting $N_{\rm re}$ from $k/(a_0 H_0)$, we obtain (see \cite{Munoz:2014eqa,German:2023} for a detailed derivation)
\begin{align}
    N_{\rm re} = \frac{4}{1-3w_{\rm re}} \left[ -N_k - \frac{1}{3}\ln\left(\frac{g_{*s}(T_{\rm re})}{g_{*s}(T_0)}\right) - \frac{1}{4}\ln\left(\frac{30}{\pi^2 g_*(T_{\rm re})}\right) - \ln\left(\frac{\rho_{\rm end}^{1/4} k}{H_k a_0 T_0}\right) \right] \, ,
    \label{eq:Nre_model_dep}
\end{align}
where $a_0$ is the scale factor today, $T_0$ is the CMB photons' temperature today, $\rho_{\rm end}$ is the total energy density of the Universe at the end of inflation, and $g_*(T)$ and $g_{*s}(T)$ are the effective numbers of relativistic degrees of freedom in energy and entropy density, respectively.
We have used $\rho_{\rm re}/\rho_{\rm end} = (a_{\rm re}/a_{\rm end})^{-3(1+w_{\rm re})}$ and the entropy conservation $a_{\rm re}^3 g_{*s}(T_{\rm re})T_{\rm re}^3 = a_0^3 g_{*s}(T_0)T_0^3$ with $g_{*s}(T_0) = 2+2\cdot (\sfrac{7}{8}) \cdot N_{\rm eff}\cdot (\sfrac{4}{11})$ where $N_{\rm eff} \simeq 3.046$ \cite{} is the effective number of SM neutrino species, and the first and second terms are the effective degrees of freedom of photon and neutrinos, respectively. 
%
The reheating temperature can be expressed (using $N_{\rm re} = \ln(a_{\rm re}/a_{\rm end}) = [3(1+w_{\rm re})]^{-1}\ln(\rho_{\rm end}/\rho_{\rm re})$) as
\begin{align}
     T_{\rm re} = \left[\frac{30 \rho_{\rm end}}{\pi^2 g_*(T_{\rm re})} \right]^{1/4} \exp\left[ -\frac{3}{4}(1+ w_{\rm re}) N_{\rm re} \right].
     \label{eq:Tre_dep}
\end{align}
For an inflationary model with specific values of  $w_{\rm re}$ and $\rho_{\rm end}$, the cosmic history can be charted using eqs.~\eqref{eq:Nre_model_dep} and \eqref{eq:Tre_dep} once $T_{\rm re}$ is given.

To calculate the CMB prediction, we need to solve for the field dynamics in Eqs.~\eqref{eq:slow_roll_Friedmann1} and \eqref{eq:slow_roll_Friedmann2} and evaluate $n_s$ in Eq.~\eqref{eq:ns} and $r$  in Eq.~\eqref{eq:r} at $N_k$ e-folds before inflation ends. 
By using Eq.~\eqref{eq:Tre_dep} to rewrite Eq.~\eqref{eq:Nre_model_dep}, we find $N_k$ in terms of other parameters as 
\begin{align}
    N_k &= \ln \left[\left(\frac{g_{*s}(T_0)}{g_{*s}(T_{\rm re})}\right)^{1/3} \left(\pi M_{\rm Pl} \sqrt{\frac{r \mathcal{A}_{s}}{2}}\right) \frac{a_0 T_0}{k T_{\rm re}} e^{-N_{\rm re}} \right] \, , \label{eq:Nk_bound_firstform}\\
    &= \ln \left[\left(\frac{g_{*s}(T_0)}{g_{*s}(T_{\rm re})}\right)^{1/3} g_*^{1/4}(T_{\rm re}) \left(\frac{\pi^6}{120}\right)^{1/4} \sqrt{r \mathcal{A}_{s}} \left(\frac{M_{\rm Pl} \, a_0 T_0}{k \rho_{\rm end}^{1/4}}\right) e^{\left(\frac{3w_{\rm re}-1}{4}N_{\rm re}\right) } \right] \, ,
    \label{eq:Nk_bound_Nre1}
\end{align}
where we express $\sqrt{2} H_k/(\pi M_{\rm Pl}) = \sqrt{\mathcal{P}_t(k)} = \sqrt{r \mathcal{P}_s(k)} = \sqrt{r \mathcal{A}_{s}}$---with $\mathcal{A}_s \simeq 2.1 \times 10^{-9}$ being the value of $\mathcal{P}_s(k)$ at CMB scale $k_{\rm CMB} = 0.05 ~ {\rm Mpc}^{-1}$ \cite{Planck:2020}--- using the slow-roll result in Eq.~\eqref{eq:ps-pt} which is valid deep inside the inflationary stage. Note that we have restored $M_{\rm Pl}$ for the moment.

Typically, $N_k$ can be chosen for any inflationary setup, i.e., for any set of $T_{\rm re}$, $\rho_{\rm end}$, and $w_{\rm re}$.
However, not all reheating scenarios are allowed by other cosmological constraints, leading to bounds on the range of possible $N_k$.
In this work, we will consider two constraints on $N_k$, coming from phenomenological bounds on the duration of the reheating phase and when it ended. See also \cite{Munoz:2014eqa,German:2023} for similar bounds.\footnote{The bounds from Ref.~\cite{German:2023} which are called the model-independent bounds are, in fact, not entirely model independent. Because the authors of \cite{German:2023} assume $w = -1$ during the inflationary stage, i.e., the inflaton is completely frozen during inflation. Therefore, our bounds in Fig.~\ref{fig:nk_bound} slightly differs from the results in \cite{German:2023}.}

\begin{figure}[t!]
    \centering
    \includegraphics[width=0.6\linewidth]{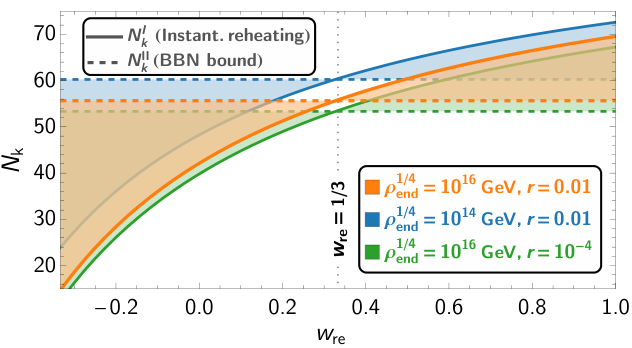}
    \vspace{-1.5em}
    \caption{Reheating bounds on the number of efolds $N_k$ from when the CMB-scale perturbation left the horizon to the end of inflation at energy scale $\rho_{\rm end}^{1/4}$. The viable range of $N_k$ is indicated by shaded regions. The solid line corresponds to the instantaneous-reheating limit in Eq.~\eqref{eq:Nk_bound_Nre} ($N_{\rm re} \geq 0$), and the dashed line shows the BBN bound in Eq.~\eqref{eq:Nk_bound_Tre} ($T_{\rm re} \geq T_{\rm BBN} = 10 \, {\rm MeV}$). Different colors are for different choices of $\rho_{\rm end}^{1/4}$ and $r$.}
    \label{fig:nk_bound}
\end{figure}

\textit{I) Instantaneous reheating bound.}---As the reheating should happen after the end of inflation, the energy density when the reheating is complete should be $\rho_{\rm re} \leq \rho_{\rm end}$, which translates to $N_{\rm re} \geq 0$.
For the extreme case of instantaneous reheating $N_{\rm re} = 0$, we obtain from Eq.~\eqref{eq:Nk_bound_Nre} a limit on $N_k$ for an inflation ending at energy scale $\rho_{\rm end}^{1/4}$ as 
\begin{align}
    N_k^I &= \ln \left[\left[\frac{g_{*s}(T_0)}{g_{*s}(T_{\rm re})}\right]^{\frac{1}{3}} g_*^{\frac{1}{4}}(T_{\rm re}) \left(\frac{\pi^6}{120}\right)^{\frac{1}{4}} \sqrt{r \mathcal{A}_{s}} \frac{M_{\rm Pl} \, a_0 T_0}{k \rho_{\rm end}^{1/4}} \right] \, ,\nonumber\\ &\simeq 55.6 + \ln\left[\left[\frac{106.75}{g_*(T_{\rm re})}\right]^{\frac{1}{12}}\sqrt{\frac{r}{0.01}} \left(\frac{10^{16} \, {\rm GeV}}{\rho_{\rm end}^{1/4}}\right)\right] \, ,
    \label{eq:Nk_bound_Nre}
\end{align}
where we plugged in parameters' values and used that $g_*(T) \approx g_{*s}(T)$ at high temperatures.
Moreover, Eq.~\eqref{eq:Nk_bound_Nre} also shows that $N_k^I$ can either be an upper or a lower limit of $N_k$ depending on $w_{\rm re}$. I.e., the bound $N_{\rm re} \geq 0$ translates to $N_k \geq N_k^I$ for $w_{\rm re} > 1/3$ and $N_k \leq N_k^I$ for $w_{\rm re} < 1/3$.

\textit{II) BBN bound on $T_{\rm re}$.}---The concordance of cosmology requires that the Universe must enter the radiation-domination era before the onset of Big Bang Nucleosynthesis (BBN), i.e., $T_{\rm re} \ge T_{\rm BBN} \approx 10$ MeV \cite{deSalas:2015glj,Kawasaki:2000en,Cyburt:2015mya}. 
Using Eqs.~\eqref{eq:Tre_dep} and \eqref{eq:Nk_bound_firstform}, one can obtain $N_k$ as a function of $T_{\rm re}$ and compute the limit of $N_k$ from BBN bound by plugging $T_{\rm re} = T_{\rm BBN}$. For a given inflationary model with known $\rho_{\rm end}$ and $w_{\rm re}$, the limit is
\begin{align}
    N_k^{II} = \ln \left[\left(\frac{g_{*s}(T_0)}{g_{*s}(T_{\rm BBN})}\right)^{1/3} \left(\pi M_{\rm Pl} \sqrt{\frac{r \mathcal{A}_{s}}{2}}\right) \frac{a_0 T_0}{k} \, T_{\rm BBN}^{\frac{1-3w_{\rm re}}{3(1+w_{\rm re})}} \left[\frac{\pi^2 g_*(T_{\rm BBN})}{30 \rho_{\rm end}}\right]^{\frac{1}{3(1+w_{\rm re})}} \right] \, .
    \label{eq:Nk_bound_Tre}
\end{align}
When $w_{\rm re} > 1/3$, the constraint $T_{\rm re} \leq T_{\rm BBN}$ leads to $N_{\rm re} \leq N_{\rm re}^{II}$, while the bound becomes $N_{\rm re} \geq N_{\rm re}^{II}$ when $w_{\rm re} < 1/3$.

\textit{The combined bound on $N_k$.}---When combining both bounds \eqref{eq:Nk_bound_Nre} and \eqref{eq:Nk_bound_Tre}, the viable range of $N_k$ is shown in Fig.~\ref{fig:nk_bound} where $N_k^I \geq N_k \geq N_k^{II}$ for $w_{\rm re} > 1/3$ and $N_k^{II} \geq N_k \geq N_k^{I}$ for $w_{\rm re} < 1/3$. 
As we shall see in Sects.~\ref{sec:alpha_attractor} and \ref{sec:axion}, each inflationary setup leads to a fixed $w_{\rm re}$ during the reheating phase and is therefore bounded within a specific range of $N_k$.
Moreover, this bound leads to the constraint on the reheating temperature $T_{\rm reh}$.
As the end of inflation is determined by the model parameters, to fix $N_k$ means a specific $T_{\rm reh}$ must be chosen.


Note that in this section $N_k$ denotes the number of e-folds for a general mode $k$ as in the standard literature, while for the rest of this work, we will use $N_{\rm cmb} \equiv N_{k_{\rm cmb}}$ to specifically denote the number of e-folds at the CMB pivot scale.

\subsection{Gravitational Wave Background}
\label{sec:gw}

Primordial tensor fluctuations at scales  smaller than the CMB scale ($k>k_{\rm CMB}$) could as well be produced during the inflationary stage
and later stayed frozen upon their horizon exit.
Once the inflation has ended, these fluctuations keep reentering the horizon and evolve along the cosmic history,
leading to GWB that spans a broad range of frequencies \cite{Starobinsky:1979, Allen:1988, Sahni:1990, Sahni:2001, Caprini:2018,Maleknejad:2025clz, Boyle:2005se, Watanabe:2006qe,Figueroa:2019paj,Simakachorn:2022yjy}.
As GW freely streams after its production, such a GWB is encoded with direct information about the inflationary physics and the entire cosmic history of the Universe 
\cite{Boyle:2005se,Watanabe:2006qe,Boyle:2007zx,Jinno:2012xb,Saikawa:2018rcs,Figueroa:2019paj,Allahverdi:2020bys,Simakachorn:2022yjy,Konings:2024zvz}.

The present frequency of GWB can be related to the size of the perturbation at the time of horizon reentry, $2 \pi f = k /a_0 = H_k a_k/a_0$, written in terms of the temperature $T$ of the radiation-dominated universe as \cite{Sahni:1990, Sahni:2001, Figueroa:2019paj}
\begin{align}
    f(T) \simeq 2.7 ~{\rm mHz}  \left[ \frac{g_*(T)}{106.75} \right]^{1/2}
    \left[ \frac{106.75}{g_{*s}(T)} \right]^{1/3}
    \left( \frac{T}{10^5 \,{\rm GeV}} \right) \, .
    \label{eq:f_T_relation}
\end{align}
The higher the frequency of the GWB signal, the earlier the time it gets produced.
Therefore, the cosmic history can be traced by reading the GWB spectrum from low to high frequencies.
Today's energy-density spectrum of GWB for the $k^{\rm th}$-mode tensor perturbation is $\Omega_{\rm GW}(k) = (k^2 a_k^2 \mathcal{P}_t(k))/(24 H_0^2)$ where $k \, a_k$ is the physical Hubble size when the mode $k$ reenters the horizon, and $H_0$ is the Hubble scale today.
In this work, we consider a power-law tensor power spectrum $\mathcal{P}_t(f) = r \mathcal{A}_s (f/f_*)^{n_t}$
with the spectral index $n_t$ obtained from the consistency relation $n_t \simeq - r/8$, whose $r$ is evaluated for a given model at the CMB scale.
The frequency corresponding to CMB pivot scale ($k = 0.05 \text{ Mpc}^{-1}$) is $f_{*} \simeq 7.7 \times 10^{-17}$ Hz.
Although the actual $\mathcal{P}_t$ varies with the scale $k$ or equivalently the frequency $f$, we discuss in App.~\ref{app:power_law_validity} that the power-law approximation is sufficient to calculate the GW signal within the windows of future-planned GW observatories.

Taking into account the reheating phase of $w_{\rm re}$, the GWB spectrum can be written as \cite{Mishra:2021, Figueroa:2019paj}
\begin{align}
    h^2 \Omega_{\rm GW}(f) &= \frac{r \mathcal{A}_{s}}{24}  \left( \frac{f}{f_{*}} \right)^{n_{t}}  h^2 \Omega_{r,0} \times \begin{cases}
    1 ~ ~ &, ~ f_{\rm eq} < f \leq f_{\rm re} \, , \\
    \left({f}/{f_{\rm re}} \right)^{2\left(\frac{3 w_{\rm re}-1}{3 w_{\rm re}+1}\right)} ~ ~ &, ~ f_{\rm re} < f \leq f_{\rm end} \, ,
    \end{cases}
    \label{eq:Omega_re}
\end{align}
where $h^2\Omega_{r,0} \simeq 2.47 \times 10^{-5} $ is the normalized fraction of energy density in radiation today with $h \simeq 0.67$ \cite{ParticleDataGroup:2018ovx}, and $f_{\rm end} = f(\rho^{1/4}_{\rm end})$ and $f_{\rm re} = f(T_{\rm re})$ are GW frequencies at the end of inflation and when the reheating phase ends, respectively. 
As Eq.~\eqref{eq:f_T_relation} is valid only during radiation era, we obtain the frequency relation during the reheating phase from $f(a) = f_{\rm re}(a_{\rm re}/a)$; that is, $f_{\rm end} = f_{\rm re} \exp(N_{\rm re})$.
We omitted discussing the signature of the matter-domination era at very low frequencies $f < f_{\rm eq}$.

\textit{GWB as a probe of reheating phase.}---The GWB spectral shape gets imprinted distinctively by the cosmic history. In particular, the non-radiation era during the reheating phase (i.e., $w_{\rm re} \neq 1/3$) leads to a non-flat spectral shape, as shown in the second line of Eq.~\eqref{eq:Omega_re}.
For $w_{\rm re} < 1/3$ (e.g., the matter-domination era $w_{\rm re} = 0$), the GWB spectrum becomes red-tilted and suppressed at frequencies $f > f_{\rm re}$, while it can be blue-tilted and enhanced significantly for a stiff era ($w_{\rm re} > 1/3$). 
Such an enhanced GWB spectrum receives a strong attention as a potential signal at future GW observatories, including the ultimate run of LIGO-Virgo-KAGRA network, LISA, ET, CE, SKA, BBO, and DECIGO.
As we shall see, for the inflationary model in Sect.~\ref{sec:axion}, a stiff reheating phase could occur and lead to a detectable GWB signal, serving as a complementary probe to the CMB observations.

\textit{$\Delta N_{\rm eff}$ constraint.}---GWB contributes to the extra radiation component of the universe, which can jeopardize the concordance of cosmology if too much of it is present. 
In terms of the effective number of neutrino species $\Delta N_{\rm eff}$, Big Bang Nucleosynthesis (BBN) and CMB provide bounds on the extra radiation. We will focus on the recent CMB constraint $\Delta N_{\rm eff} \leq 0.17$ \cite{Calabrese:2025}.
The total GWB energy density produced before the onset of CMB must satisfy,
\begin{align}
\label{Neff-bound}
\int_{f_{\rm CMB}}^{f_{\rm end}} \frac{df}{f} \Omega_{\rm GW}(f) \leq \frac{7}{8} \left(\frac{4}{11}\right)^{4/3} \Omega_{r,0} \Delta N_{\rm eff} \lesssim 2.1 \times 10^{-6} \, ,
\end{align}
leading to a rough upper bound $\Omega_{\rm GW}h^2 \leq 9.52 \times 10^{-7}$.

With the theoretical and observational bounds discussed in this section, we now proceed to apply them to specific inflationary models.
We will investigate the viability of the $\alpha$-attractor and natural inflations within the $K$-inflation framework, examining their consistency with CMB observations and their potential for generating detectable GW signals.

\section{$\alpha$-Attractor T-Model}
\label{sec:alpha_attractor}

We consider the $\alpha$-attractor T-model potential \cite{Kallosh:2013a, Kallosh:2013b} with the $K$-inflation coupling
\begin{equation}
    V(\phi) = V_0 \tanh^n\left( \frac{\phi}{\sqrt{6\alpha}M_{\rm Pl}} \right) ~ ~ , ~ ~  G(\phi) = -e^{\beta \phi/M_{\rm Pl}} \, .
    \label{eq:model_alpha_attractor}
\end{equation}
In the standard inflationary scenario ($\beta=0$), this model with a smaller $n$ exhibits greater tension with the recent Planck-ACT joint data \cite{Iacconi:2025}.
Despite many attempts to alleviate this discrepancy by invoking $w_{\rm re} > 1/3$ during the reheating, $w_{\rm re}$ cannot be chosen freely, as discussed in Sect.~\ref{sec:reheating} and extensively in \cite{Iacconi:2025}. 
In particular, its value must be obtained by solving the inflaton dynamics after inflation ends, e.g., using Eq.~\eqref{eq:w_integral}.
While a larger $n$ can lead to a stiffer $w_{\rm re}$ and be more consistent with the observational data, a smaller $n$ still suffers from the tension.
In this section, we investigate how much the $K$-inflation framework can improve the consistency of the small-$n$ $\alpha$-attractor model, in particular $n=2$, to explain the recent CMB data.

\begin{figure*}[t!]
    \centering \includegraphics[width=.5\columnwidth]{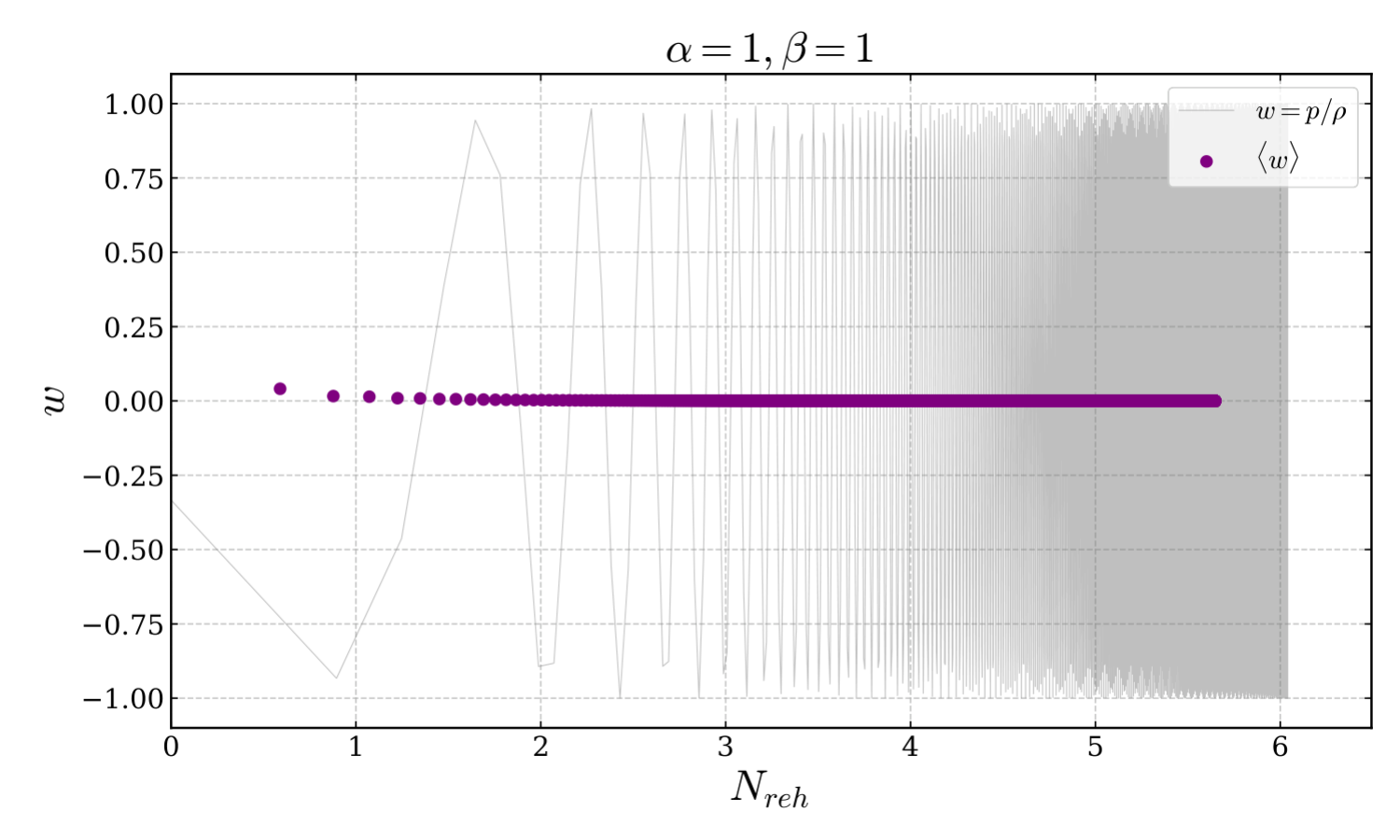}\hfill \includegraphics[width=.5\columnwidth]{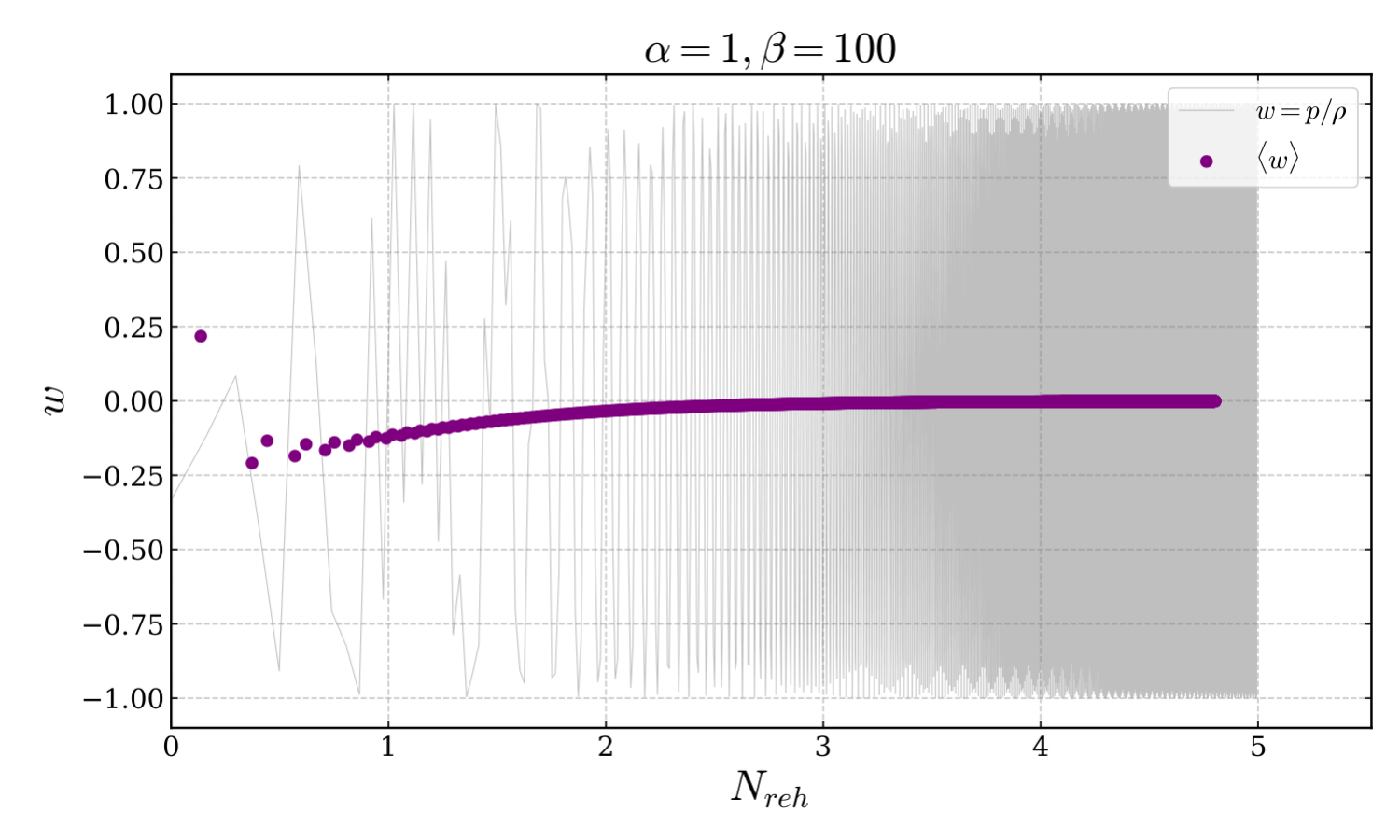}\\[-1.75em]
    \caption{The evolution of the instantaneous equation of state parameter in Eq.~\eqref{eq:w_re_instant} (gray lines) and its time-averaged value $w_{\rm re}$ in Eq.~\eqref{eq:w_integral} (purple dots) during the reheating phase for the T-model \eqref{eq:model_alpha_attractor} with $n=2$. With $\alpha = 1$, the left and right panels correspond to $\beta=1$ and $\beta=100$, respectively. In both cases, the reheating phase converges to the matter domination, $w_{\rm re} \to 0$.}
    \label{fig-TM-wre}
\end{figure*}

%

The effect of $K$-inflation is tunable through $\beta$ in Eq.~\eqref{eq:model_alpha_attractor}, i.e., the magnitude of the kinetic coupling modifies the friction of inflaton dynamics, shifting the scalar field values at both the pivot scale ($\phi_*$) and the end of inflation ($\phi_{\rm end}$). The shift in $\phi_*$ directly impacts the $n_s$ and $r$ predictions, while the change of $\phi_{\rm end}$ modifies the energy density at the end of inflation ($\rho_{\rm end}$) and the reheating scenario (required for a fixed $N_{\rm cmb}$).
Hence, for a set of $\{\alpha, \beta\}$, we solved the inflaton's dynamics deep into the reheating phase.
Illustrated in Fig. \ref{fig-TM-wre}, the instantaneous $w_{\rm re}$ in gray (from Eq.~\eqref{eq:w_re_instant}) is shown against $w_{\rm re}$ averaged over oscillation (from Eq.~\eqref{eq:w_integral}) in purple.
We see that $w_{\rm re}$ asymptotically approaches 0, which is the analytical expectation\footnote{Since $\tanh(\phi) \approx \phi$ near the minimum, the potential effectively becomes a power law with index $n$ at late times.} for the T-model with $n=2$.



\begin{figure*}[t!]
    \centering
    \includegraphics[width=0.49\textwidth]{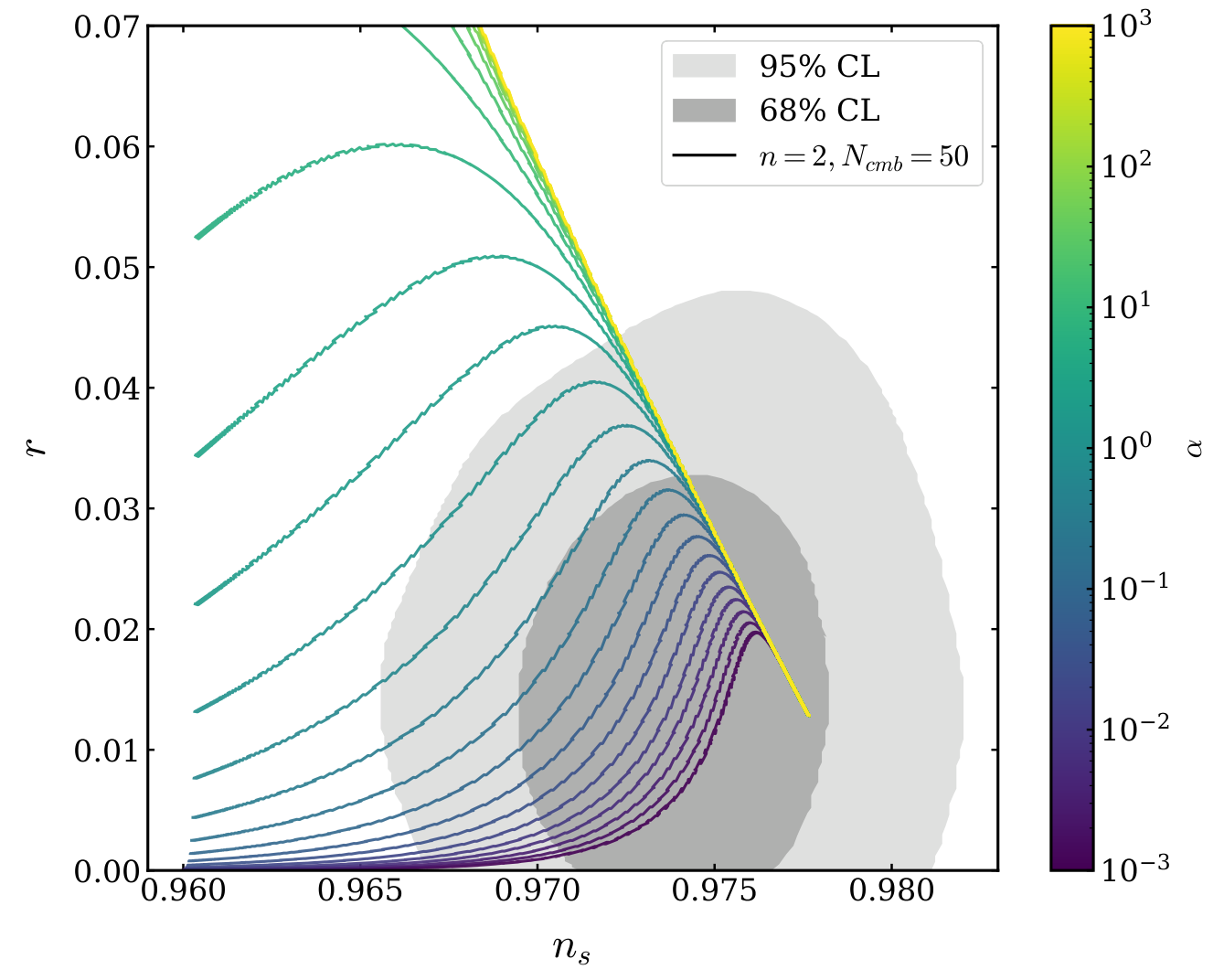}\hfill
    \includegraphics[width=0.49\textwidth]{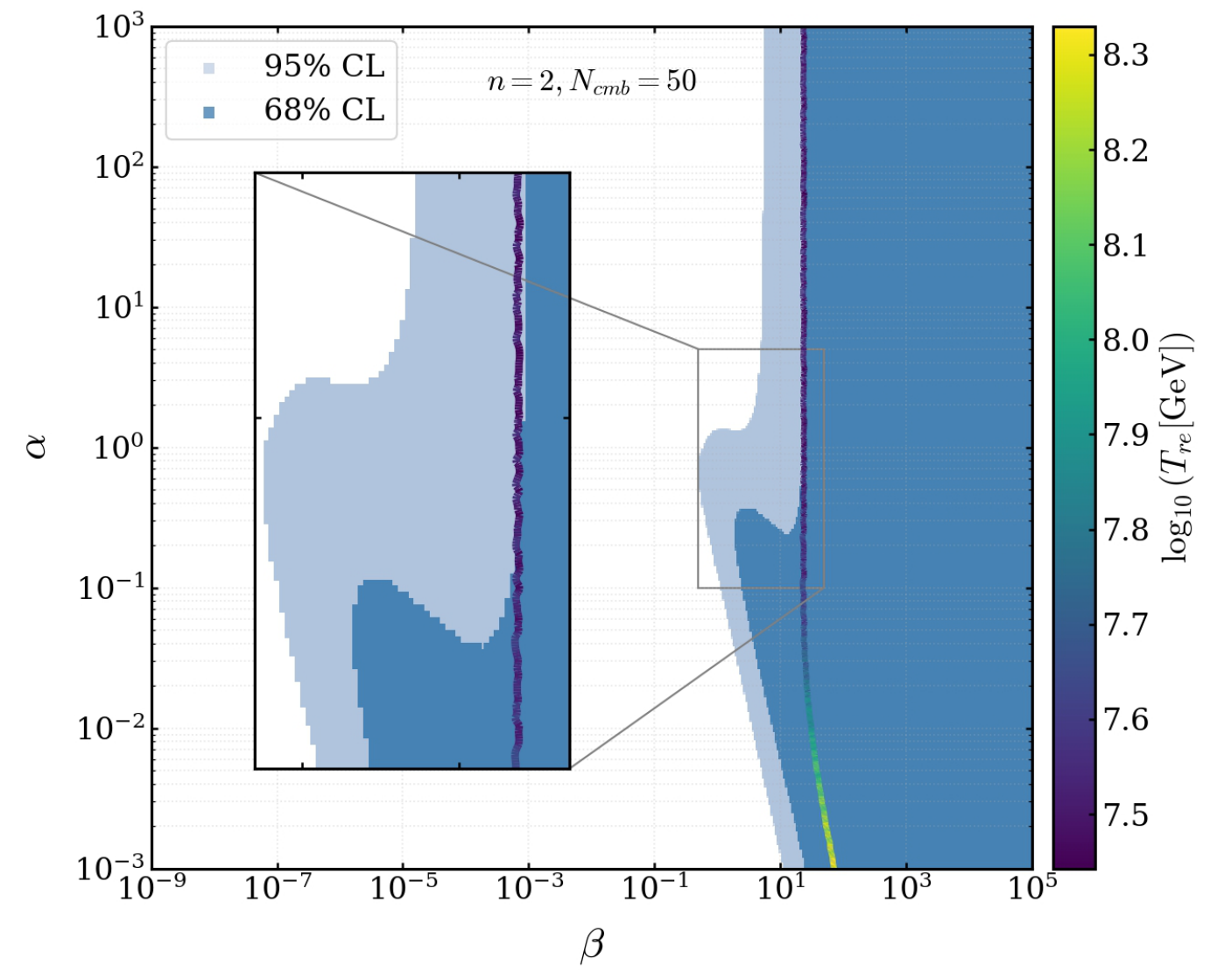}\\[-0.25em]
    \includegraphics[width=0.49\textwidth]{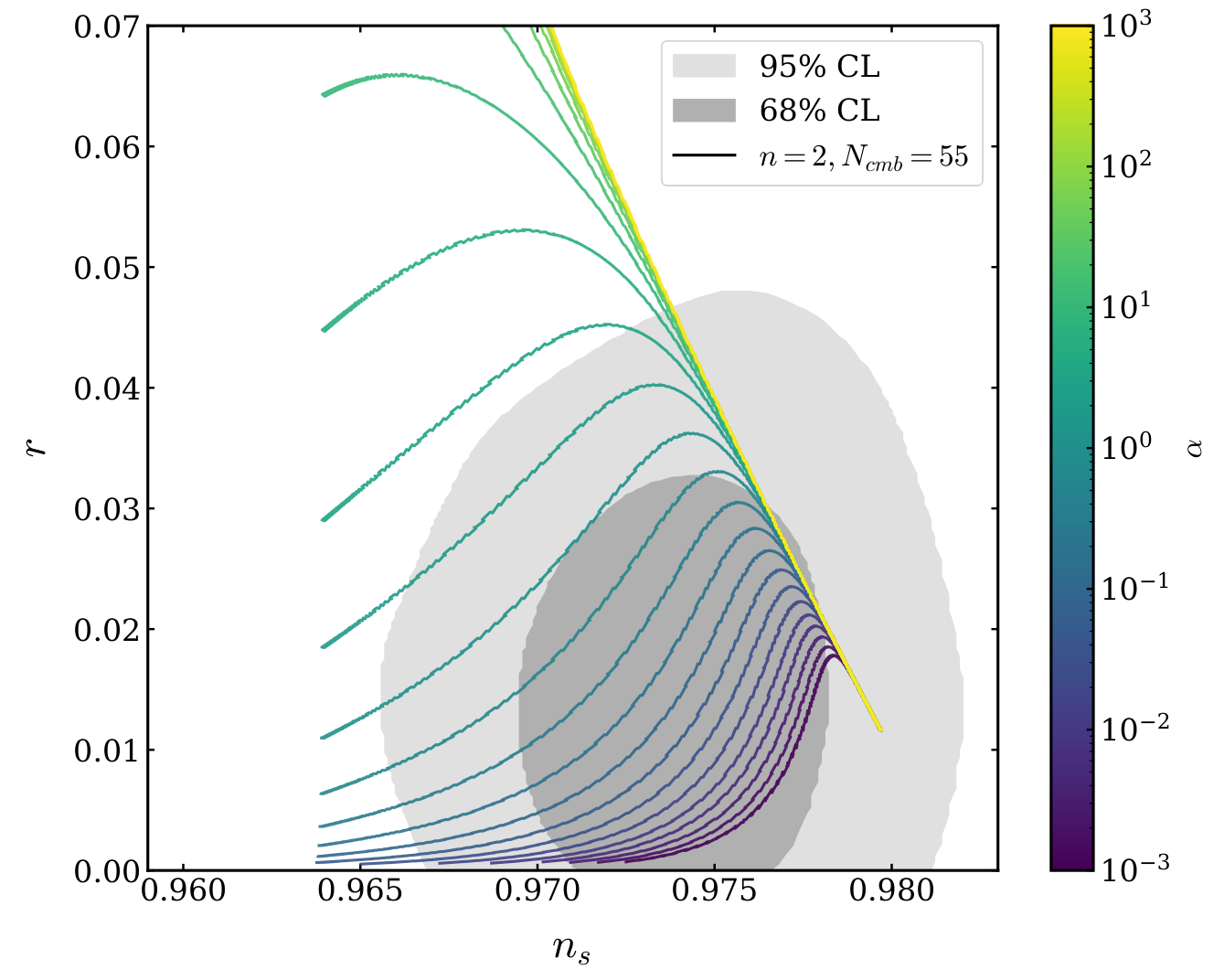}\hfill
    \includegraphics[width=0.49\textwidth]{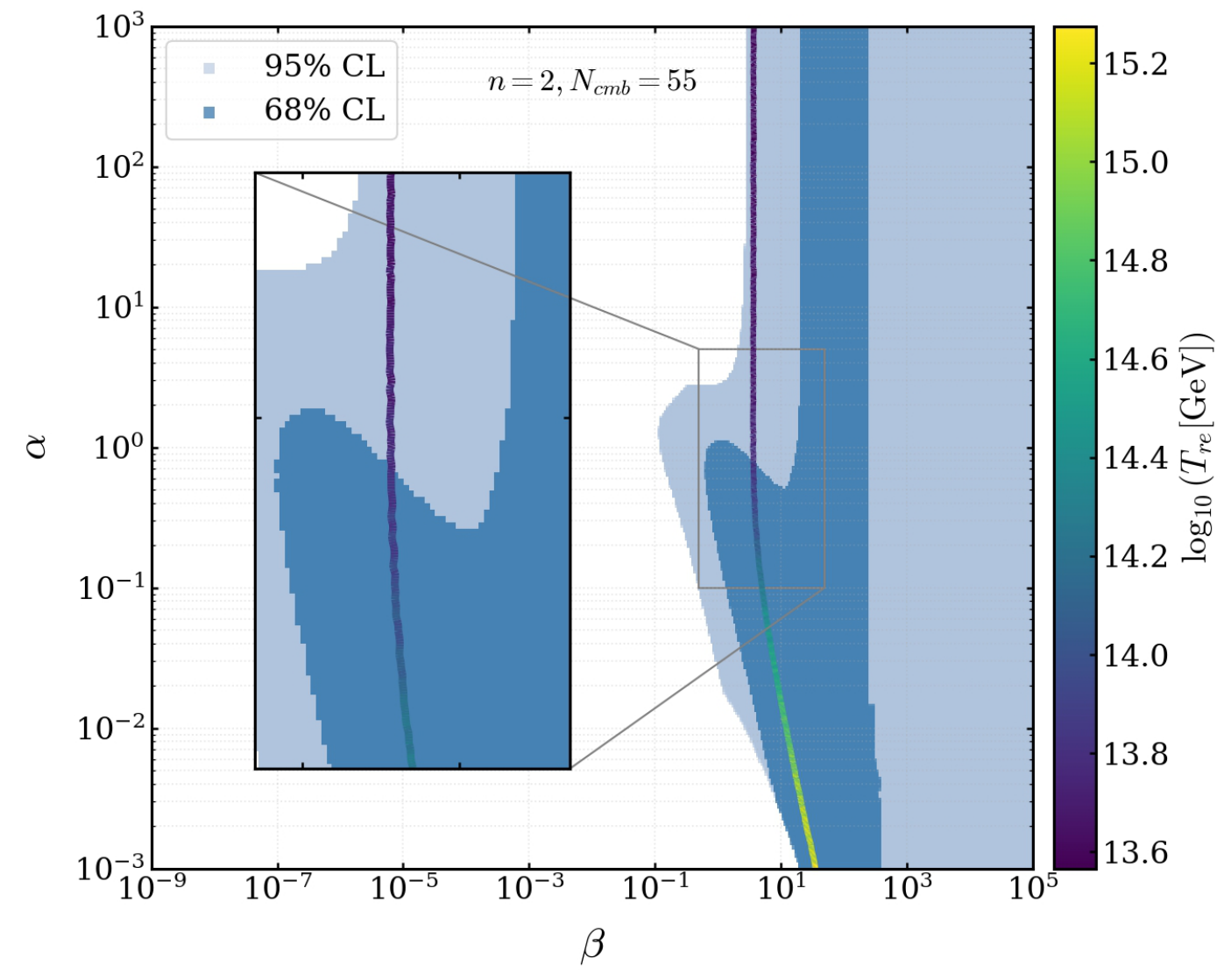}\\[-1.75em]
    \caption{\textit{Left:} Each colored line represents the $r-n_s$ predictions  for the T-model ($n=2$) for a fixed $\alpha$ with varying $\beta$ (from 0 at the smallest $n_s$ to $10^6$ at the largest $n_s$), compared against 68\% and 95\% C.L. posteriors from CMB observations (gray regions). \textit{Right:} The parameter spaces in the $(\alpha,\beta)$ plane, compatible with the $(n_s,r)$ posteriors, are shown in blue. The colored line corresponds to $\{\alpha,\beta\}$ that gives the best-fitted $n_s = 0.974$ \cite{Calabrese:2025} when the reheating temperature $T_{\rm re}$ is varied along the line.}
    \label{TM-back}
\end{figure*}

As discussed in Sect.~\ref{sec:reheating}, the number of inflationary efolds is bounded when the inflationary model is specified. In our case with $w_{\rm re} \approx 0$, the viable range is approximated to be $N_{\rm cmb} \in \left[ 42, 56 \right]$. 
The left column of Fig.~\ref{TM-back} shows the predicted $n_s$-$r$ (colored lines) at the CMB scale for $N_{\rm cmb} = 50$ and $55$ against the recent CMB constraints (gray regions) \cite{Louis:2025, Calabrese:2025}.
Each colored curve corresponds to a constant $\alpha$ and a varying $\beta$---ranging from 0 at the smallest $n_s$ to $10^6$ at the largest $n_s$. 
The left ends of all curves (i.e., $\beta = 0$) would converge to the standard inflationary prediction, where a decreasing $\alpha$ leads to a smaller $r$; this is not compatible with the observational data.
Conversely, the $K$-inflation framework with a larger $\beta$ can improve the model's consistency with the recent observations for $\alpha \in [10^{-3}, 10^{3}]$, especially for a large enough $\beta$, $r$ gets enhanced without strongly increasing $n_s$.
A large $\beta$ creates a strong friction term in Eq.~\eqref{eq:slow_roll_Friedmann2} which fixes $\phi$ in place during inflation. In this regime, the kinetic coupling $G(\phi)$ dictates the values of $n_s$ and $r$, regardless of $\alpha$, and prevents the suppression of $r$ for small $\alpha$, which happens for the standard inflation scenario ($G(\phi) = 0$).

Interested in the viability of the model parameter space, we mapped the ($n_s-r$)-posterior distributions into regions in the $\alpha-\beta$ plane, shown in blue in the right panel of Fig.~\ref{TM-back}.
E.g., each point in the dark blue region means the corresponding $\{\alpha,\beta\}$ leads to $n_s$ and $r$ predictions within the 68\% C.L. region (dark gray in the left panel of Fig.~\ref{TM-back}).
A more sophisticated analysis could be done via Bayesian inference to obtain the posterior distribution directly in the $\alpha-\beta$ plane; however, this is beyond the scope of this work.
Furthermore, the colored line  corresponds to $\alpha$ and $\beta$ that leads to the best-fitted $n_s = 0.974$ and $r\leq 0.032$ \cite{Calabrese:2025}, 
when the reheating temperature $T_{\rm reh}$ varies along the line.
This variation occurs because, when $N_{k}$ is fixed, the smaller $T_{\rm re}$ requires a smaller $\rho_{\rm end}$, which corresponds to a fixed set of $\{\alpha, \beta\}$.
For the best-fitted $n_s$, the left panel of Fig.~\ref{TM-back} show that the requiring reheating temperature for the 68\% C.L. region satisfies the reheating bound discussing in Sect.~\ref{sec:reheating}, i.e., $T_{\rm reh} \gtrsim 3 \times 10^7$~GeV for $N_{\rm cmb}=50$ and $T_{\rm reh} \gtrsim 6 \times 10^{13}$~GeV for $N_{\rm cmb}=55$.
For the two cases considered here, we see that this model can explain the best-fitted result with $\beta \sim \mathcal{O}(10)$.


\begin{figure*}[t!]
    \centering
    \includegraphics[width=0.49\textwidth]{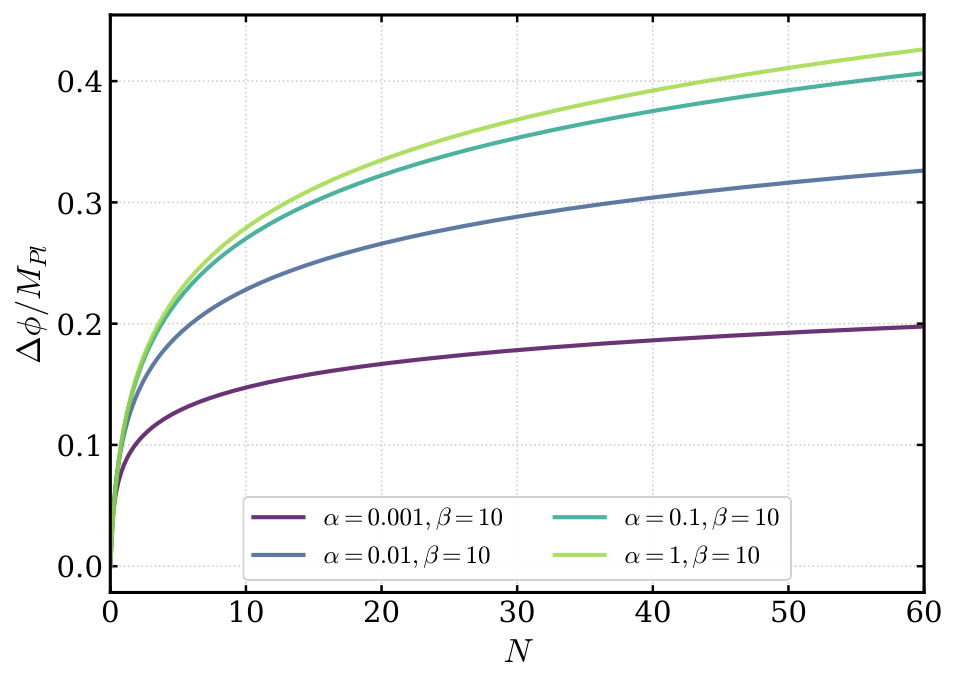}\hfill
    \includegraphics[width=0.485\textwidth]{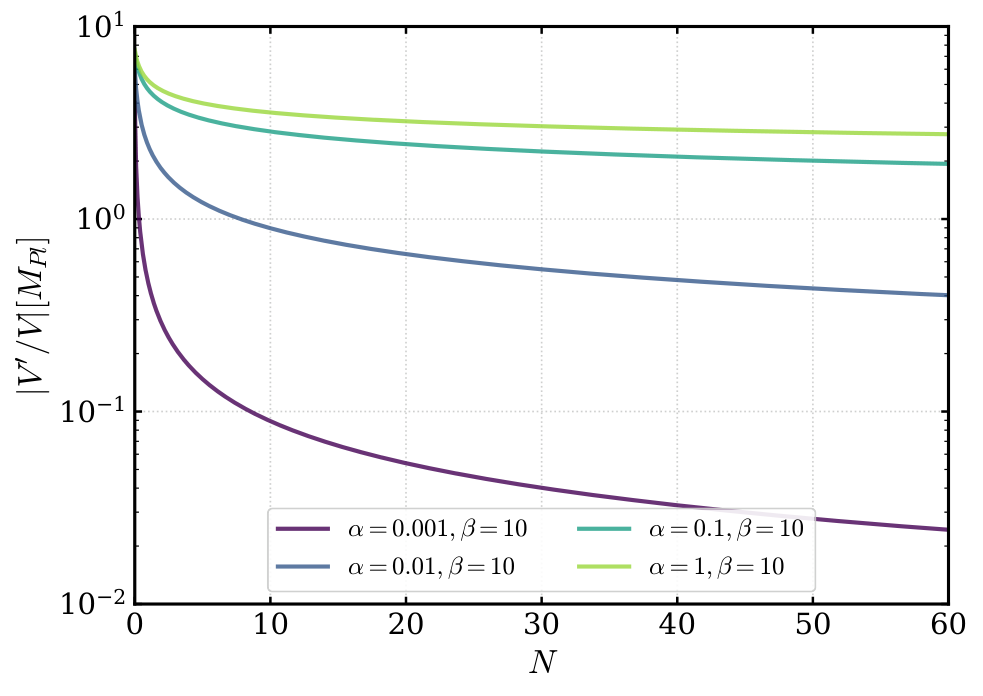}\\[-1.75em]
    \caption{Assuming $\alpha$-attractor T-model ($n=2$) in $K$-inflation, i.e., Eq.~\eqref{eq:model_alpha_attractor}, the left and right panels show the inflaton excursion $\Delta \phi/M_{\rm Pl}$ and $|V'/V|$, which are subjected to the Swampland criteria in Eqs.~\eqref{sdc-conj} and \eqref{eq:ds_conj}, respectively.
    Focusing on $\beta = 10$ that explains the best-fit result, but over a range of $\alpha$, their evolutions are plotted as a function of the e-folding number before inflation ends.}
\label{fig:Tmodel_Swampland}
\end{figure*}


\textit{Swampland criteria.}---Fig.~\ref{fig:Tmodel_Swampland} presents the inflaton excursion $\Delta \phi$ and $|V'/V|$, which are subjected to the distance and de Sitter conjectures in Eqs.~\eqref{sdc-conj} and \eqref{eq:ds_conj}, respectively. We observe that the $\alpha$-attractor T-model ($n=2$) in K-inflation can explain the CMB results while well complying with these Swampland criteria.
Notably, both $\Delta \phi$ and $|V'/V|$ saturate when $\alpha$ increases, ensuring that the criteria are satisfied in the large-$\alpha$ limit. However, for small $\alpha \lesssim 10^{-3}$, the potential gradient $|V'/V|$ drops significantly and can violate the de Sitter conjecture. While K-inflation is compatible with CMB data across a wide range of $\alpha$, a strict adherence to the Swampland criteria favors $\alpha \gtrsim \mathcal{O}(10^{-3})$.



\textit{GW signature.}---Since the reheating phase proceeds with the matter domination ($w_{\rm re} \approx 0$), its signature in the  GWB spectrum gets suppressed (see Eq.~\eqref{eq:Omega_re}) well below the sensitivities of even future observatories like BBO and DECIGO.
While $K$-inflation successfully rescues the T-model in the context of CMB observations, this model remains \emph{invisible} to future GW observatories.

%

\section{Natural Inflation}
\label{sec:axion}
Consider an extended natural inflationary potential \cite{Kitabayashi:2023vfe,Zhang:2018wbn}, modified by the $K$-inflation coupling,
\begin{align}\label{eq:Natural}
    V(\phi) = \Lambda^4 \left[ 1 \pm \cos\left( \frac{\phi}{\alpha M_{\rm Pl}} \right) \right]^n \, , \quad G(\phi) = -\left(\frac{\phi}{M_{\rm Pl}}\right)^{\beta} \, .
\end{align}
The standard single-field slow-roll inflation ($G(\phi) = 0$) with the natural potential~\eqref{eq:Natural} predicts an excessively red-tilted $n_s$, which is unfitted to Planck and ACT data, unless $n < 1$ \cite{Zhang:2018wbn}.
We will later see that $K$-inflation can significantly improve the consistency with observations, even for $n > 1$.
As an example, we shall focus on the potential with a \emph{plus} sign in the main text, while we show in App.~\ref{app:natural_inflation_minus} that the natural inflation with a negative cosine can also be consistent with the CMB observations with the help of $K$-inflation, although $\beta \sim 1-100$ is required.


Although increasing $n$ in Eq.~\eqref{eq:Natural} exacerbates the red-tilted $n_s$'s problem \cite{Zhang:2018wbn}, a larger $n$ leads to a stiffer equation of state during the reheating phase ($w_{\rm re} > 1/3$), which could leave a signature in GWB; see Sect.~\ref{sec:gw}.
We will see that the $K$-inflation could help ease this tension, while this scenario remains testable via GW. 
As illustrated in Fig. \ref{fig-AX-wre} for $n=2,~3,~4$ and $5$, the averaged equation of state $w_{\rm re}$ converges to $1/3,~1/2,~3/5$, and $2/3$, respectively.\footnote{Around the minimum of the natural potential, the potential \eqref{eq:Natural} effectively scales as $V \propto \phi^{2n}$. Using the scalar virial theorem, the averaged equation of state is $w_{\rm re} = (n-1)/(n+1)$, which agrees well with our numerical results.}
Evidently from Fig.~\ref{fig-AX-wre}, $w_{\rm re}$ does not reach its asymptotic value instantaneously but undergoes a relaxation phase lasting a few efolds.
We have checked that the duration of the reheating stage using the fully numerical $w_{\rm re}$ yields a similar result from using the asymptotic $w_{\rm re}$ up to $\lesssim\mathcal{O}(1)$ efolds.
Given that this correction is subdominant compared to observational uncertainties, $w_{\rm re}$ can be approximated by its asymptotic value throughout the reheating phase for computational efficiency.
Since we are interested in the GW signature, we will focus on quartic ($n=4$) and quintic ($n=5$) cases where $w_{\rm re}$ is stiff enough to generate detectable GW signals.\footnote{For $n=3$, we also checked that the associated GWB can be detected at CE for $N \gtrsim 60.55$.}

}


\begin{figure*}[t!]
    \centering
        \includegraphics[width=0.49\textwidth]{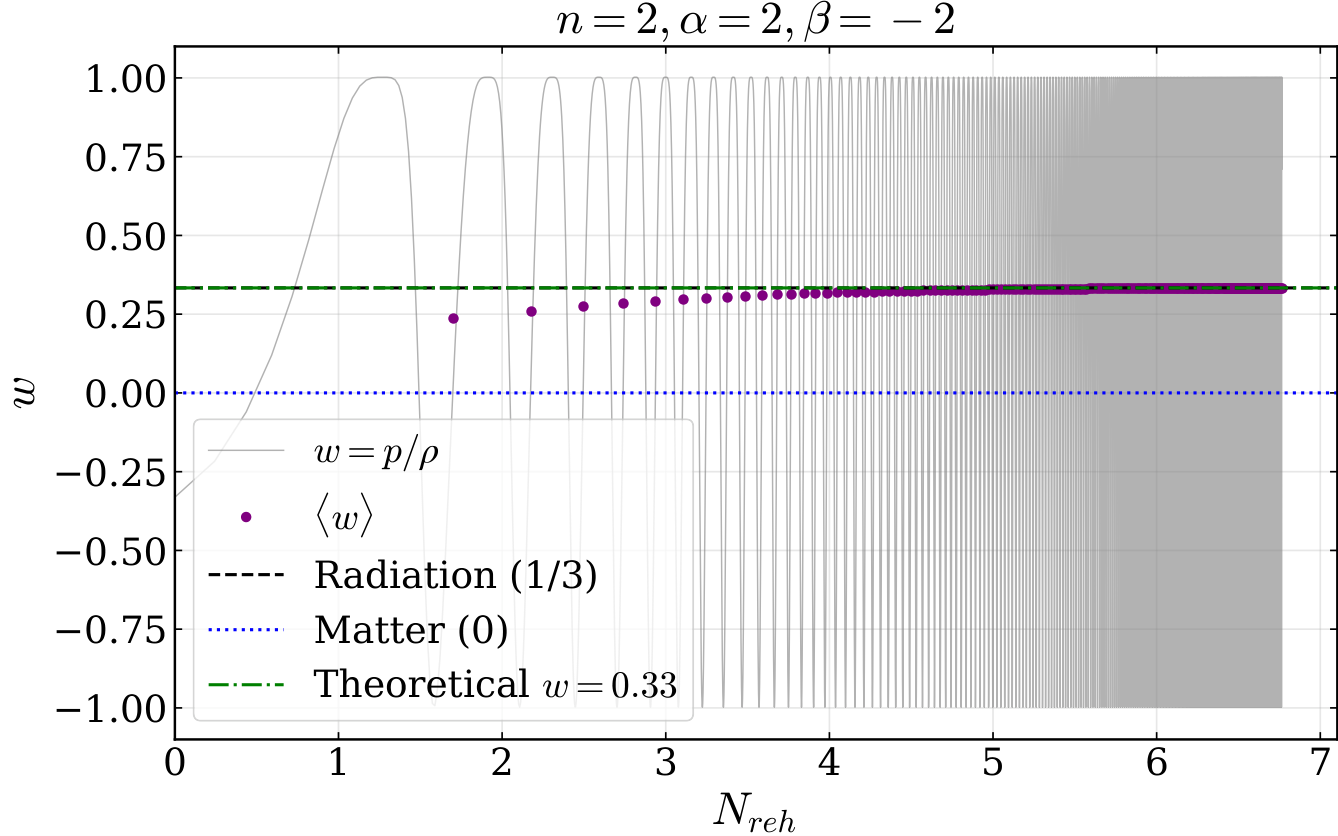}\hfill
        \includegraphics[width=0.49\textwidth]{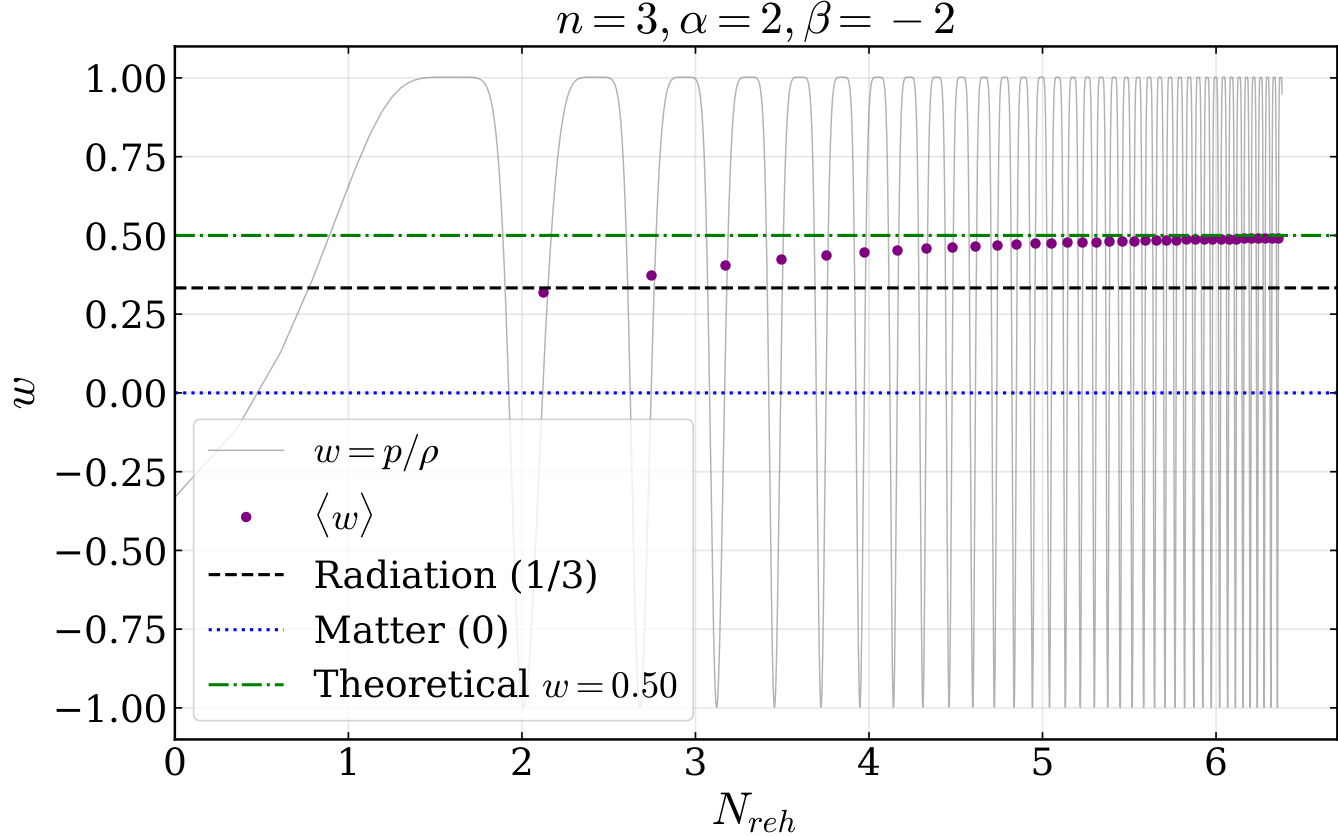}\\[0.25em]
        \includegraphics[width=0.49\textwidth]{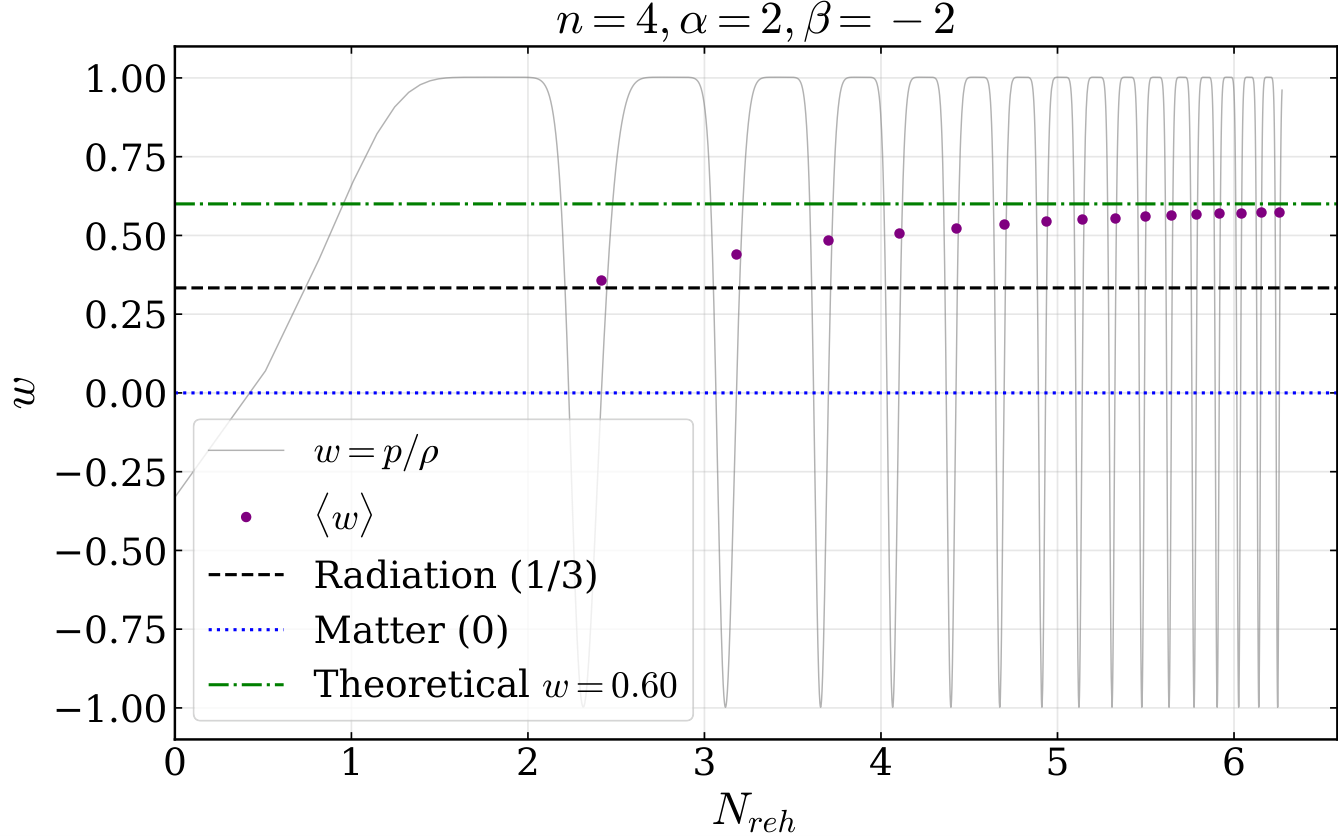}\hfill
        \includegraphics[width=0.49\textwidth]{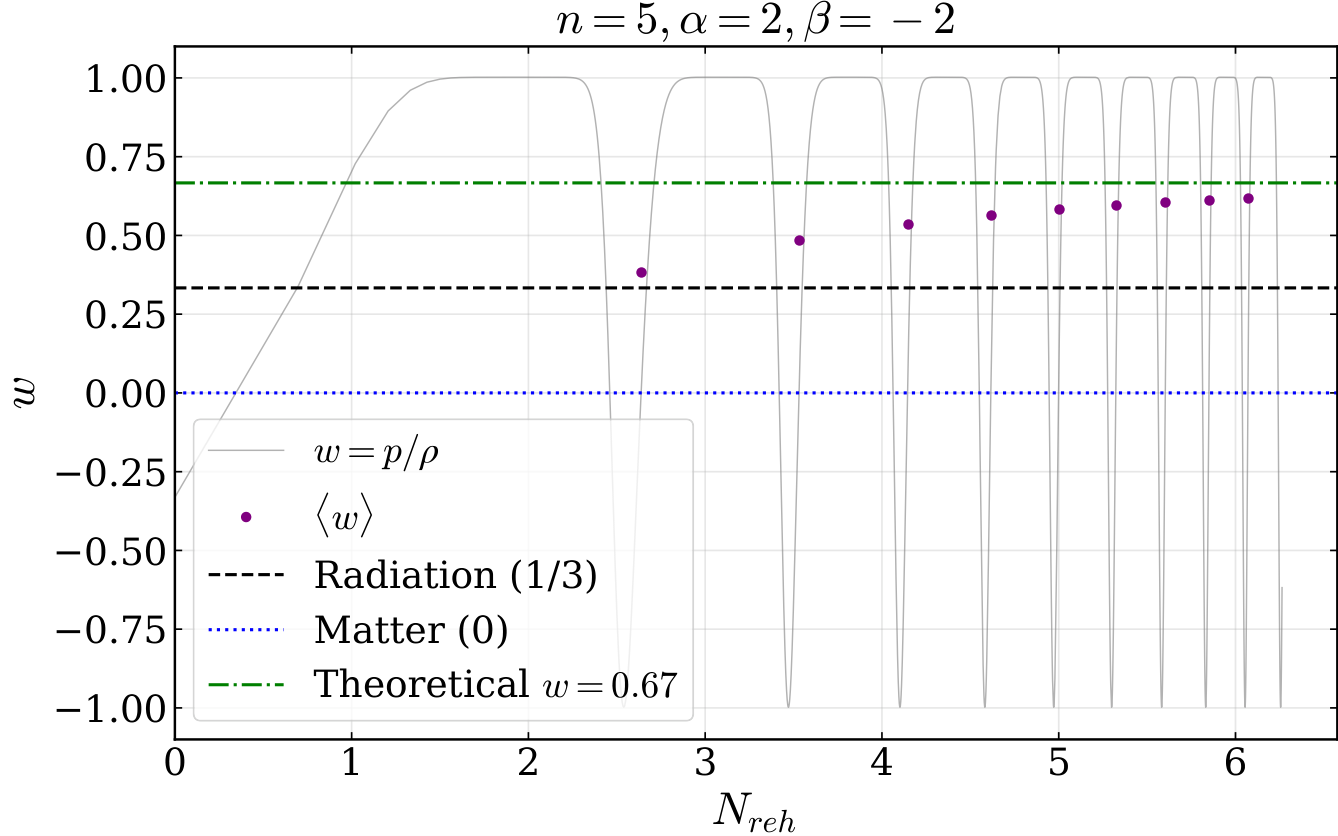}\\[-1.5em]
    \caption{The evolution of the equation of state parameter $w=P/\rho$ (gray lines) and its time-averaged value $w_{\rm re}$ (purple dots) during the reheating phase for the natural potential with $n=2,3,4$ and 5. The green dashed line represents the asymptote when}
    \label{fig-AX-wre}
\end{figure*}


\subsection{Quartic case ($n=4$)}

\begin{figure}[p!]
    \centering
    \includegraphics[width=0.495\textwidth]{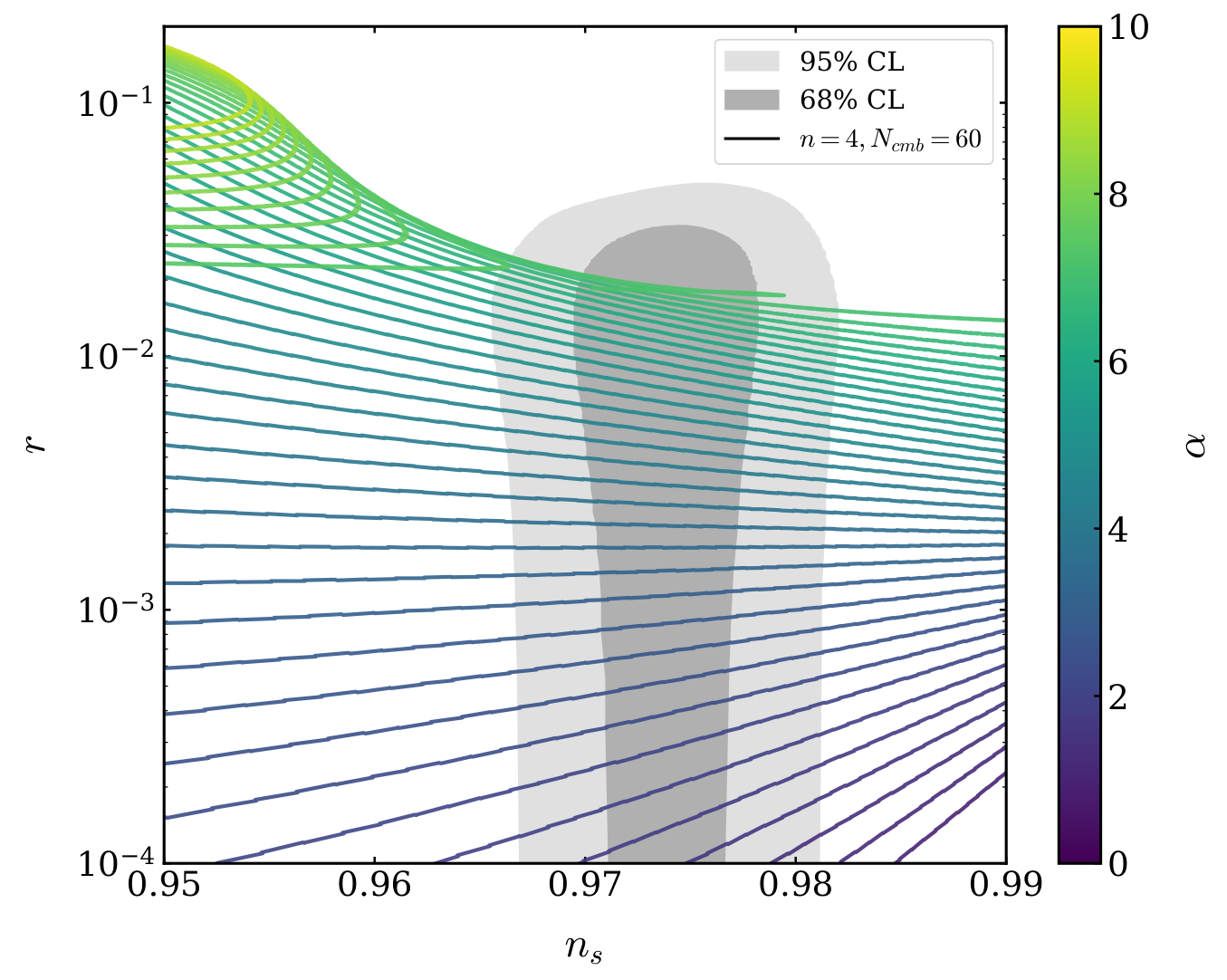}\hfill
    \includegraphics[width=0.47\textwidth]{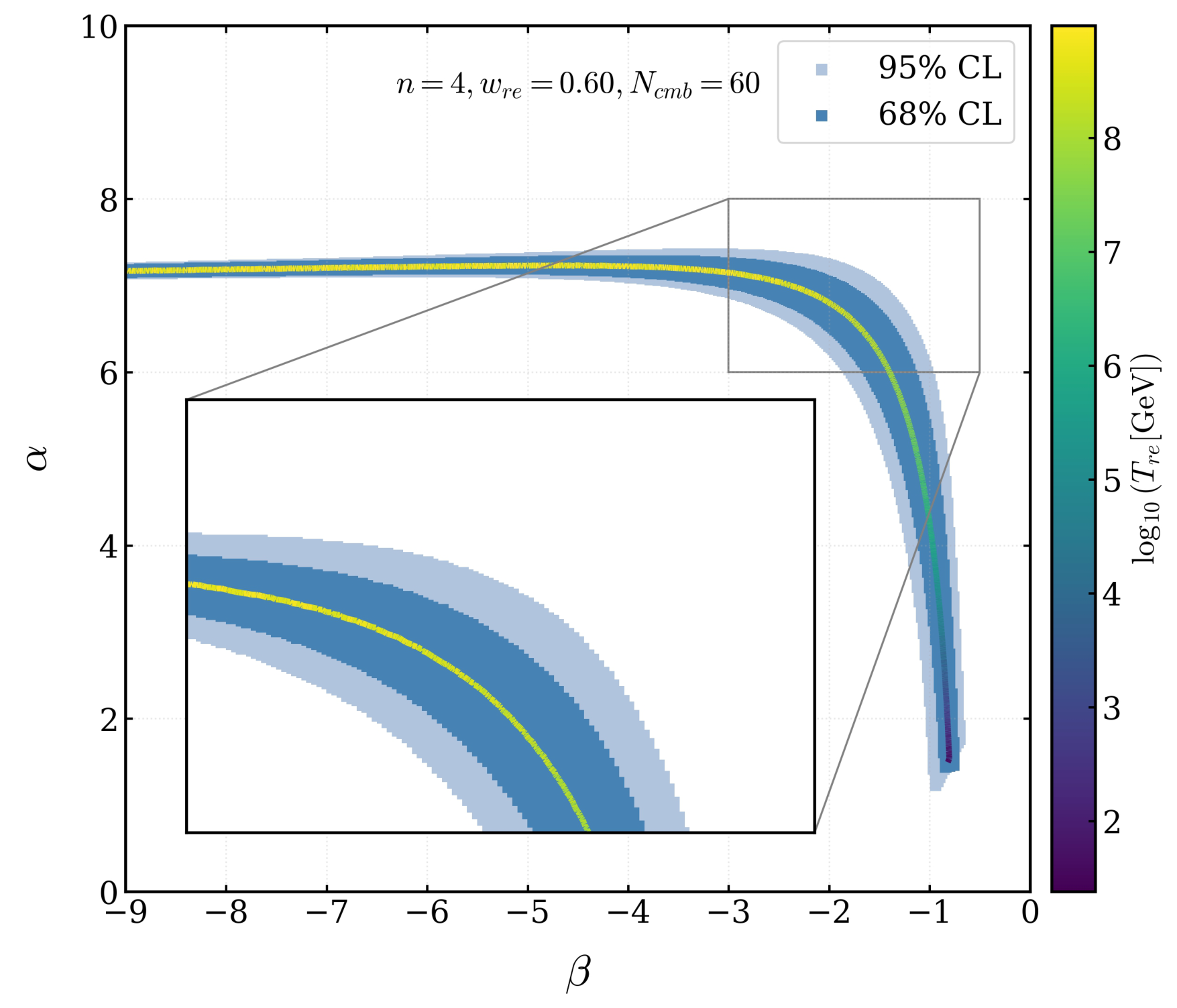}\\
    \includegraphics[width=0.495\textwidth]{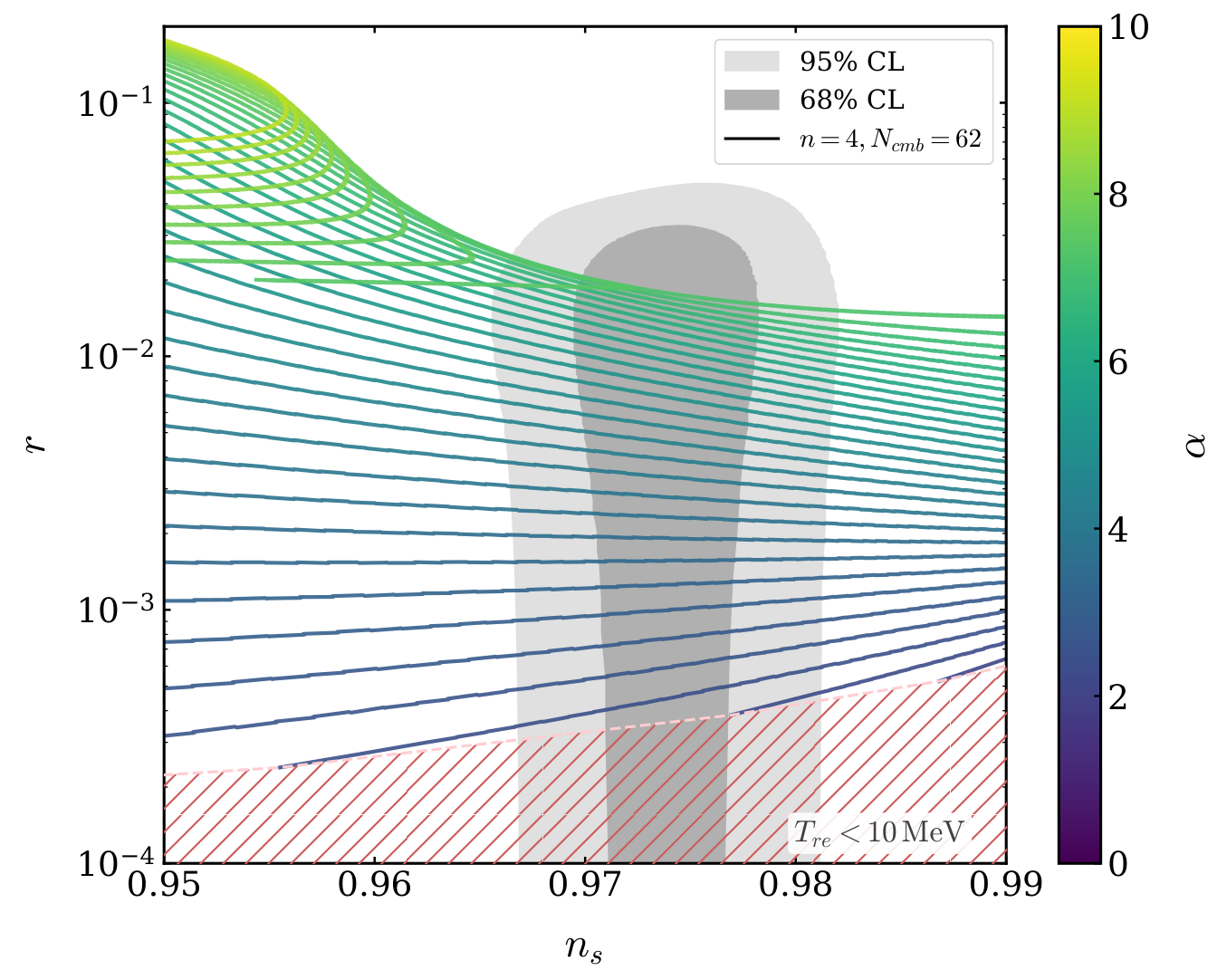}\hfill
    \includegraphics[width=0.48\textwidth]{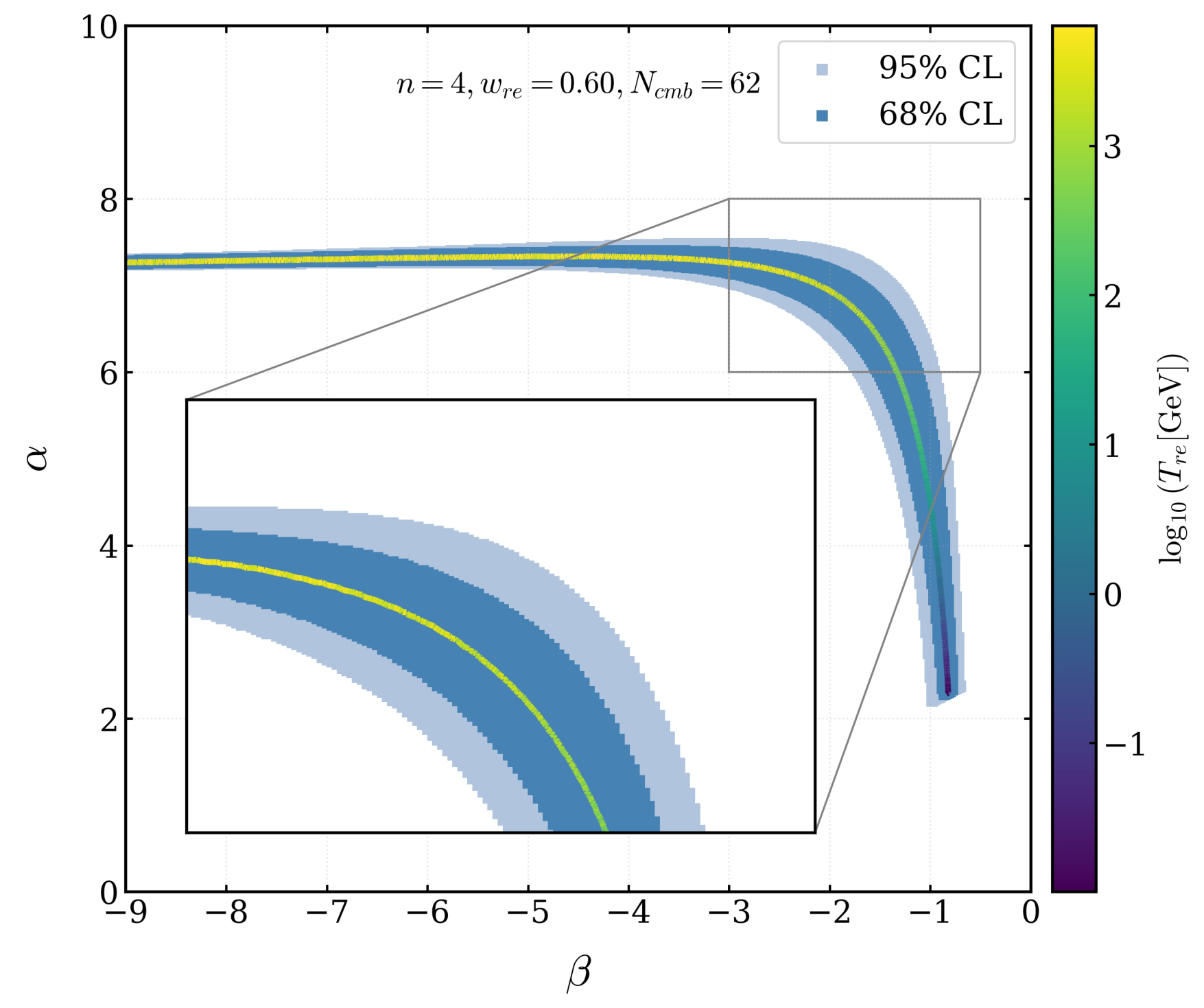}\\
    \includegraphics[width=0.495\textwidth]{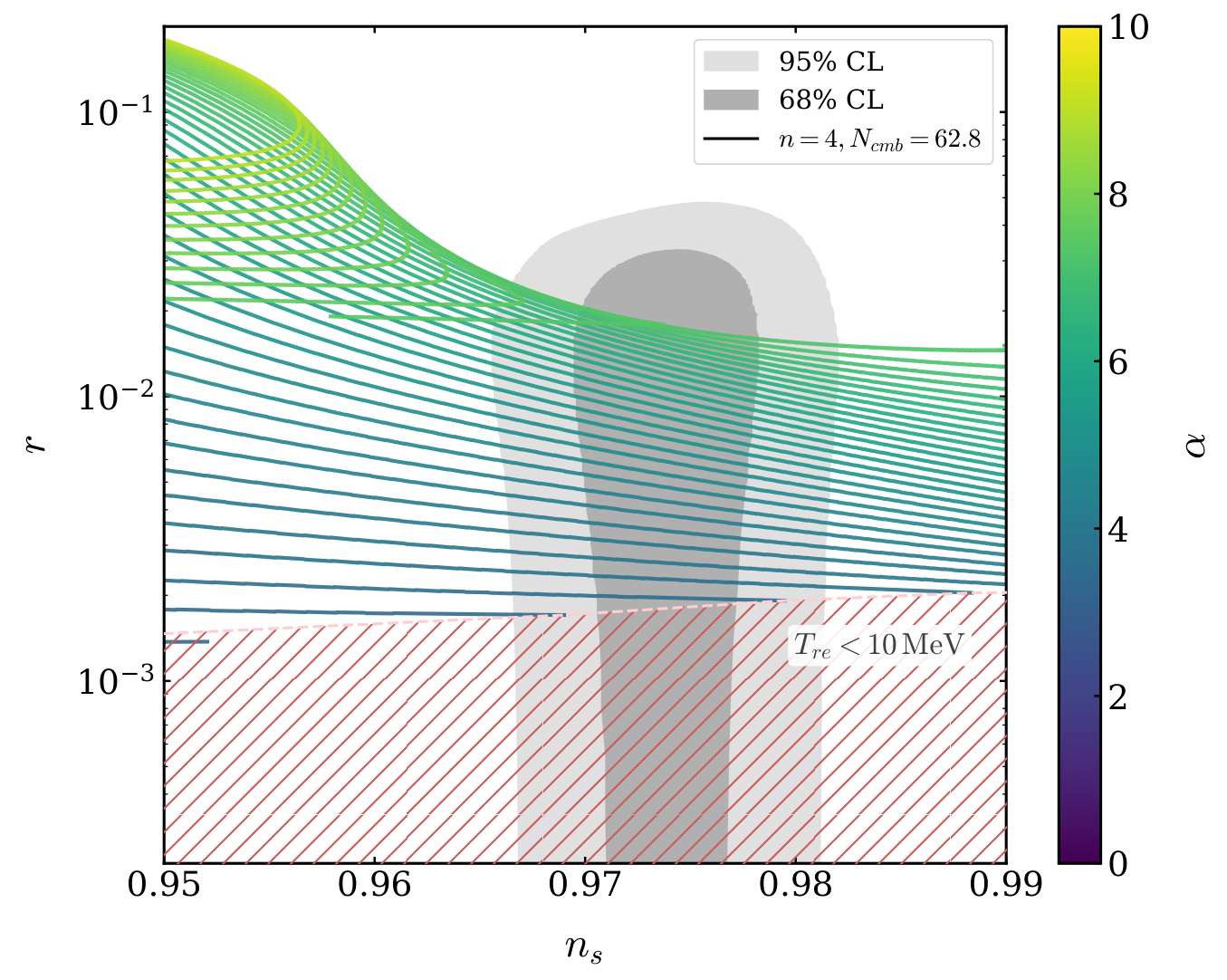}\hfill
    \includegraphics[width=0.48\textwidth]{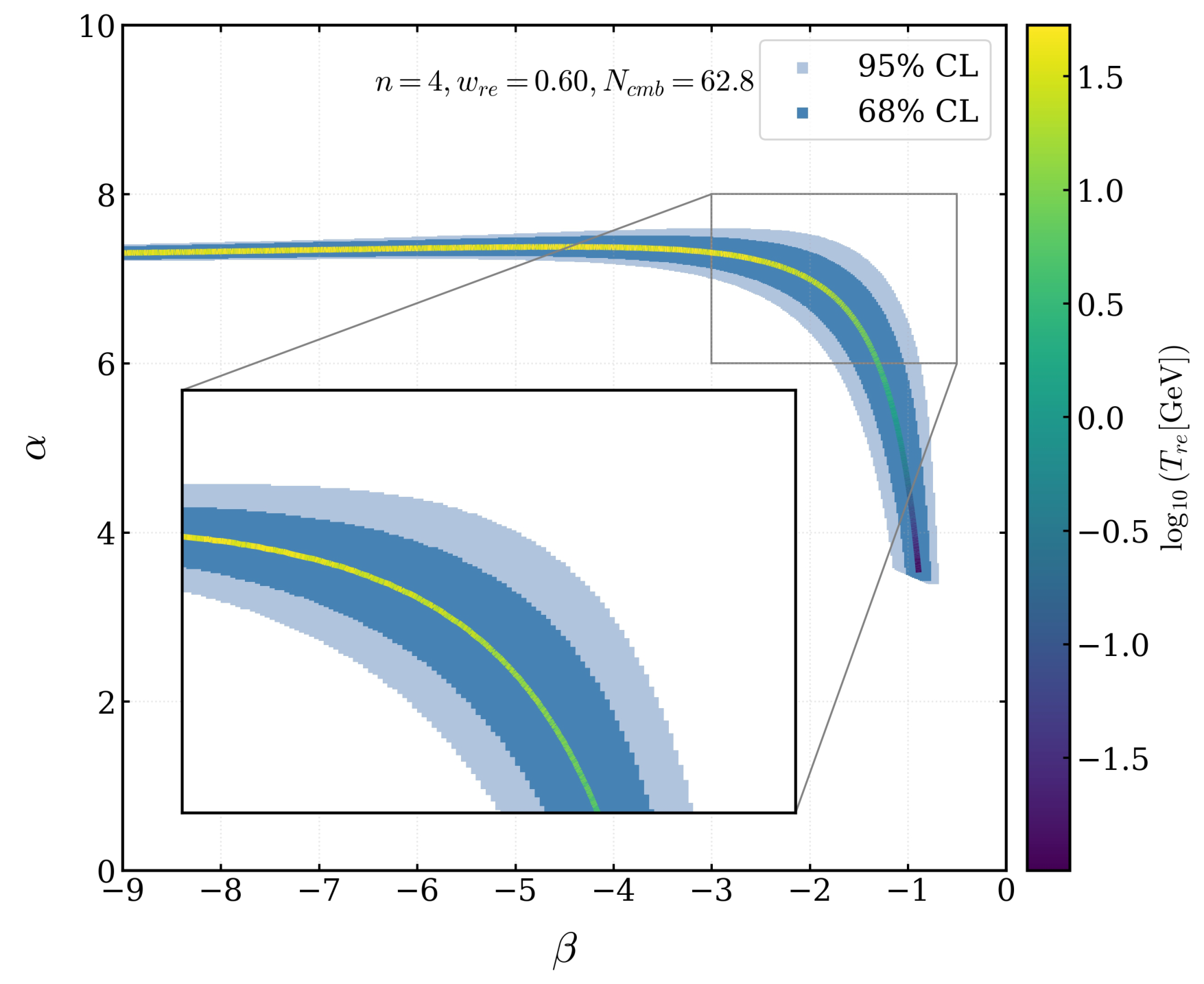}\\[-1em]
        \caption{\textit{Left:} Each colored line represents the $(r,n_s)$ prediction for the natural inflation ($n=4$) with a fixed $\alpha$ value, compared against 68\% and 95\% C.L. posteriors from CMB observations (gray regions). 
        For each line, $\beta$ varies from $-9$ at the largest $n_s$ to $0.1$ at the smallest $n_s$, except those with turnover points whose $\beta$ decreases with a smaller $r$.  
        The red hatched region is excluded by the BBN bound $T_{\rm reh} \lesssim 10 ~ {\rm MeV}$. \textit{Right:} The parameter spaces in the $(\alpha,\beta)$ plane, compatible with the $(n_s,r)$ posteriors, are shown in blue. The colored line corresponds to $\{\alpha,\beta\}$ that gives the best-fitted $n_s = 0.974$ \cite{Calabrese:2025} when the reheating temperature $T_{\rm re}$ is varied along the line.}
        \label{fig-Ax-Back4}
\end{figure}

\begin{figure*}[t!]
    \centering
        \includegraphics[width=0.32\textwidth]{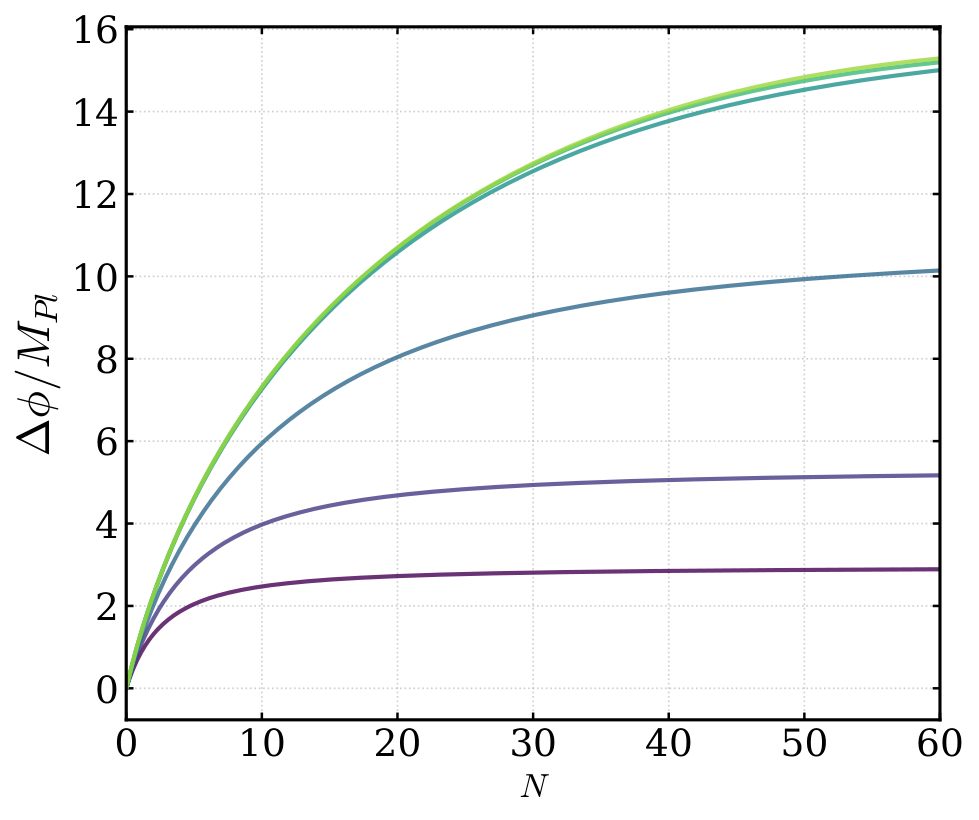}\hfill
        \includegraphics[width=0.333\textwidth]{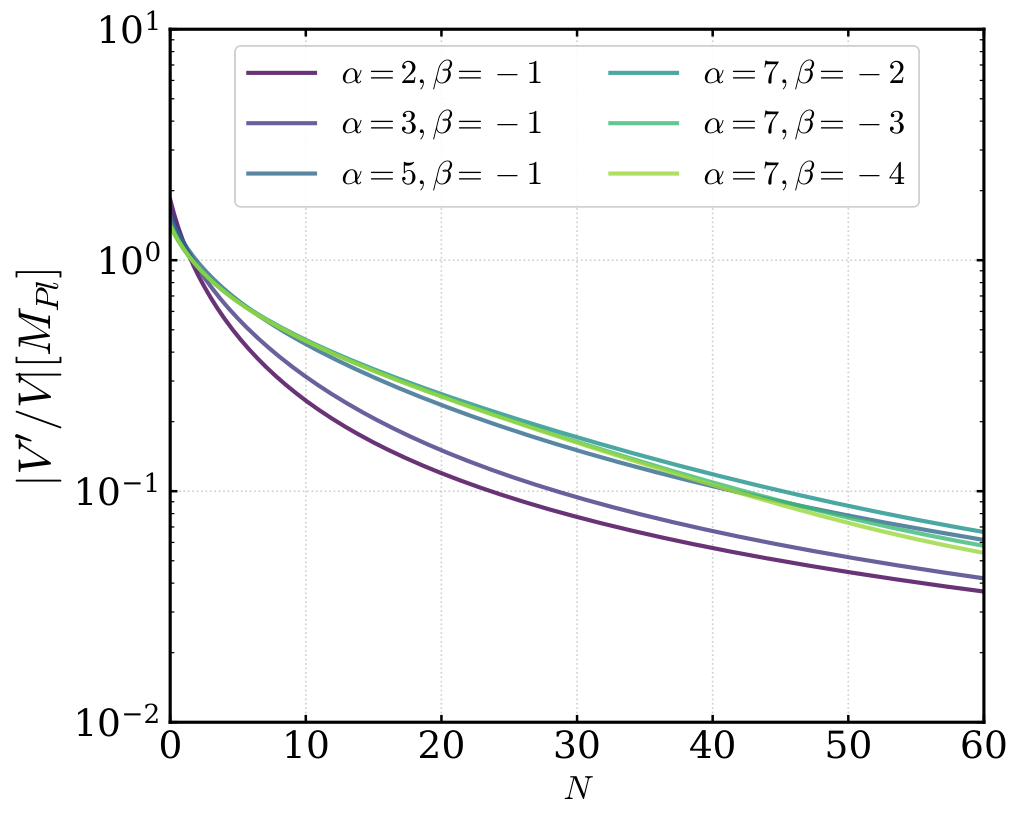}\hfill
        \includegraphics[width=0.333\textwidth]{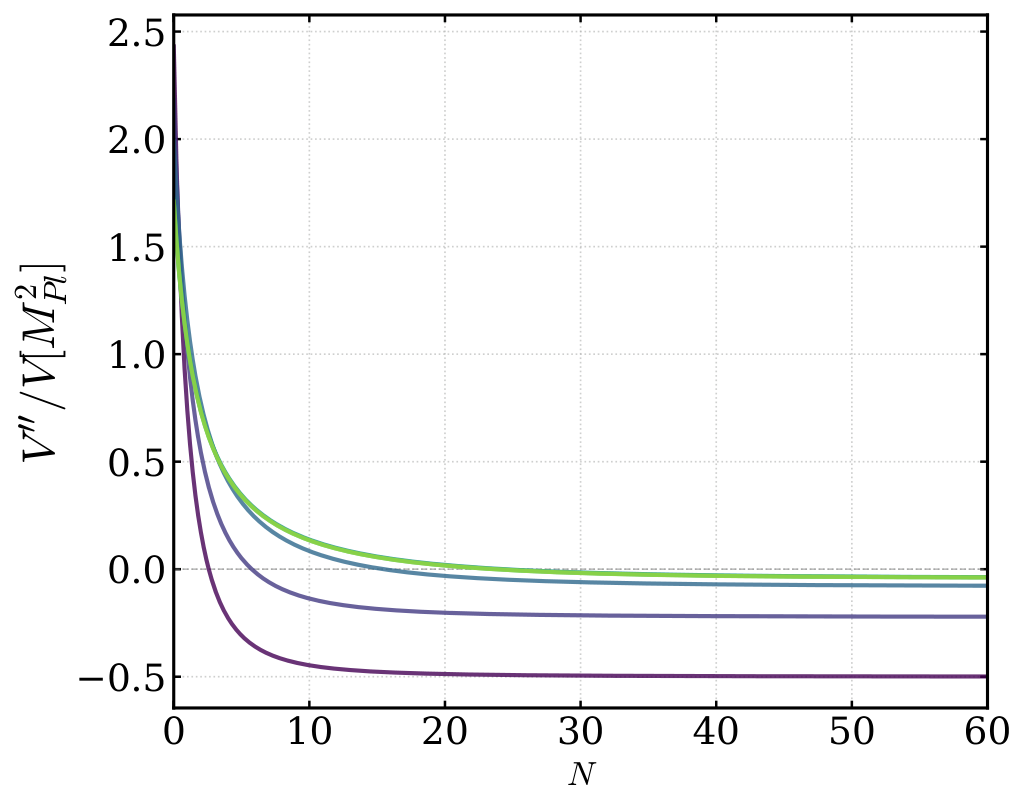}\\[-1.75em]
    \caption{
    Assuming $K$-inflation with ($n=4$) natural potential \eqref{eq:Natural}, the left, middle, and right panels show $\Delta \phi/M_{\rm Pl}$, $|V'/V|$, and $V''/V$ during inflation, respectively, which are subjected to the Swampland criteria in Sect.~\ref{sec:swampland}.
    From the left plot, the Swampland distance conjecture \eqref{sdc-conj} is strongly violated, i.e., $\Delta \phi \gtrsim \mathcal{O}(10 M_{\rm Pl})$, for $\alpha \gtrsim 5$.
    Regarding the de Sitter conjecture \eqref{eq:ds_conj}, $|V'/V|$ does not strongly satisfied the first condition, while $V''/V$ can be consistent for $\alpha \lesssim 5$.
    The legend is common to all panels.}
    \label{fig:Natural_Swampland}
\end{figure*}

The colored lines in the left column of Fig.~\ref{fig-Ax-Back4} shows the predicted $n_s$ and $r$ for the $K$-inflation with ($n=4$) natural potential where the number of efolds are 60, 62, and 62.8, chosen from the allowed range $N_{\rm cmb} \in [56, 62.8]$; see Eqs.~\eqref{eq:Nk_bound_Nre} and \eqref{eq:Nk_bound_Tre} in Sect.~\ref{sec:reheating}.
These colored curves assume fixed values of $\alpha$ with $\beta \in [-9,0.1]$ varying from smaller-to-larger $n_s$.
We also show the red hatched region, which is excluded by the BBN bound $T_{\rm re} \lesssim 10 ~{\rm MeV}$.
The right panel of Fig.~\ref{fig-Ax-Back4} demonstrates the $(n_s-r)$-posterior translated into the $\alpha-\beta$ parameter space.
For the best-fitted $n_s=0.974$ \cite{Calabrese:2025}, the colored curve represents different sets of $\{\alpha,\beta\}$ require to realize scenarios with different reheating temperatures $T_{\rm re}$; specifically, $T_{\rm re}$ increases with both $\alpha$ and $|\beta|$.
We find that the $K$-inflation with ($n=4$) natural potential is compatible with the CMB result in gray up to $\alpha \approx 7$.
As shown in the left panel of Fig.~\ref{fig-Ax-Back4}, the predicted $n_s$ for $\alpha \gtrsim 7$ exhibits a turning point and reverts to smaller values when $\beta$ is sufficiently negatively large. This turning occurs around $\beta \simeq -9$ and $-2$ for $\alpha \simeq 7$ and $8$, respectively.
Since the inflaton's field value at the CMB scale becomes super-Planckian for $\alpha \gtrsim 7$ (see Fig.~\ref{fig:Natural_Swampland}-left), a more negative $\beta$ leads to a smaller $G(\phi)$.
Therefore, in this large-$\alpha,|\beta|$ regime, the effect of $G$ in the third term in the $n_s$ prediction \eqref{eq:ns} decouples and makes $n_s$ inevitably reverted to the prediction of the standard inflationary scenario.

\begin{figure}[t!]
\centering
\includegraphics[width=0.5\textwidth]{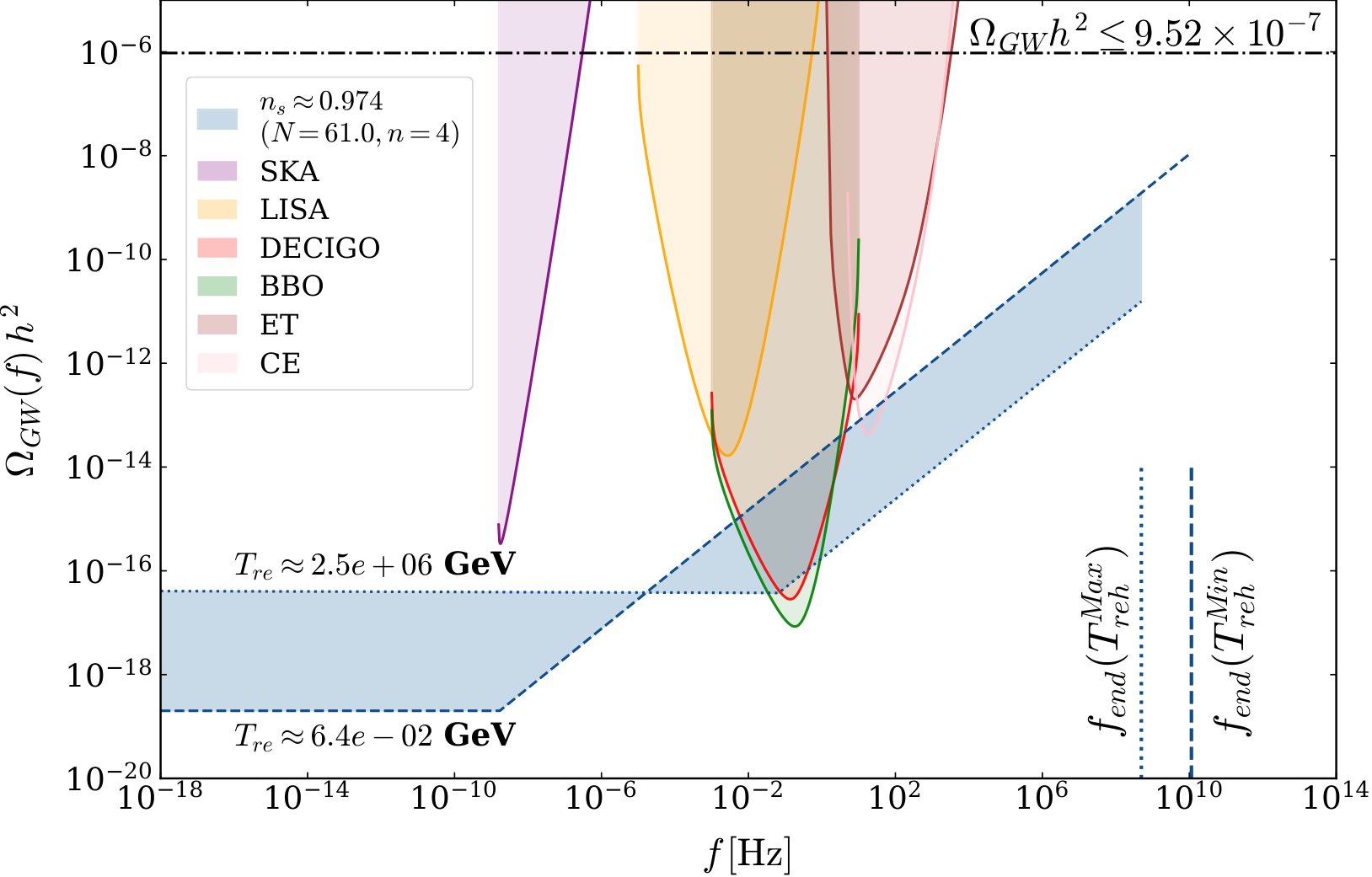}\hfill
\includegraphics[width=0.5\textwidth]{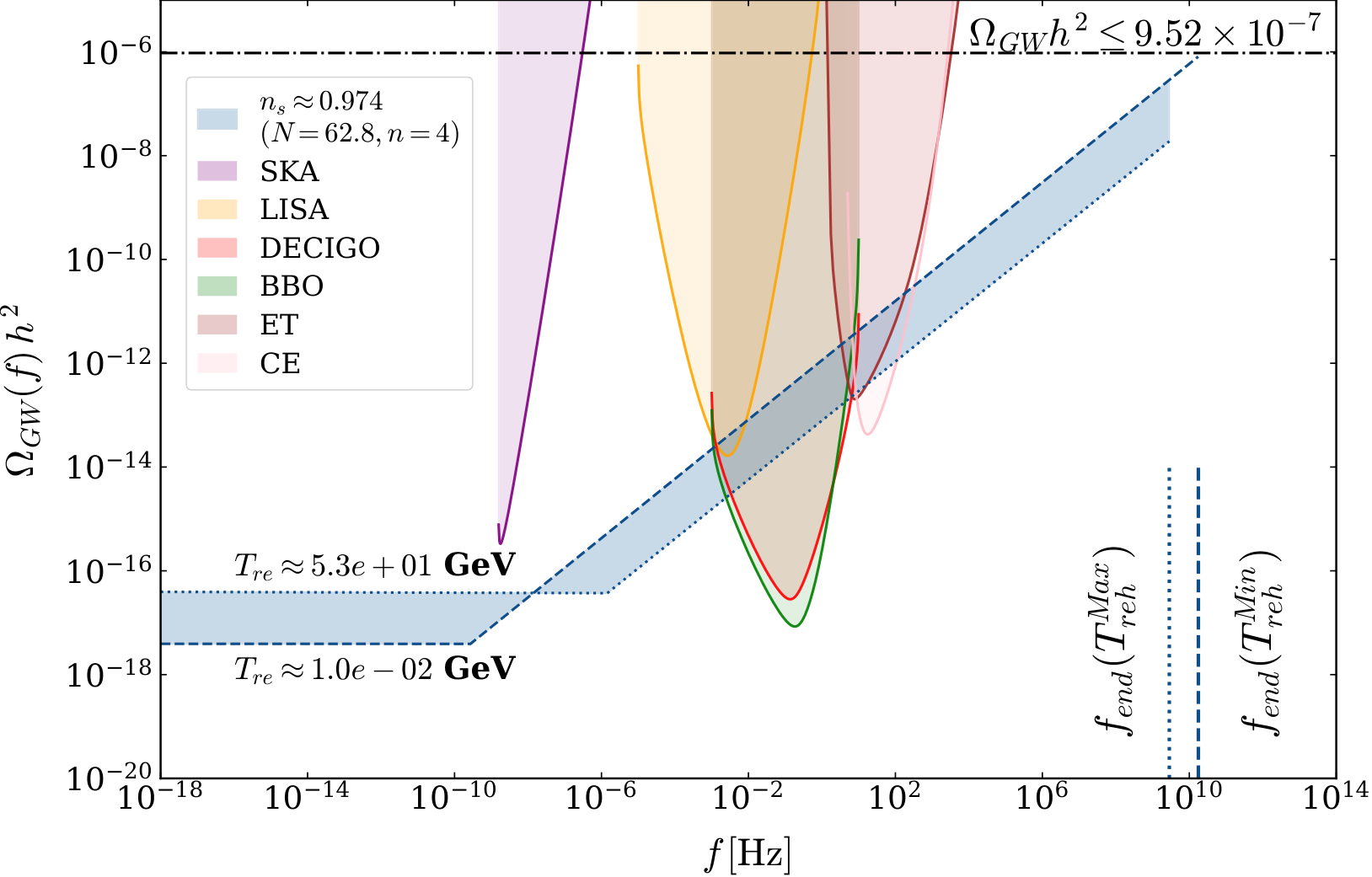}\\[-1.75em]
\caption{Spectra of inflationary GWB from the $K$-inflation with $(n=4)$ natural potential \eqref{eq:Natural}.
Assuming $N_{\rm cmb} = 61$ (left) and $62.8$ (right), we show results for the largest (dotted) and smallest (dashed) reheating temperatures, compatible with the best-fitted $n_s = 0.947$.
The spectral enhancement due to a stiff reheating phase occurs from  $f(T_{\rm re})$ [ Eq.~\eqref{eq:f_T_relation}], associated with the end of reheating, to $f_{\rm end}$ associated with the end of inflation and indicated by the small vertical line.
Other colored regions show the power-law-integrated sensitivity curves for LISA, ET, CE, DECIGO, BBO, and SKA.
The region above the horizontal dashed line is excluded by the $\Delta N_{\rm eff}$ bound~\eqref{Neff-bound}.
}
\label{fig-AX-GWspec}
\end{figure}

\begin{figure}[p!]
\centering
\includegraphics[width=0.43\textwidth]{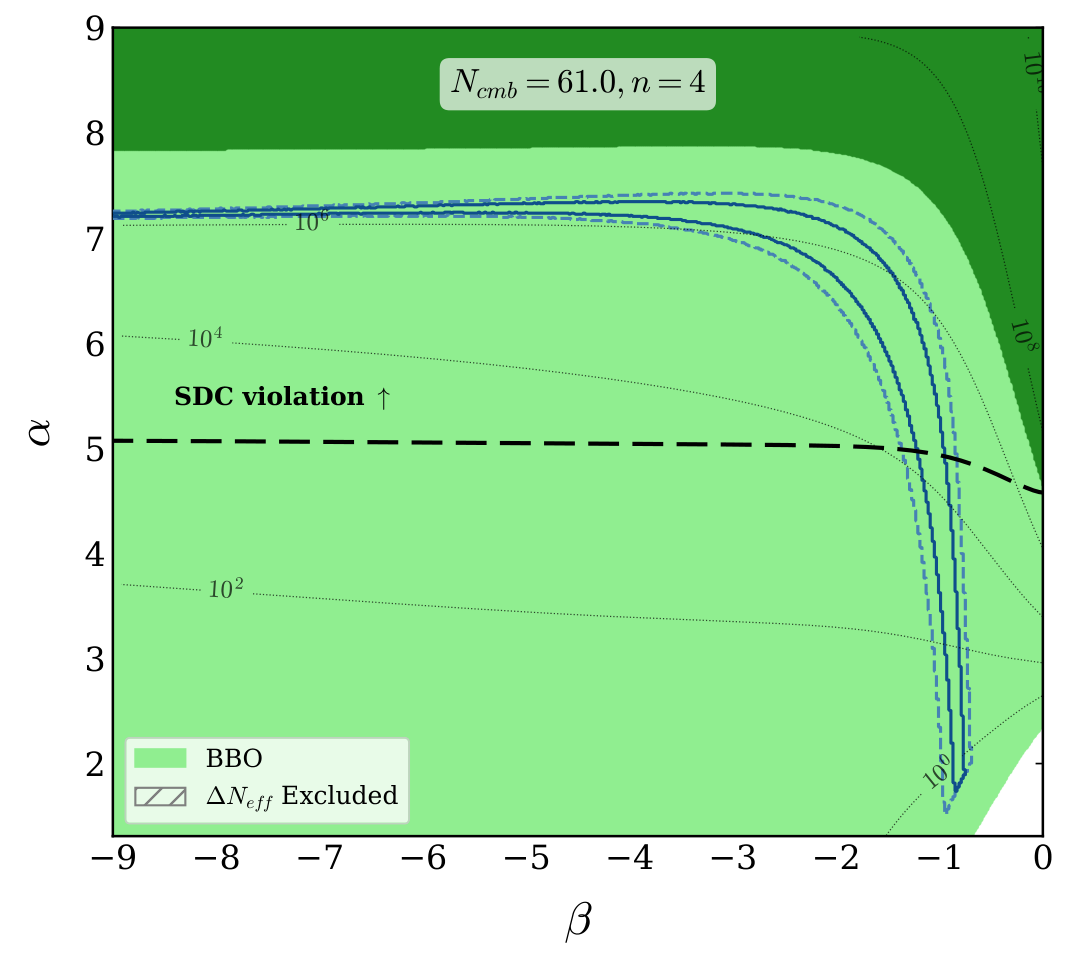}\hspace{0.5cm}
\includegraphics[width=0.43\textwidth]{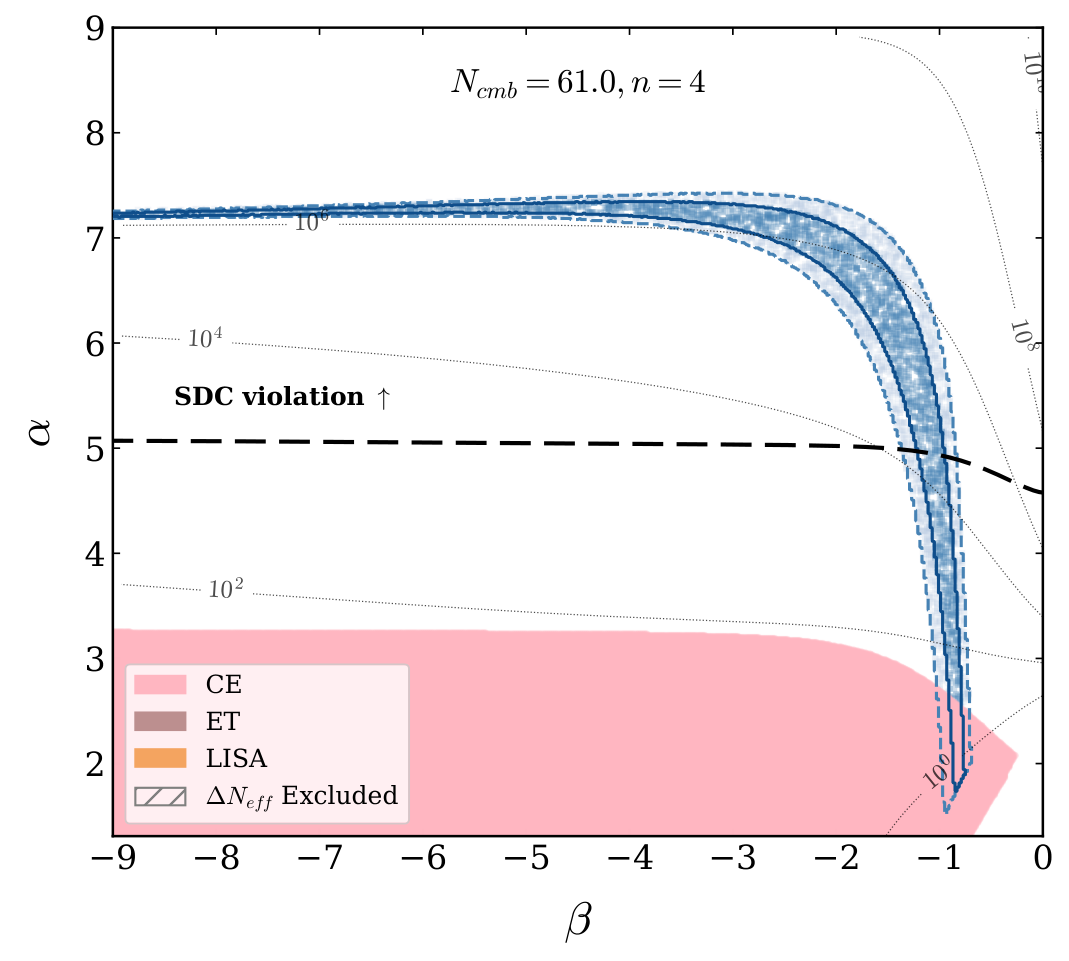}\\[-0.5em]
\includegraphics[width=0.43\textwidth]{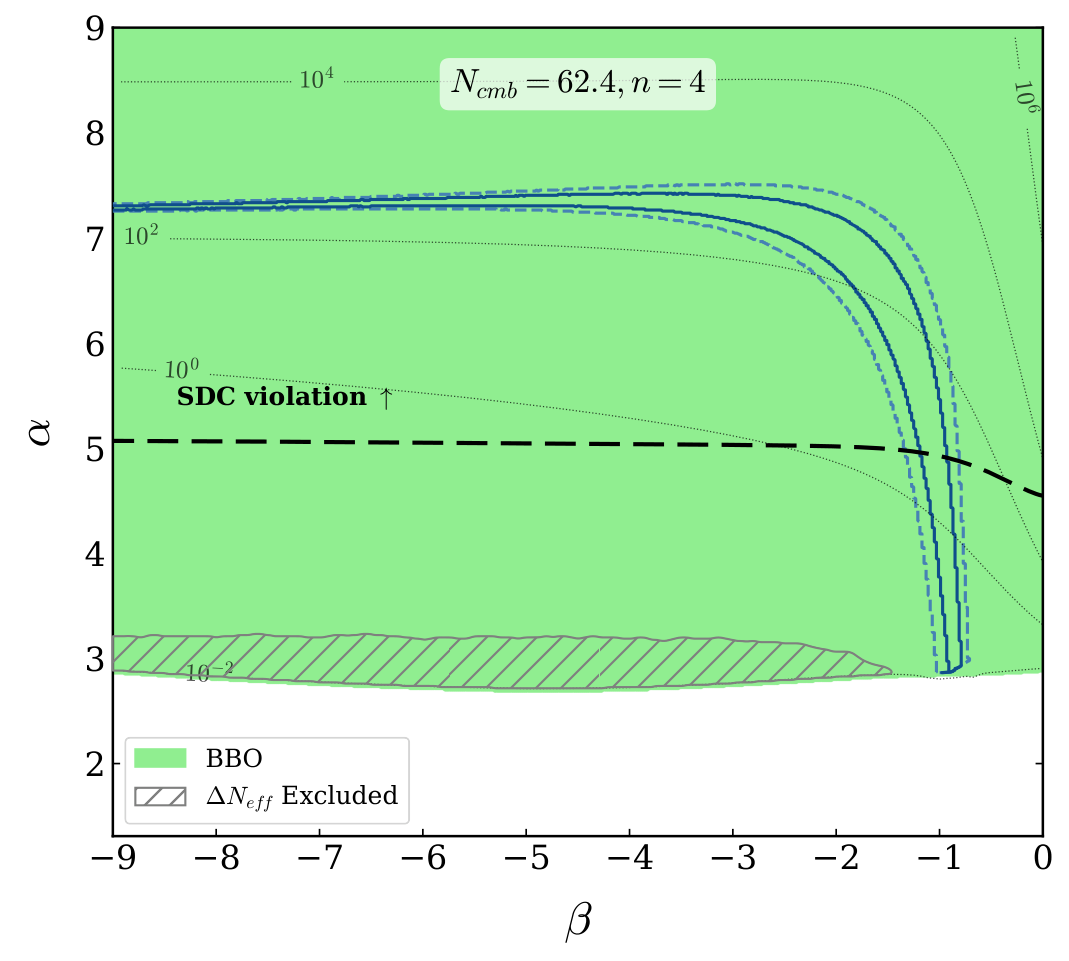}\hspace{0.5cm}
\includegraphics[width=0.43\textwidth]{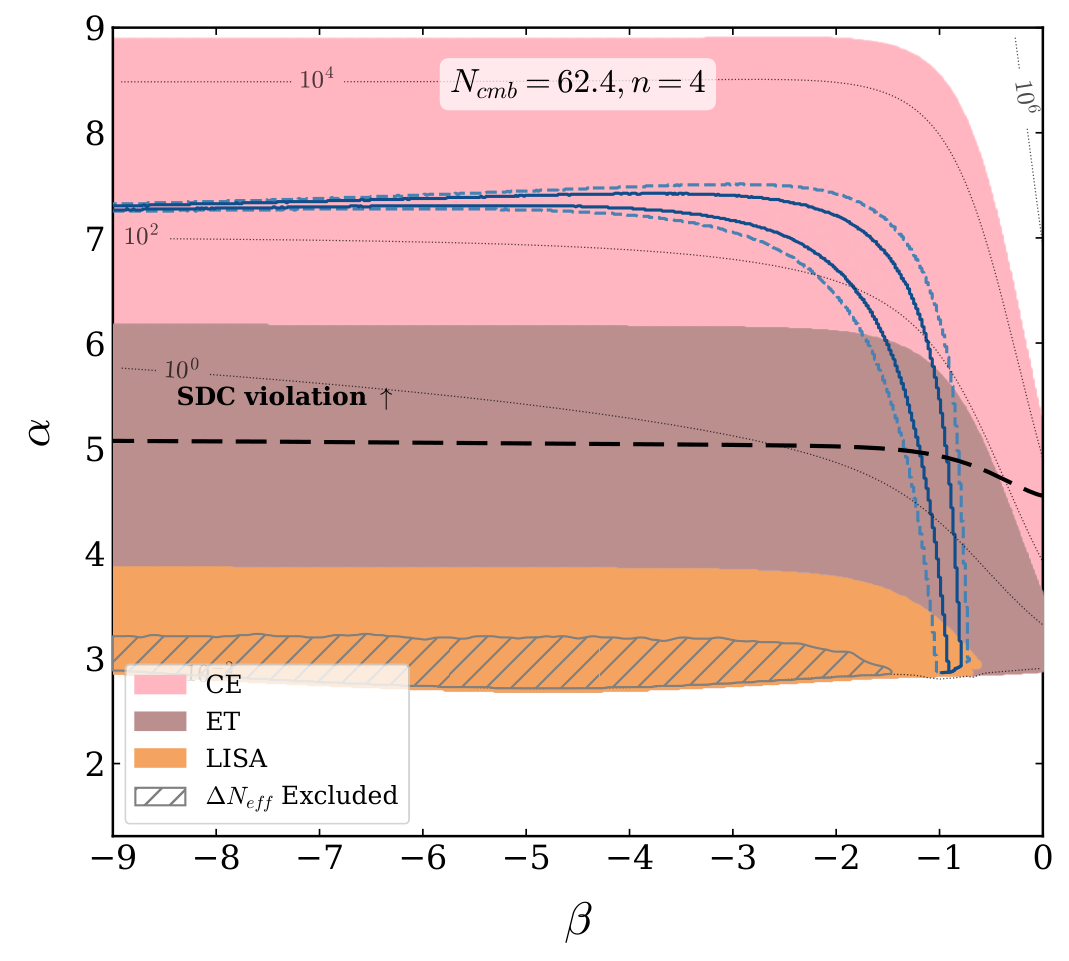}\\[-0.5em]
\includegraphics[width=0.43\textwidth]{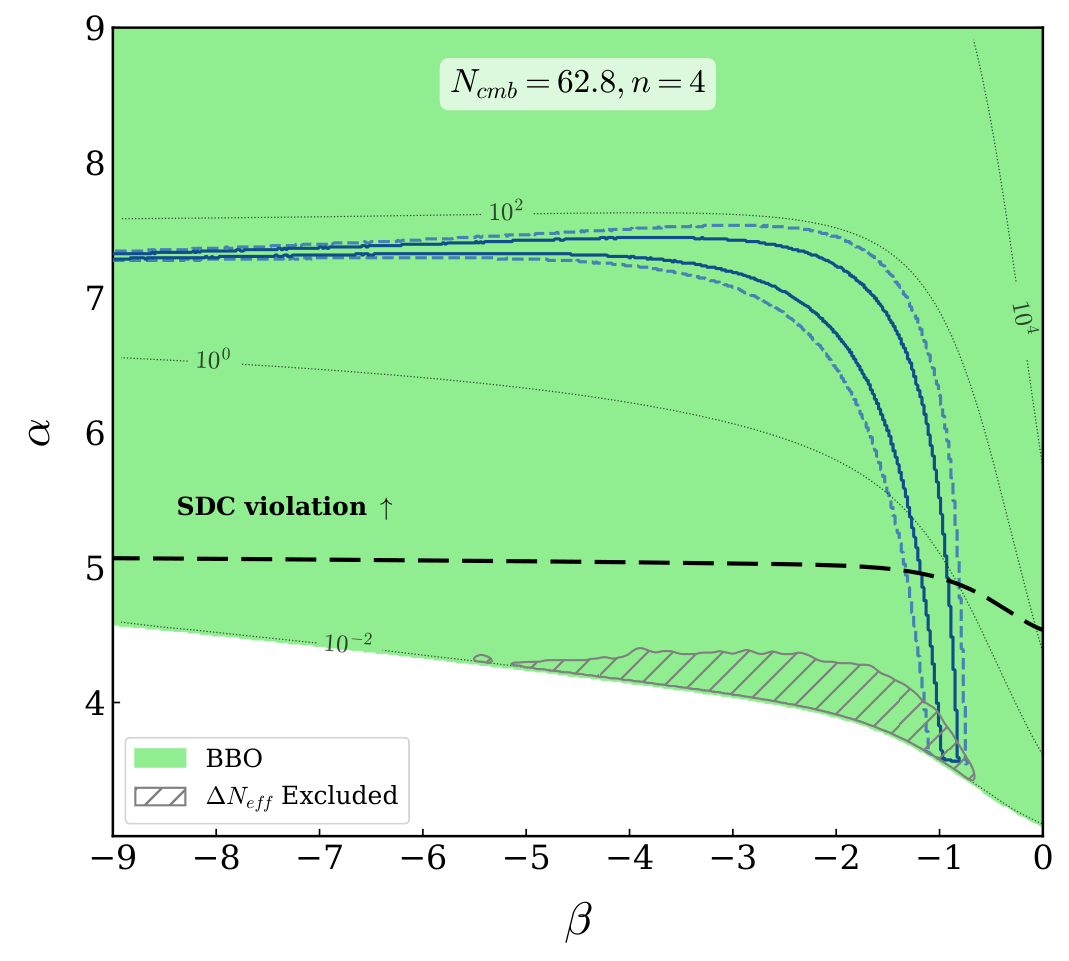}\hspace{0.5cm}
\includegraphics[width=0.43\textwidth]{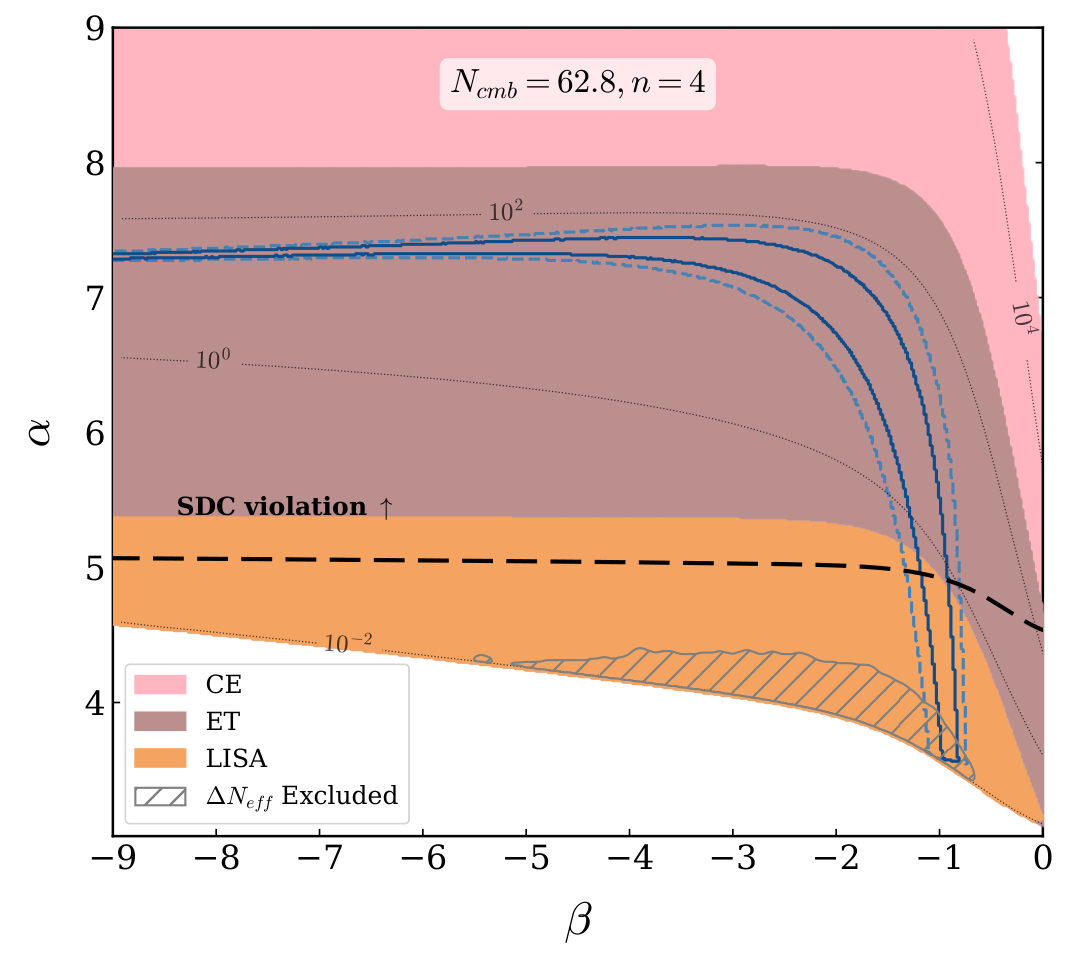}\\[-1.5em]
\caption{Detectability of inflationary GWB in the $K$-inflation model with $(n=4)$ natural potential \eqref{eq:Natural} is shown in the $\alpha-\beta$ plane, assuming $N_{\rm cmb} = 61, 62.4,$ and $62.8$.
The left column shows the parameter space that BBO probed in green, while the right column shows the regions for CE, ET, and LISA in pink, brown, and orange, respectively.
For all colored regions, the signature of $\Omega_{\rm GW} \propto f^{4/7}$ is detectable with a signal-to-noise ratio larger than 10, except the dark green region where BBO only observes a flat spectrum, i.e., frequencies at $f < f(T_{\rm re})$.
The blue contours represent the translated CMB result, already shown in the right panel of Fig.~\ref{fig-Ax-Back4}.
The gray dotted curves denote the reheating temperature $T_{\rm re}$ in GeV, and the hatched regions are excluded by $\Delta N_{\rm eff}$ bound \eqref{Neff-bound}.
The region above the black dashed line is incompatible with the Swampland distance conjecture (SDC).}
\label{fig-Nat4-GWsDet}
\end{figure}

\textit{Swampland criteria.}---Fig.~\ref{fig:Natural_Swampland} presents the model's consistency with the Swampland conjectures, discussed in Sect.~\ref{sec:swampland}. 
We see that the de Sitter conjecture \eqref{eq:ds_conj}---which can be satisfied if one of its two conditions is met---is consistent with the ($n=4$) natural inflation through the $V''/V$ condition with $\alpha \lesssim 5$, although the condition on $|V'/V|$ does not strongly satisfy.
In contrast, the distance conjecture \eqref{sdc-conj} is satisfied only marginally, and it is strongly violated for $\alpha \gtrsim 5$, i.e., $\Delta \phi/M_{\rm Pl} \gtrsim \mathcal{O}(10)$.
Both de Sitter and distance conjectures are strongly violated simultaneously for $\alpha \gtrsim 5$, signaling the need for extensions beyond the string landscape to explain most of the CMB-consistent region in Fig.~\ref{fig-Ax-Back4}-right-panel. See also the black dashed line in Fig.~\ref{fig-Nat4-GWsDet}.  

\textit{GW signature.}---Since the ($n=4$) natural potential leads to the approximated equation of state of $w_{\rm re} = 3/5$ during the reheating phase, as discussed earlier, the inflationary GWB  thus gets enhanced for modes reentering the horizon before the reheating is completed, i.e., at $f > f_{\rm re}$; see Eqs.~\eqref{eq:f_T_relation} and ~\eqref{eq:Omega_re} in Sect.~\ref{sec:gw}.
For a primordial tensor power spectrum with $n_t = 0$, which is a sufficient approximation in this model, the GWB spectrum has a signature of $\Omega_{\rm GW} \propto f^{4/7}$. 
Fig.~\ref{fig-AX-GWspec} shows the spectra of inflationary GWB for given choices of $\{N_{\rm cmb},T_{\rm re}\}$, i.e., these correspond to specific sets of $\{\alpha,\beta\}$.
For a fixed $N_{\rm cmb}$, we consider the largest and smallest $T_{\rm re}$ that yield the best-fitted $n_s = 0.974$.
Our result shows that the parameter space where this model can explain the CMB results yields observable GW signatures at many future GW observatories, such as ET, CE, BBO, DECIGO, and SKA.
Note that we have neglected the effect of astrophysical foregrounds (see e.g., \cite{Adams:2013qma,Cornish:2017vip,Robson:2018ifk,Babak:2023lro,Staelens:2023xjn,Hofman:2024xar,Boileau:2025jkv,Perego:2025bif}) which can degrade the reconstruction of such signatures.
The larger $N_{\rm cmb}$ leads to a larger signal, as the reheating phase can be longer.
In particular, ET and LISA can detect the inflationary model with $N_{\rm cmb} \gtrsim 61.5$ and $\gtrsim 62.4$, respectively.
We also checked that $N_{\rm cmb}$ as low as $60$ can still be probed by DECIGO and BBO.

For completeness, as shown in Fig.~\ref{fig-Nat4-GWsDet}, we scanned over the $\alpha-\beta$ plane and charted the parameter space in color where the GWB can be observed.
Each point in green, pink, brown, and orange regions has a detectable signature of the reheating phase, i.e., a blue-tilted spectrum $\Omega_{\rm GW} \propto f^{4/7}$, at BBO, CE, ET, and LISA, respectively. Note that, shown as dark green, BBO can also observe the flat GW spectrum corresponding to the radiation era at frequencies $f < f({T_{\rm re}})$.
We checked that the results for DECIGO are essentially similar to those of BBO, whereas SKA can only probe a flat-spectrum region, which is not compatible with CMB results.
Our analysis shows that BBO, ET, and CE can observe a blue-tilted GW signature for the whole CMB-compatible region when $N_{\rm cmb} \gtrsim 61, ~62,$ and $~62.6$, respectively.
LISA can probe, however, the signature in the CMB-compatible region that does not violate the swampland distance conjecture, i.e., even at the largest possible $N_{\rm cmb}$; see Fig.~\ref{fig-Nat4-GWsDet}-bottom-right.
In addition, the $\Delta N_{\rm eff}$ constraint does not severely exclude the CMB-compatible parameter space in the $K$-inflation model with ($n=4$) natural potential.
 Only with the largest $N_{\rm cmb} = 62.8$, it rules out the CMB region of $\alpha \lesssim 4$. As we shall see in the case with $n=5$, the $\Delta N_{\rm eff}$ bound becomes more stringent due to a stiffer reheating phase.

\begin{figure}[t!]
    \centering
    \includegraphics[width=0.75\textwidth]{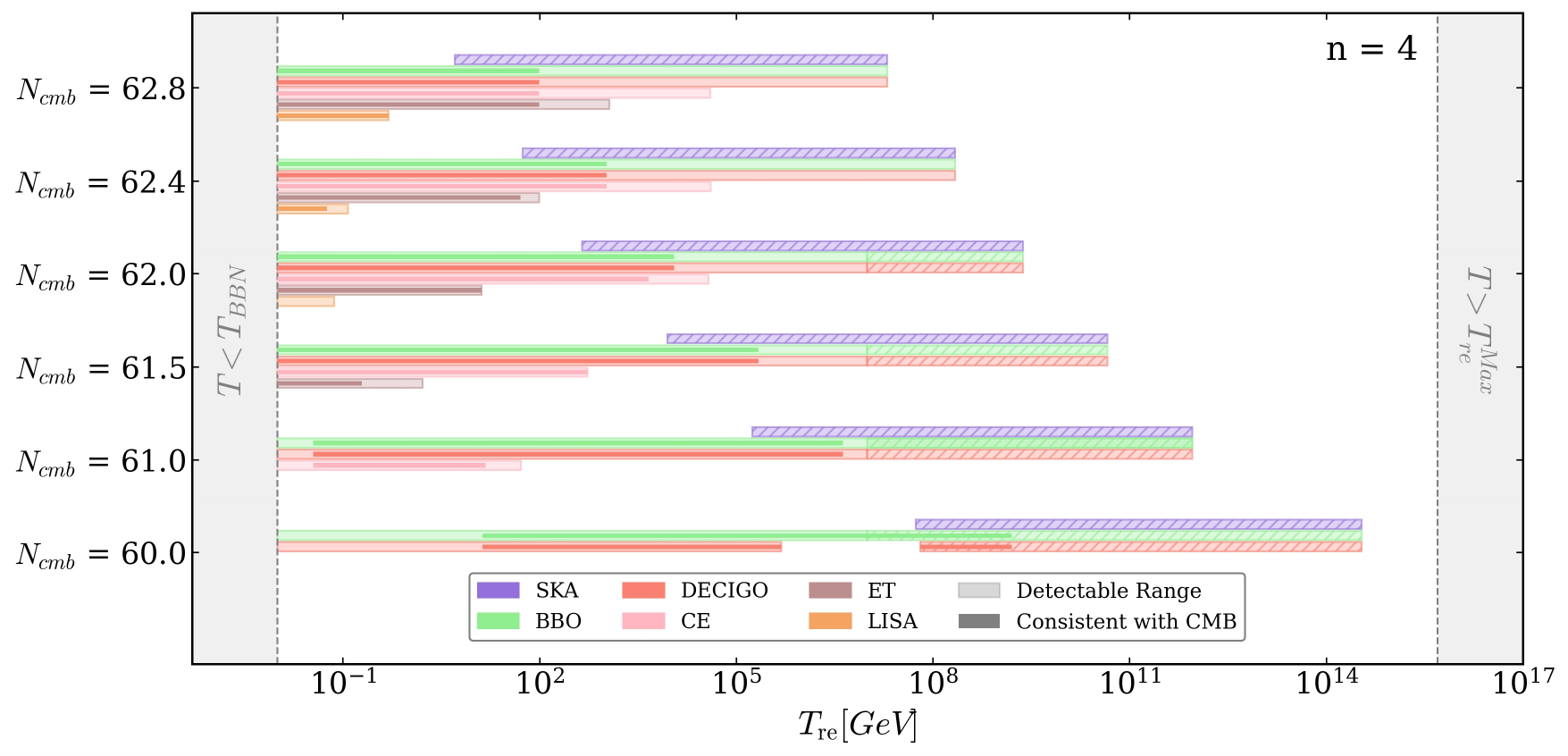}\\[-1.5em]
    \caption{The range of reheating temperatures $T_{\mathrm{re}}$ accessible to future gravitational wave observatories (SKA, BBO, DECIGO, CE, ET, LISA) for distinct values of the e-folding number $N$ with $n=4$. The light bars indicate the total detectable range derived from the parameter scan over $0 < \alpha < 9$ and $-9 < \beta < 0.1$. The Dark bars represent the subset of these ranges that remains consistent with Planck 2018 constraints on the scalar spectral index $n_s$ and tensor-to-scalar ratio $r$ at the $2\sigma$ level. The hatched regions correspond to scenarios probing the flat plateau of the stochastic GW background. The vertical gray shaded regions mark the BBN lower bound ($T_{\mathrm{re}} \lesssim 10$ MeV) and the inflation energy scale upper bound.}
    \label{fig:barchart-n4}
\end{figure}

To summarize the detectability of this $K$-inflation with $(n=4)$ natural potential, Fig.~\ref{fig:barchart-n4} shows the possible range of reheating temperatures for different $N_{\rm cmb}$ that can be probed at future GW observatories with $0\leq \alpha \leq 9$ and $-9\leq \beta \leq 0.1$.
We also show the results of SKA and DECIGO, which are not included in Fig.~\ref{fig-Nat4-GWsDet}.
As shown in a darker color, there is a wide range of reheating temperatures that this model can be consistent with CMB, and simultaneously provides a test via GW observations.
The lighter color indicates the region that can be probed by GW, but does not match the CMB results. 
For high reheating temperatures shown in hatched regions, the flat spectrum of the GWB---corresponding to the cosmological evolution after reheating---lies within the sensitivity of GW detectors; such an observation cannot be clearly identified, as this model can be confused with other scale-invariant slow-roll inflation results.

 \begin{figure}[h!]
    \centering
        \includegraphics[width=0.495\textwidth]%
        {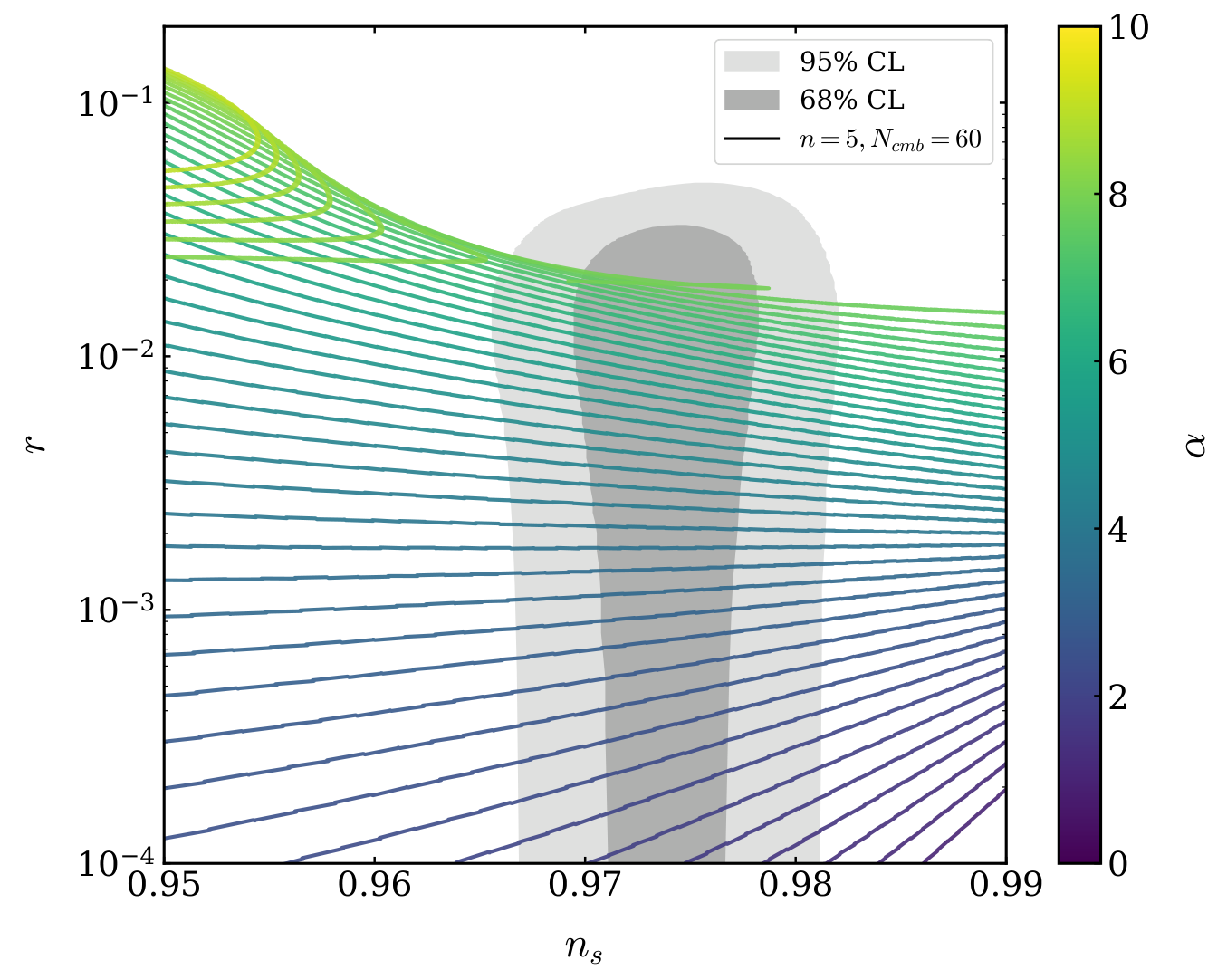}\hfill
        \includegraphics[width=0.48\textwidth]%
        {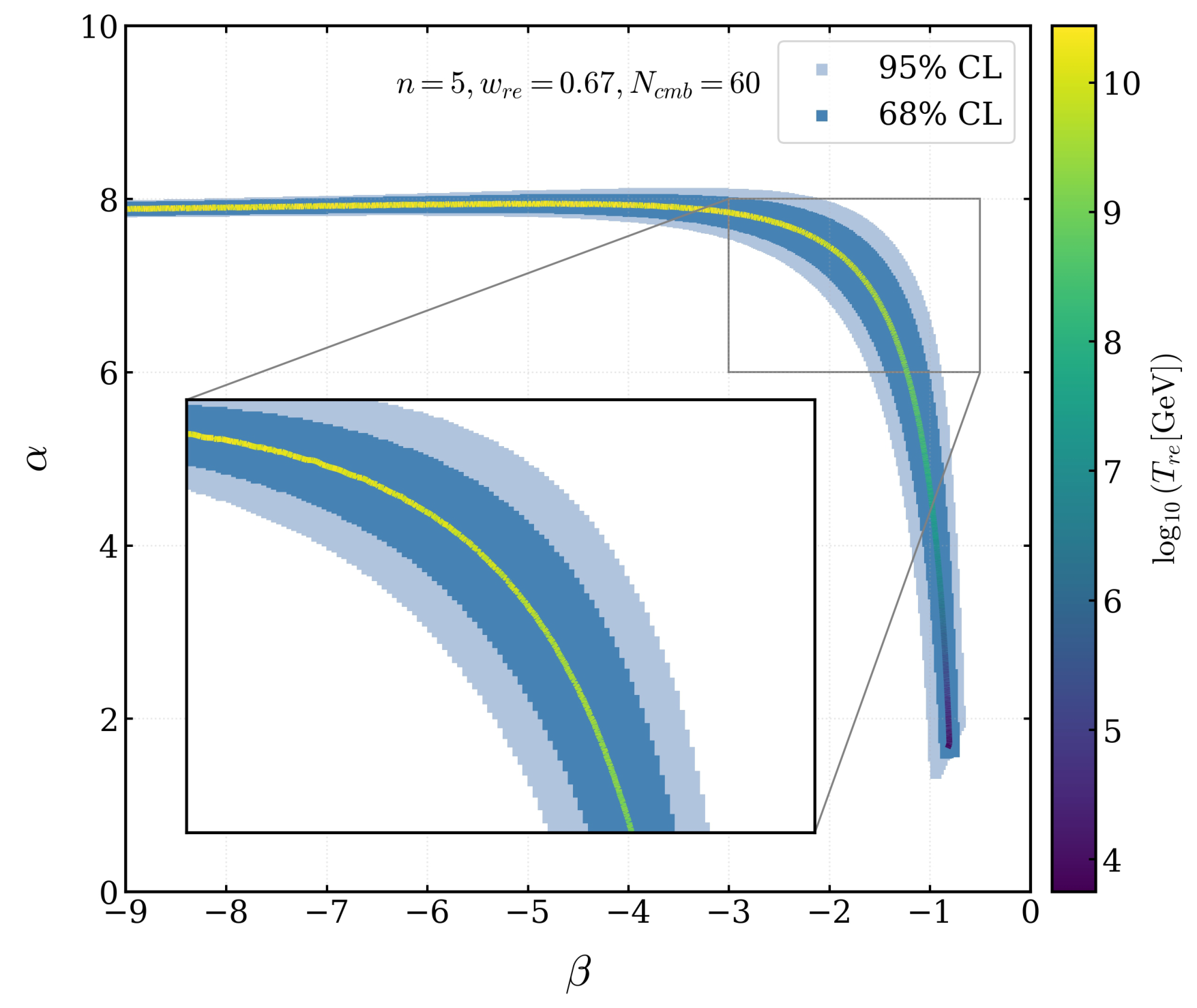}\\[-0.5em]
        \includegraphics[width=0.495\textwidth]%
        {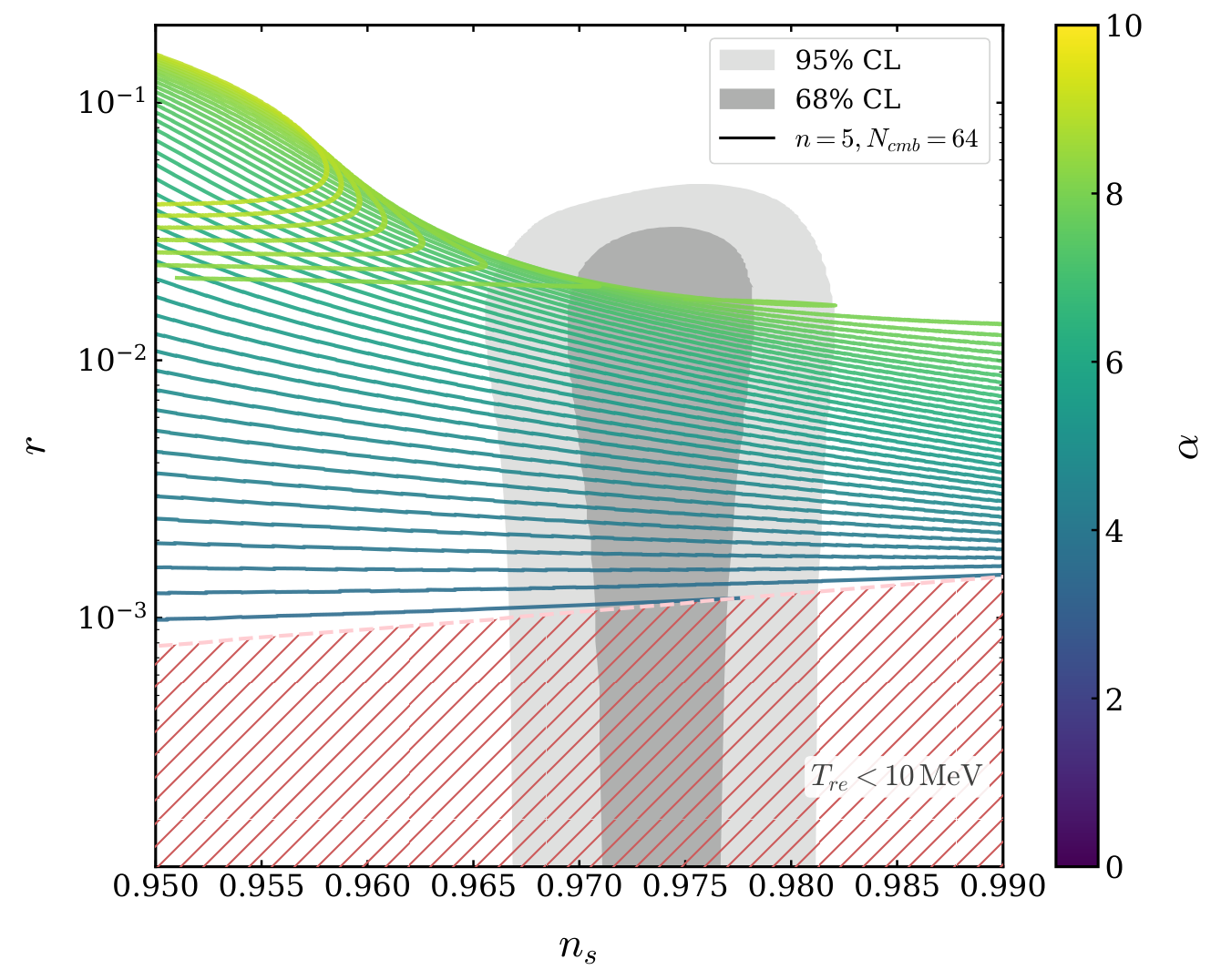}\hfill
        \includegraphics[width=0.48\textwidth]%
        {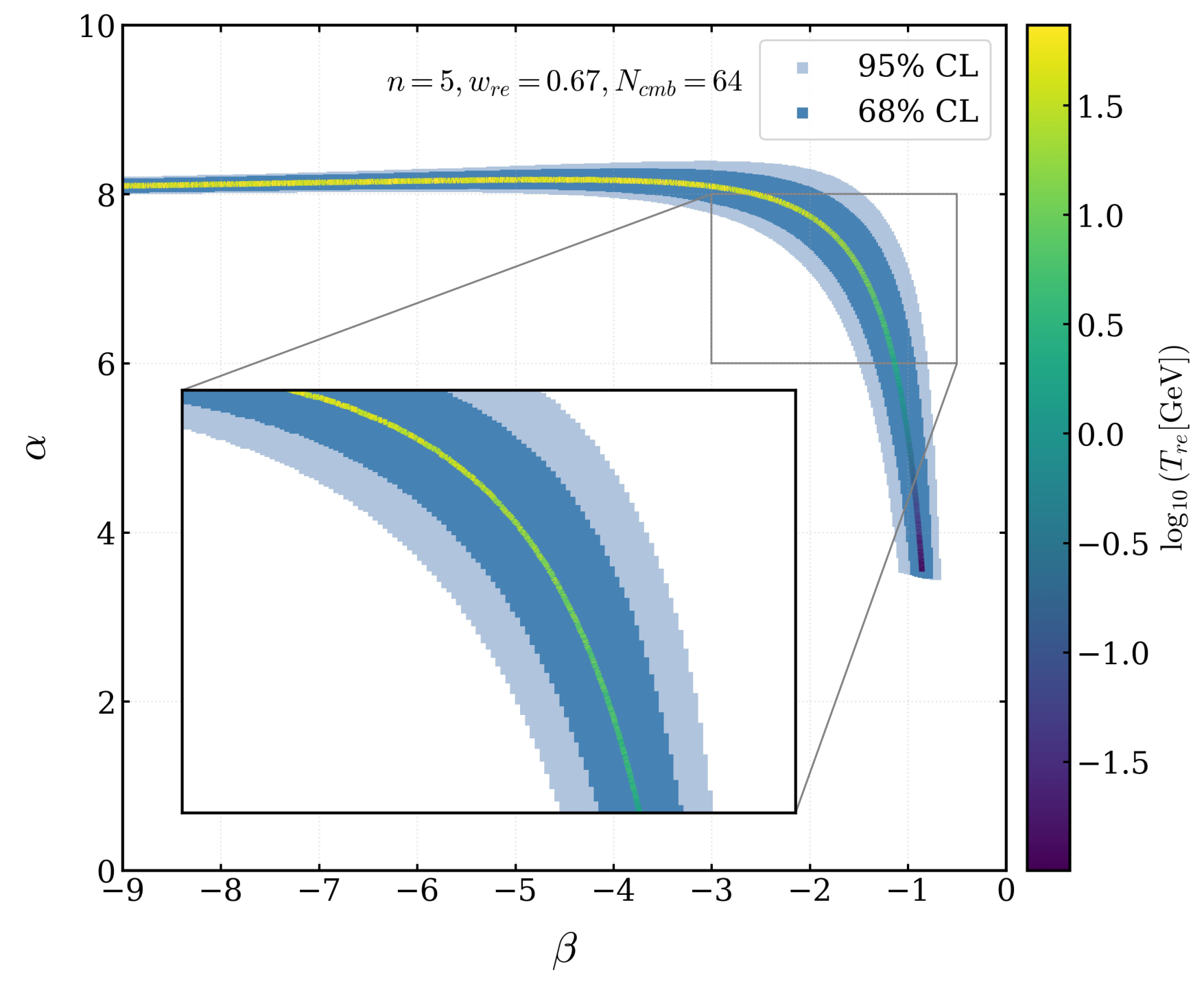}\\[-1.5em]
    \caption{\textit{Left:} Each colored line represents the $(r,n_s)$ predictions for the natural inflation ($n=5$) with a fixed $\alpha$, compared against 68\% and 95\% C.L. posteriors from CMB observations (gray regions). 
        For each line, $\beta$ varies from $-9$ at the largest $n_s$ to $0.1$ at the smallest $n_s$, except those with turnover points whose $\beta$ decreases with a smaller $r$.
        The red hatched region is excluded by the BBN bound $T_{\rm reh} \lesssim 10 ~ {\rm MeV}$. \textit{Right:} The parameter spaces in the $(\alpha,\beta)$ plane, compatible with the $(n_s,r)$ posteriors, are shown in blue. The colored line corresponds to $\{\alpha,\beta\}$ that gives the best-fitted $n_s = 0.974$ \cite{Calabrese:2025} when the reheating temperature $T_{\rm re}$ is varied along the line.}
    \label{Axion-Back}
\end{figure}

\subsection{Quintic case ($n=5$)}

\begin{figure}[t!]
\centering
\includegraphics[width=0.5\textwidth]{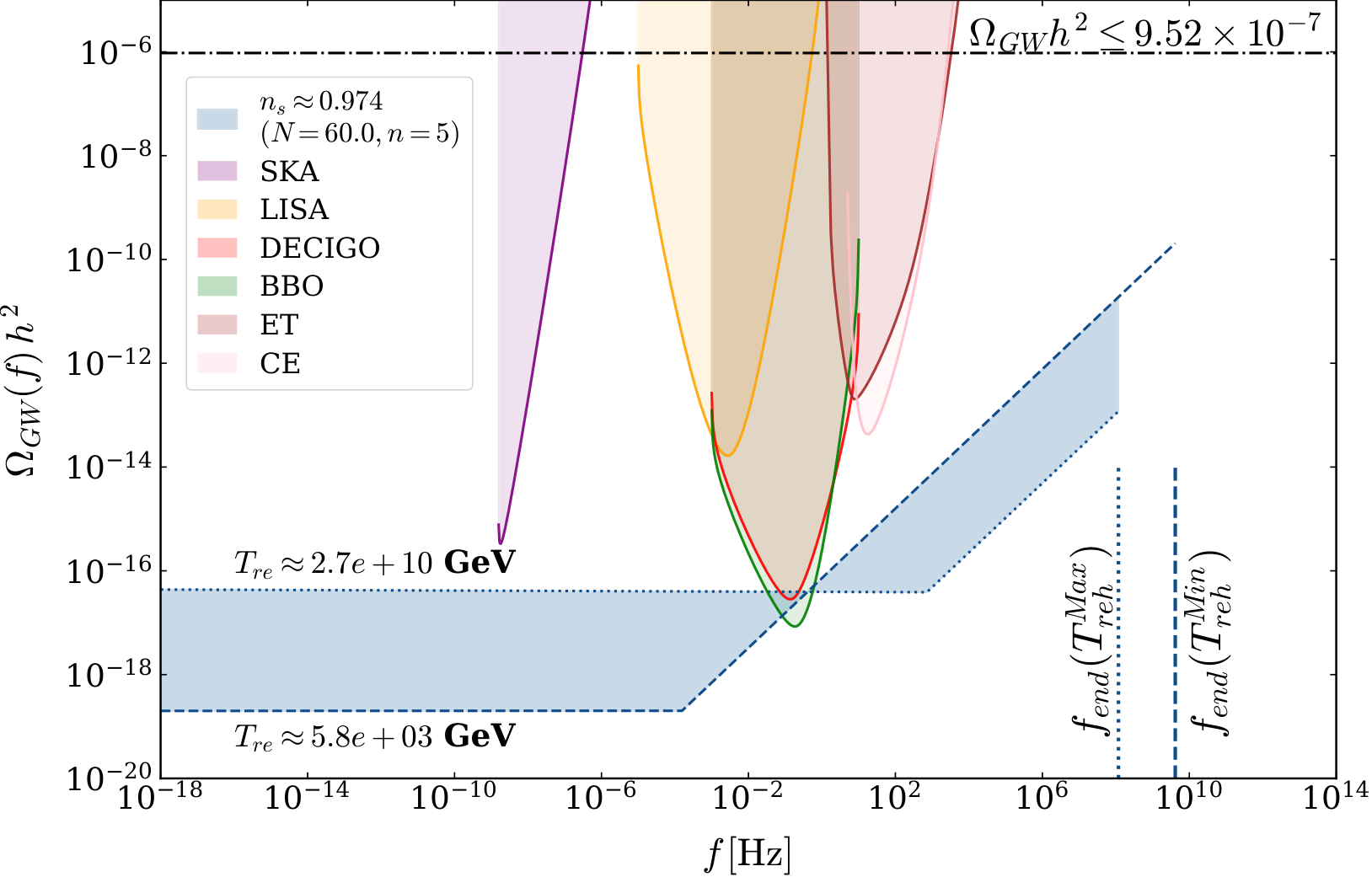}\hfill
\includegraphics[width=0.5\textwidth]{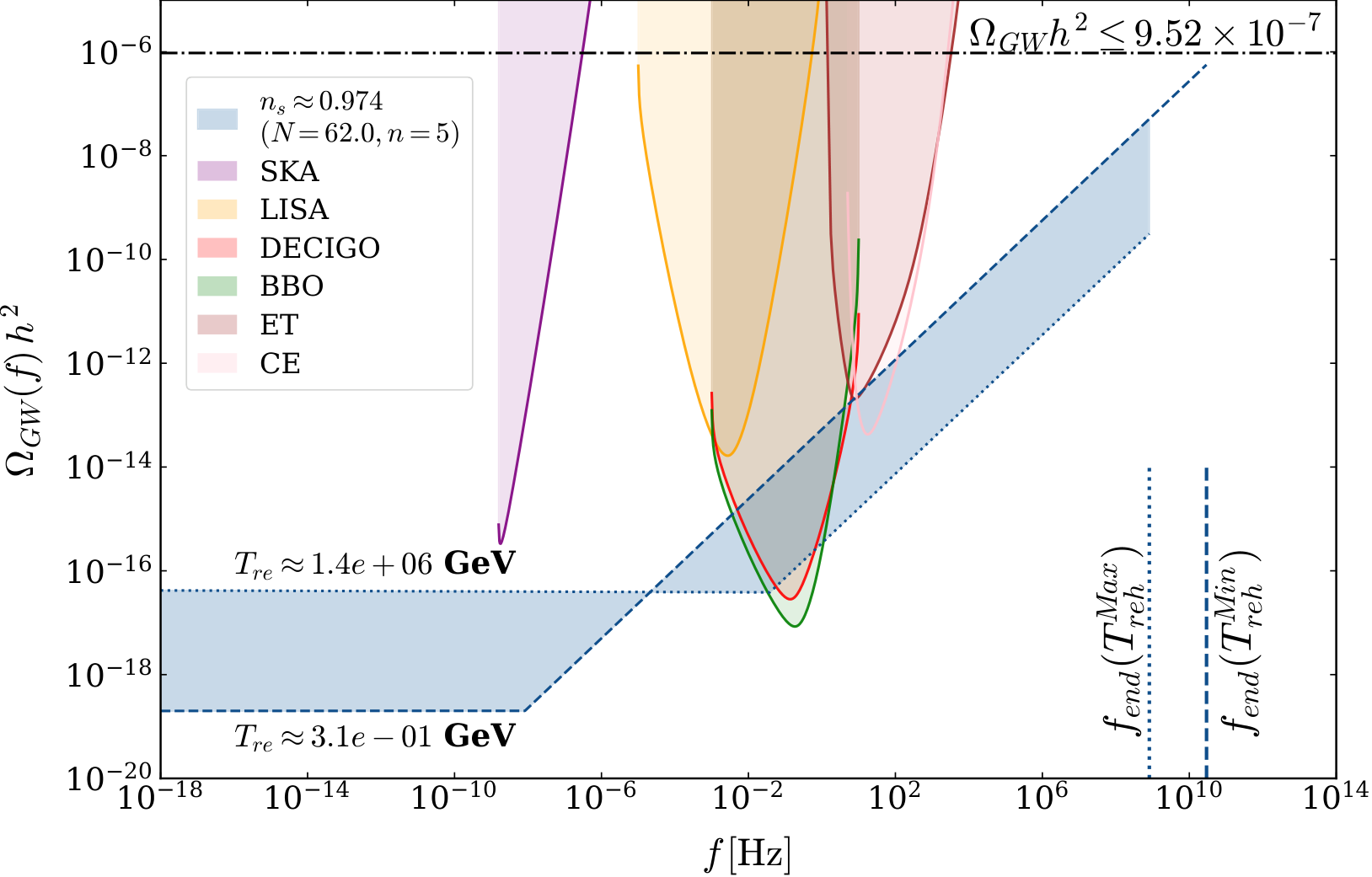}\\[-1.75em]
\caption{Similar to Fig.~\ref{fig-AX-GWspec}. 
Spectra of inflationary GWB from the $K$-inflation with $(n=5)$ natural potential \eqref{eq:Natural}.
Assuming $N_{\rm cmb} = 60$ (left) and $62$ (right), we show results for the largest (dotted) and smallest (dashed) reheating temperatures that are compatible with the best-fit $n_s = 0.947$.}
\label{fig-AX-5-GWspec}
\end{figure}

As $K$-inflation can improve the consistency of the natural inflation potential with CMB results, we now consider the case of $n=5$ whose equation of state during reheating (converging to $w_{\mathrm{re}} \approx 0.67$) is stiffer than that of the $n=4$ case and would lead to a more promising GW signals.
Similarly to the $n=4$ case, Fig.~\ref{Axion-Back} shows how $K$-inflation's parameter $\beta$ modifies $n_s-r$ predictions for different $\alpha$ values, allowing them to fit into the CMB's region, over the allowed range of $N_{\rm cmb} \in [56, 64]$.
The prominent effect in the $n=5$ case is the enhanced GWB with the slope of $\Omega_{\rm GW} \propto f^{2/3}$, as shown in Fig.~\ref{fig-AX-5-GWspec}.

\begin{figure}[p!]
\centering
\includegraphics[width=0.43\textwidth]{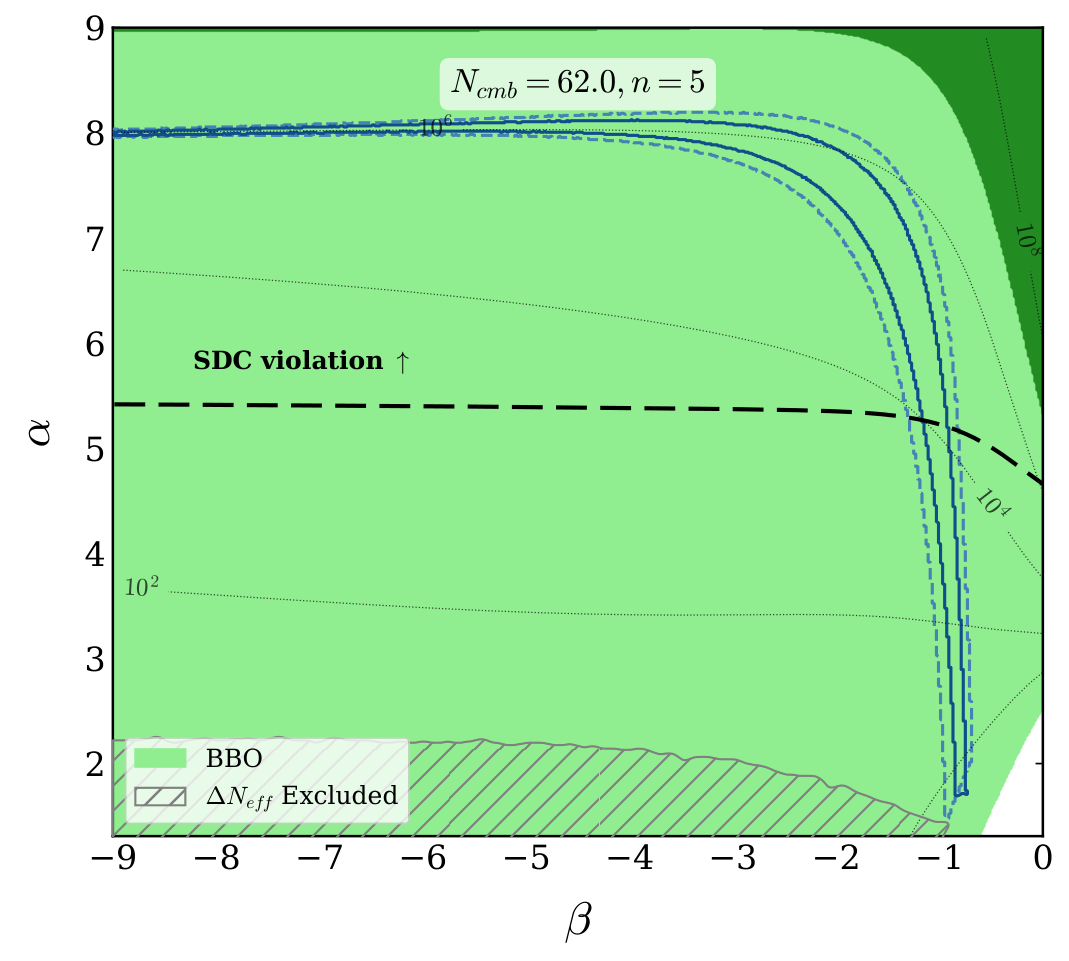}\hspace{0.5cm}
\includegraphics[width=0.43\textwidth]{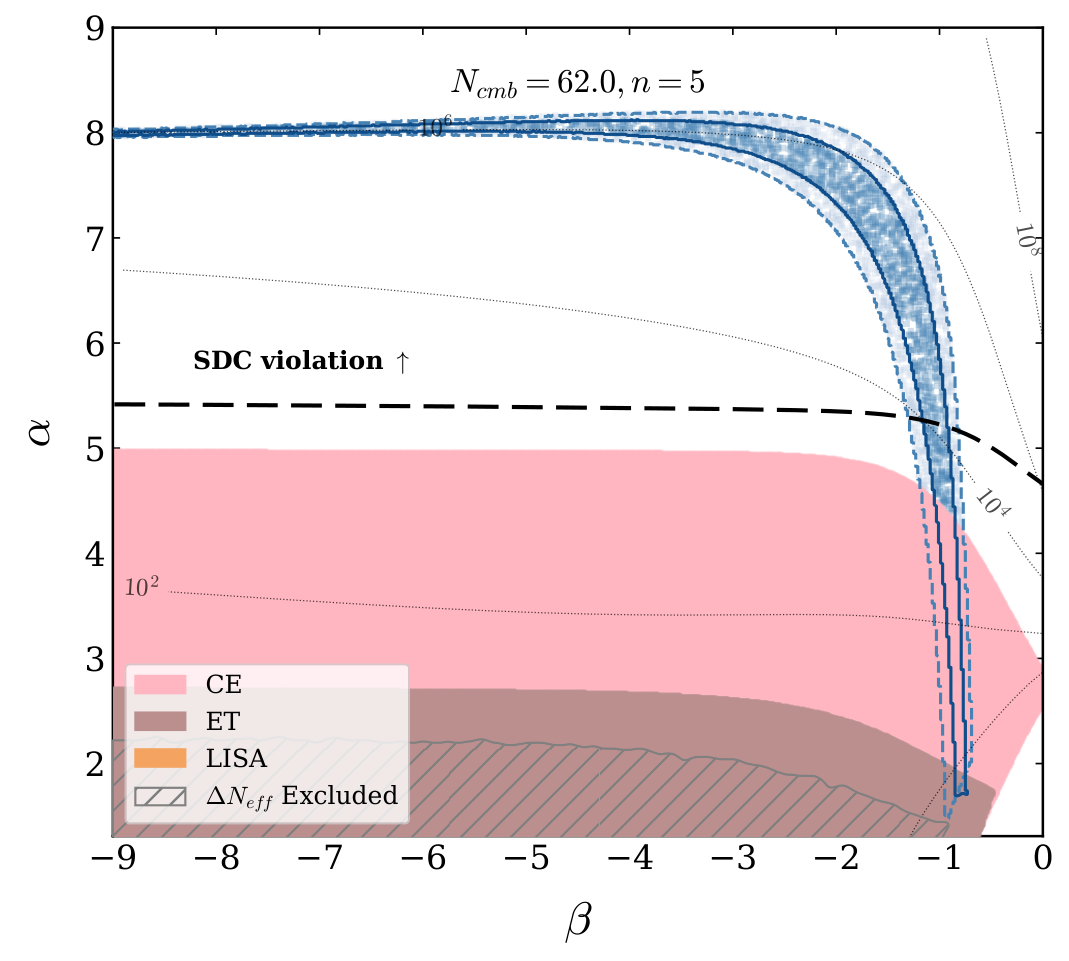}
\\
\includegraphics[width=0.43\textwidth]{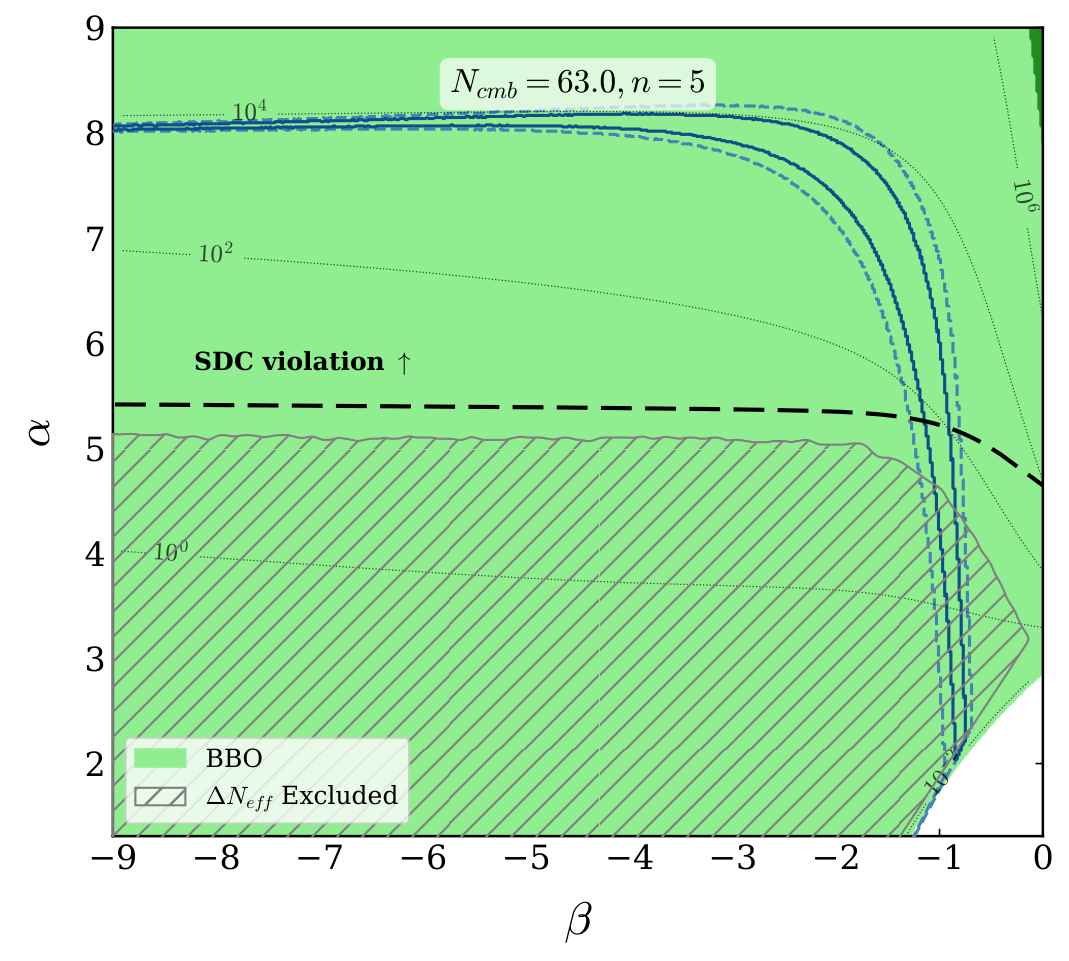}\hspace{0.5cm}
\includegraphics[width=0.43\textwidth]%
{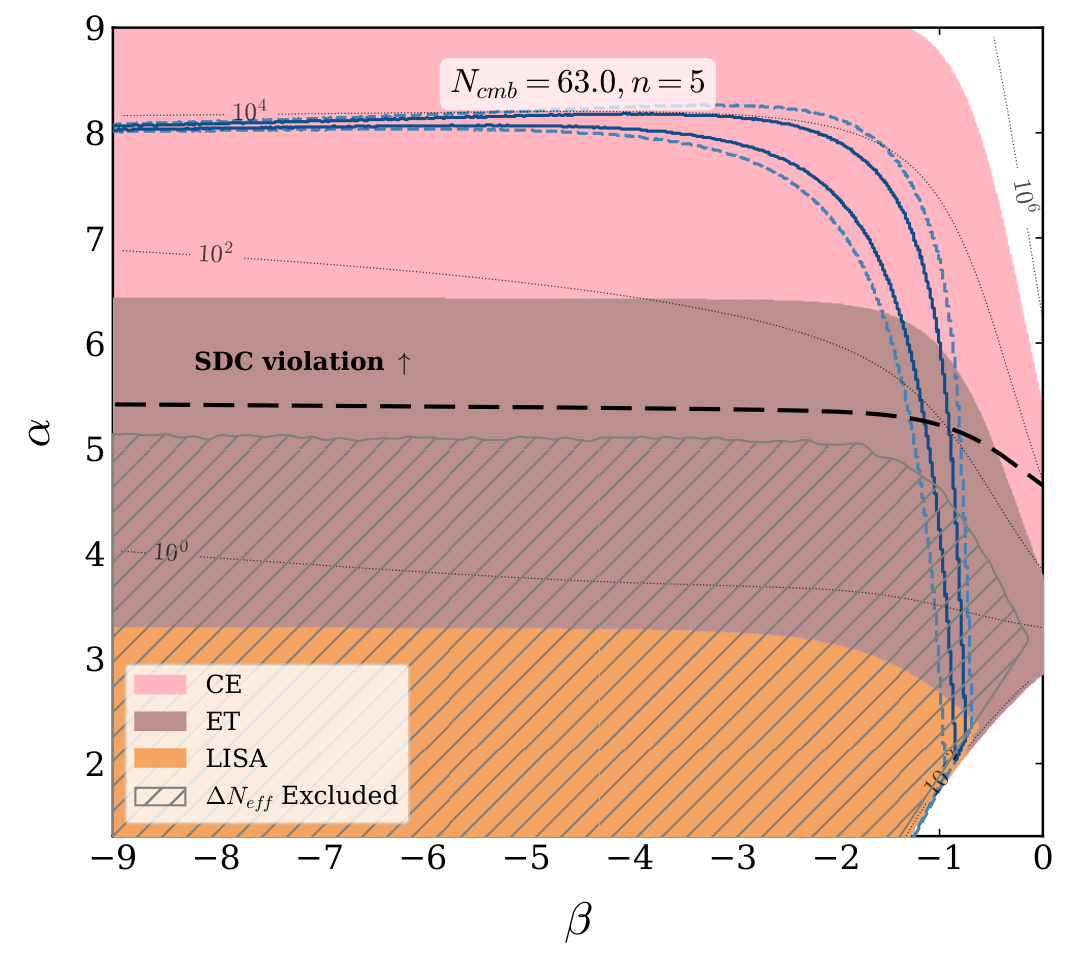}
\\
\includegraphics[width=0.43\textwidth]{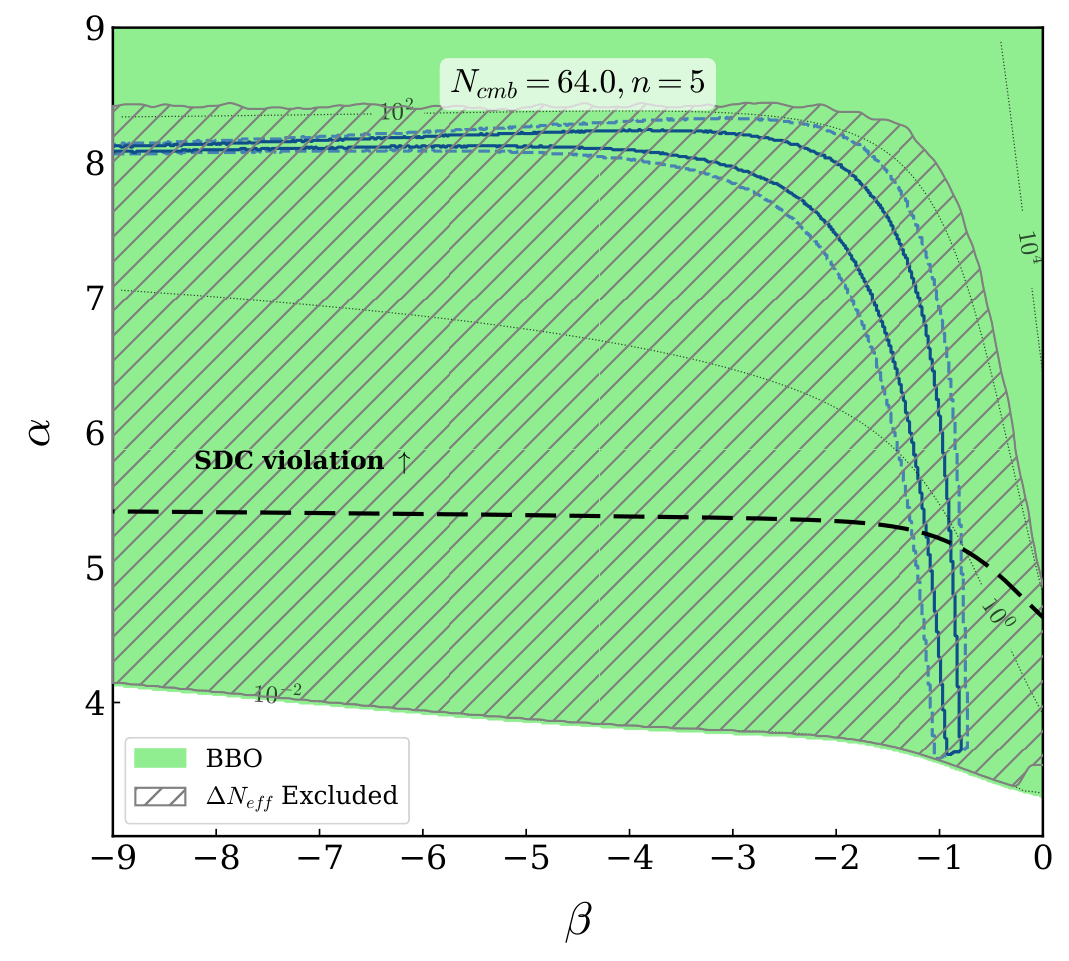}\hspace{0.5cm}
\includegraphics[width=0.43\textwidth]{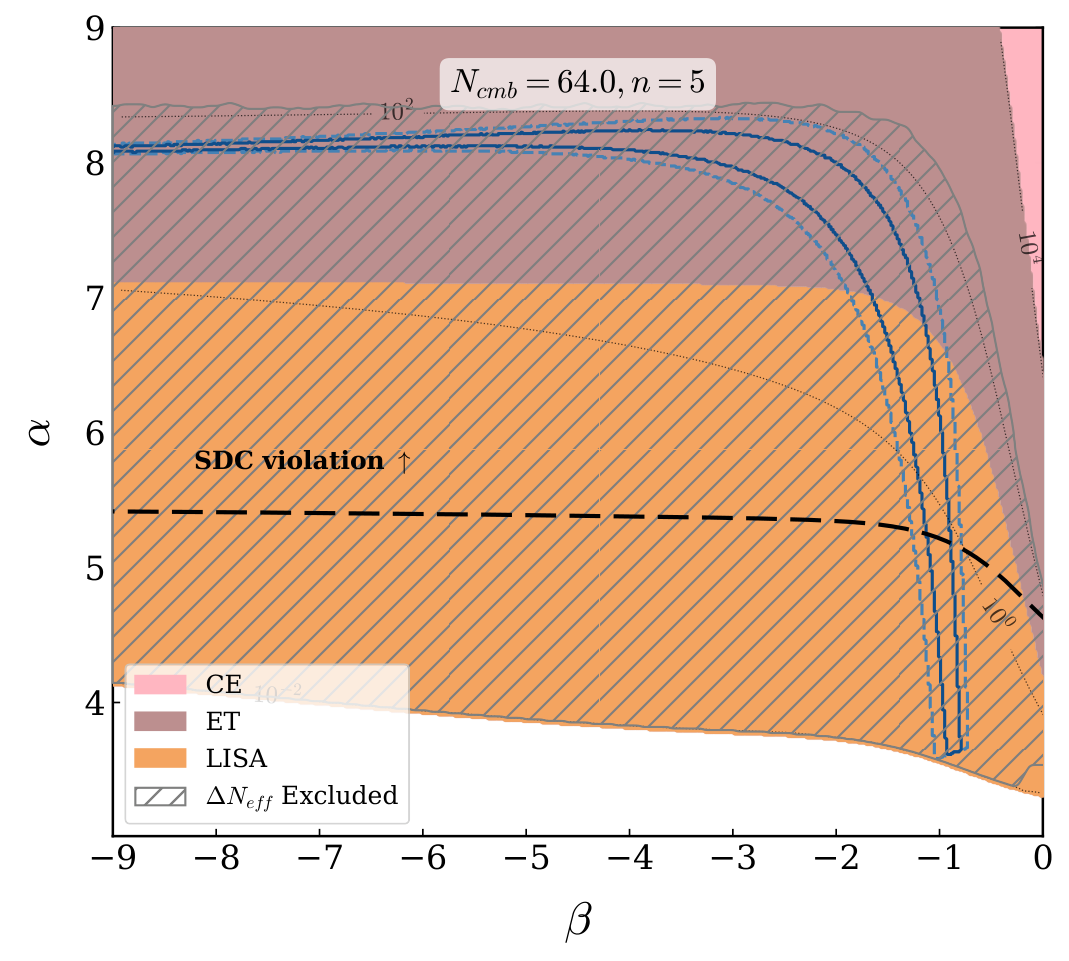}\\[-1.5em]
\caption{Same as Fig.~\ref{fig-Nat4-GWsDet}, but for the natural Inflation potential with $n=5$. The rows correspond to $N_{\rm cmb} = 60,~62,~63,$ and $64$ (from top to bottom). Note that for higher $N_{\rm cmb}$, the $\Delta N_{\rm eff}$ constraint (gray hatched region) significantly restricts the parameter space available for GW detection, ruling out the LISA-detectable region entirely.}
\label{fig-Nat5-GWsDet}
\end{figure}

Despite its strongly detectable GW signal and its ability to match the CMB's results, this scenario worsens the overproduction of GW problem (Eq.  \eqref{Neff-bound}) when $N_{\rm CMB} \gtrsim 62$.
This tension becomes more evident in Fig.~\ref{fig-Nat5-GWsDet} where the $\Delta N_{\rm eff}$ bound \eqref{Neff-bound} rules out the $(\alpha-\beta)$ phase space substantially and starts cutting the CMB-consistent region when $N_{\rm CMB} \gtrsim 62$. This bound is so strong that it rules our all CMB compatible region when $N_{\rm CMB} \gtrsim 64$.
This bound clearly rules out any detectable signal at LISA but still permits detection at ET, CE, and BBO.
Additionally, we display the Swampland distance conjecture bound as black dashed lines, suggesting that the case that can explain the CMB result and have a detectable GW signal might lie either in the string Landscape or the Swampland.
Similar to the $n=4$ case, those with $\alpha \gtrsim 5$ fall into the Swampland for $N_{\rm CMB}$ up to 64 efolds.

\begin{figure}[t!]
    \centering
    \includegraphics[width=0.75\textwidth]{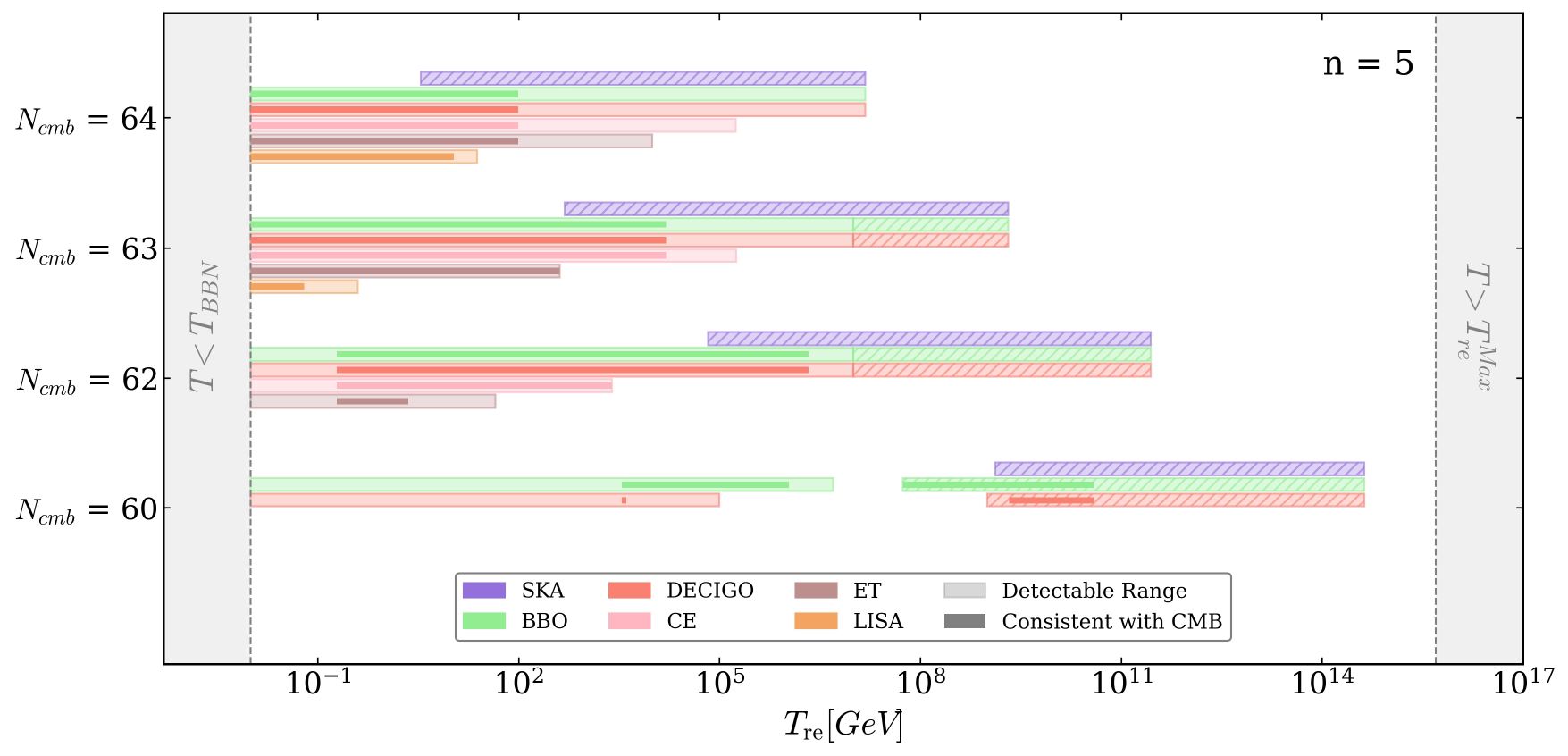}\\[-1.5em]
    \caption{The range of reheating temperatures $T_{\mathrm{re}}$ accessible to future gravitational wave observatories for distinct values of the e-folding number $N$ with $n=5$. The conventions for light/dark bars, hatched regions, and shaded bounds are the same as in Fig. \ref{fig:barchart-n4}.
}
    \label{fig:barchart-n5}
\end{figure}

Lastly, we summarize the detectability of the $n=5$ case via GW in Fig.~\ref{fig:barchart-n5}.
Comparing it to the $n=4$ case in Fig.~\ref{fig:barchart-n4}, we notice that the detectability of a larger $n$ is lower despite having a stronger GW signal.
This result suggests that a scenario with a larger exponent $n>5$ in its potential \eqref{eq:Natural} would be even more restricted, agreeing with the previous result of a non-detectable GW signal from the kination-like reheating (i.e., when $n\to\infty$) at future GW observatories.
For the $K$-inflation with the natural potential, we find that the $n=4$ case deems the optimal scenario, producing a strongly detectable GW signal, improving its consistency with CMB results, and not significantly violating the $\Delta N_{\rm eff}$ bound.

\section{Discussion and Conclusion}
\label{sec6}

The ACT DR6 result shifts away the $
\{n_s,r\}$ parameter space from the region previously consistent with predictions from some standard slow-roll inflationary scenarios, i.e., it suggests a larger value of the scalar spectral index $n_s$.
In this paper, we revisit two classes of inflationary models---namely, the $\alpha$-attractor T-model and natural inflation---and extend them to the K-inflation scenario, where the kinetic term becomes non-canonical and depends on a coupling function $G(\phi)$.
The main reason why the K-inflation can improve the consistency of these models with the recent CMB observations is due to an additional friction in the inflaton's dynamics caused by $G(\phi)$ in Eq.~\eqref{eq:slow_roll_Friedmann2}.
This friction modifies the inflaton's dynamics and helps bring the $r-n_s$ predictions from both models into the observed regions of the joint Planck-ACT-LB-BK18 dataset over a wide range of K-inflation parameters.
However, we show that not all regions that can explain the CMB data are allowed by reheating constraints. 
We then consider the Swampland criteria for identifying which regions require an inflationary model in the string landscape or the Swampland.
Lastly, since the inflaton dynamics in these models lead to a specific equation of state during a reheating era, the GWB from inflation gets imprinted with a signature that could be probed by future GW observatories.


For the $\alpha$-attractor T-model \eqref{eq:model_alpha_attractor} with $n=2$, the recent CMB data can be explained with $\beta \sim \mathcal{O}(10)$ over a wide range of $\alpha$ and $N_{\rm cmb} \in [42,56]$. Furthermore, the region consistent with the Swampland criteria requires $\alpha \gtrsim \mathcal{O}(10^{-3})$, ensuring that the K-inflation extension of this model is consistent with quantum gravity principles.
As the effective potential behaves quadratically near the minimum, a matter-like reheating phase with $w_{\rm re} \approx 0$ occurs after inflation ends.
The reheating signature in GWB is red-tilted and therefore cannot be probed at near-future GW detectors.
Although we discuss only the example of $n=2$, which clearly shows the improvement over K-inflation, we have checked that larger-$n$ cases can still explain the CMB result and generate a stiff reheating phase that could be probed via the blue-tilted signature in GWB.
We leave the task of distinguishing $\alpha$-attractor models with different $n$ using the GW signature to future work.


For natural inflation potential \eqref{eq:Natural}, we focus on cases of $n>0$, where the K-inflation is required for consistency with CMB results.
In particular, the $n=4$ and $n=5$ cases can explain the data when $\alpha \lesssim 7$ and $\alpha \lesssim 8$, respectively, and $\beta \lesssim -1$ for both cases.
The inflaton oscillating on this potential with $n>2$ leads to a stiff reheating phase, i.e., for $n = 4$ ($n = 5$), the asymptotic equation of state is $w_{\rm re} \approx 3/5$ ($w_{\rm re} \approx 2/3$), which has the blue-tilted GW spectrum of slope $\Omega_{\rm GW} \propto f^{4/7}$ ($\Omega_{\rm GW} \propto f^{2/3}$) at frequencies $f_{\rm re}< f < f_{\rm end}$ (see Eq.~\eqref{eq:Omega_re}).
Our scan over K-inflation parameter space in Figs.~\ref{fig-Nat4-GWsDet} and \ref{fig-Nat5-GWsDet} shows the existence of $\{\alpha,\beta\}$ range where the model is consistent with CMB data, satisfies BBN and $\Delta N_{\rm eff}$ bounds, and produces a GW signal detectable at LISA, ET, CE, DECIGO, and BBO.
We also summarize in Figs.~\ref{fig:barchart-n4} and \ref{fig:barchart-n5} the range of reheating temperatures that these GW experiments could probe, with and without consistency with the CMB results.
Moreover, we find that while the de Sitter conjecture is generally satisfied, the Swampland distance conjecture is realized only when $\alpha \lesssim 5$.
The detection of the GW signal will allow us to pin down the range of $\{\alpha,\beta\}$ that K-inflation can explain CMB observations; when combined with the Swampland criteria, this could hint at the class of UV completions needed to describe our Universe.



In conclusion, the K-inflation framework provides a working mechanism to revive simple inflationary potentials in light of the new CMB data and yields a testable signal through GW observations.
Lastly, we note that our results are independent of the reheating mechanism, and we treat the reheating temperature as a free parameter. 
With specific reheating details (e.g., couplings between the inflaton and other particles), the range of K-inflation parameters would be more tightly constrained.
Both CMB and GW observations in this era of precision cosmology will not only guide us toward the UV completion of the underlying theory of the Universe, but also explain how this theory evolves into the low-energy regime and how the Universe as we know it emerges.



\section*{Acknowledgment}
This work has received scholarship under the Post Doctoral Training for Frontier Research from Khon Kaen University, Thailand (Grant No. PD2568-03-13).


\appendix
\section{Validity of the Power-Law $\mathcal{P}_t$ Approximation}
\label{app:power_law_validity}

\begin{figure}[t!]
    \centering
    \includegraphics[width=0.6\columnwidth]{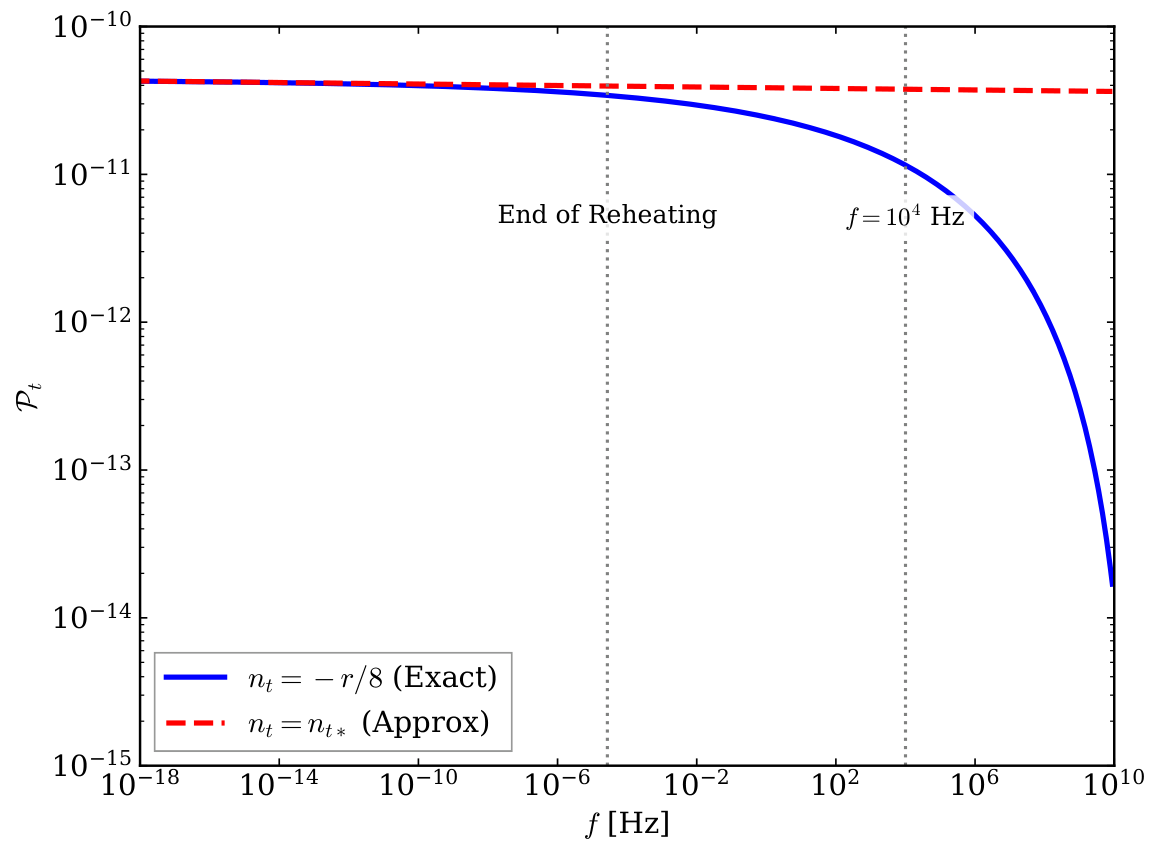}
    \caption{Comparison of the primordial tensor power spectra calculated from the numerical result [using the consistency relation $n_t = -r/8$] of the inflaton's dynamics (blue) versus the standard power-law approximation with fixed $n_t = n_{t*}$ (red), where $n_{t*}$ is the spectral index at the pivot scale. We show only the case of $n=4$, $\alpha=7$, and $\beta=-2$, corresponding to the ACT-favored spectral index $n_s \approx 0.974$ and a reheating temperature $T_{\rm re} \approx 10^8$~GeV. 
    We have checked that the results are similar for other sets of parameters.
Compared to the power-law approximation, the exact result is suppressed less than one order of magnitude up to $f \sim 10^4$~Hz.}
    \label{fig:appendix_nt_comparison}
\end{figure}

As discussed in Sect.~\ref{sec:gw}, we parameterized the \emph{power-law} primordial tensor power spectrum, $\mathcal{P}_t(k) \propto (k/k_*)^{n_t}$, with $n_t$ fixed at the CMB pivot scale.
Fig.~\ref{fig:appendix_nt_comparison} shows that such a power-law assumption closely approximates the exact result obtained by computationally intensive solving of the inflaton dynamics and the use of the consistency relation. 
Within the windows of future GW observatories (i.e., up to $f \simeq 10^{4}$~Hz), the difference is less than one order of magnitude, which is much smaller than the spectral enhancement (up to 4 orders of magnitude) from the stiff reheating phase.
Therefore, we adopt the power-law approximation for computational efficiency when scanning the model parameter space and computing the GW spectrum at each point.
We perform this comparison using the benchmark parameters $n=4$, $\alpha=7$, and $\beta=-2$, which yields the CMB-best-fitted scalar spectral index $n_s \approx 0.974$, while we obtained similar results for other sets of parameters that are considered in the main text.
Specifically, Fig.~\ref{fig:appendix_nt_comparison} shows the case of the largest allowed $\alpha$ (see Fig.~\ref{fig-Ax-Back4}); we have checked that the deviation between numerical and approximated results for smaller $\alpha$ happens higher frequencies.
Lastly, although the power-law form of $\mathcal{P}_t$ is a sufficient approximation when considering the GW signal up to the interferometer windows ($f \simeq 10^4$~Hz), one has to adopt the fully numerical result when calculating the signal in the ultrahigh frequency ($\gtrsim 10$~kHz) regime; see e.g., \cite{Pi:2024kpw,Wang:2026pff}.

\section{Natural Inflation with Negative Cosine}
\label{app:natural_inflation_minus}
For completeness, we also consider the natural inflation potential given by Eq.~\eqref{eq:Natural} with the \emph{negative} sign,
\begin{align}\label{eq:Natural-negative}
    V(\phi) = \Lambda^4 \left[ 1 - \cos\left( \frac{\phi}{\alpha M_{\rm Pl}} \right) \right]^n \, , \quad G(\phi) = -\left(\frac{\phi}{M_{\rm Pl}}\right)^{\beta} \, .
\end{align}

\begin{figure}[t!]
    \centering
        \includegraphics[width=0.485\textwidth]{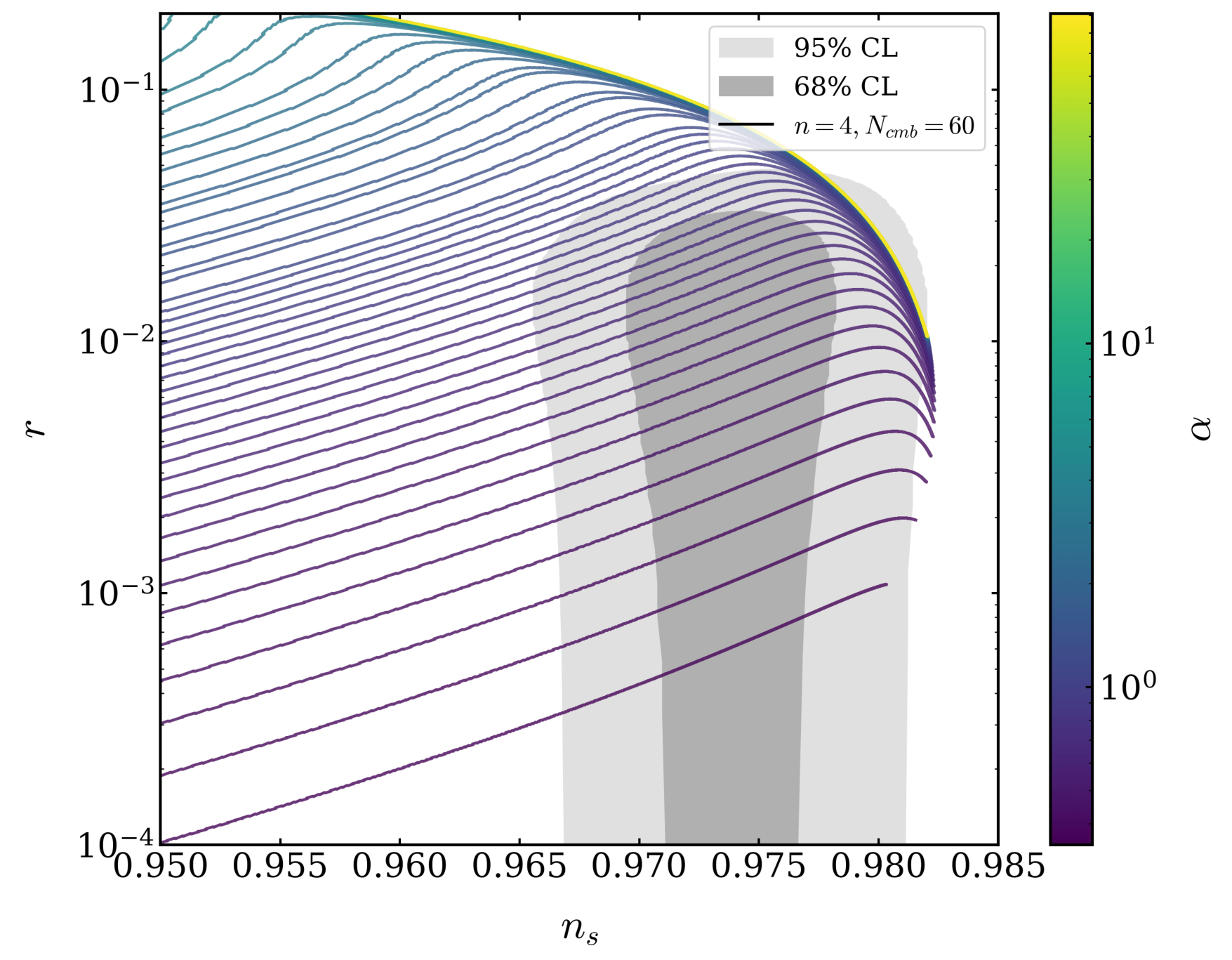}\hspace{0.25cm}
        \includegraphics[width=0.475\textwidth]{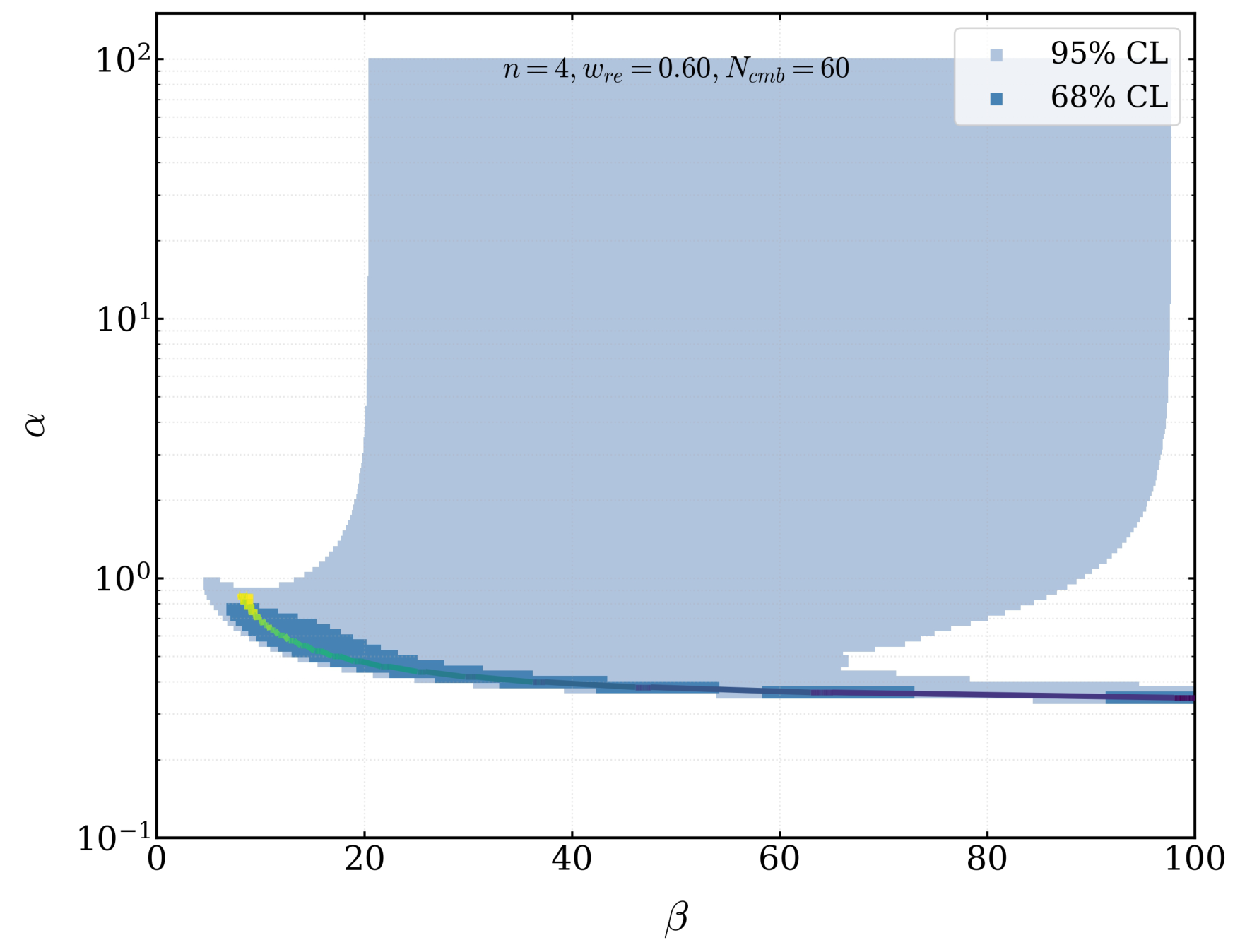}\\
        \includegraphics[width=0.485\textwidth]{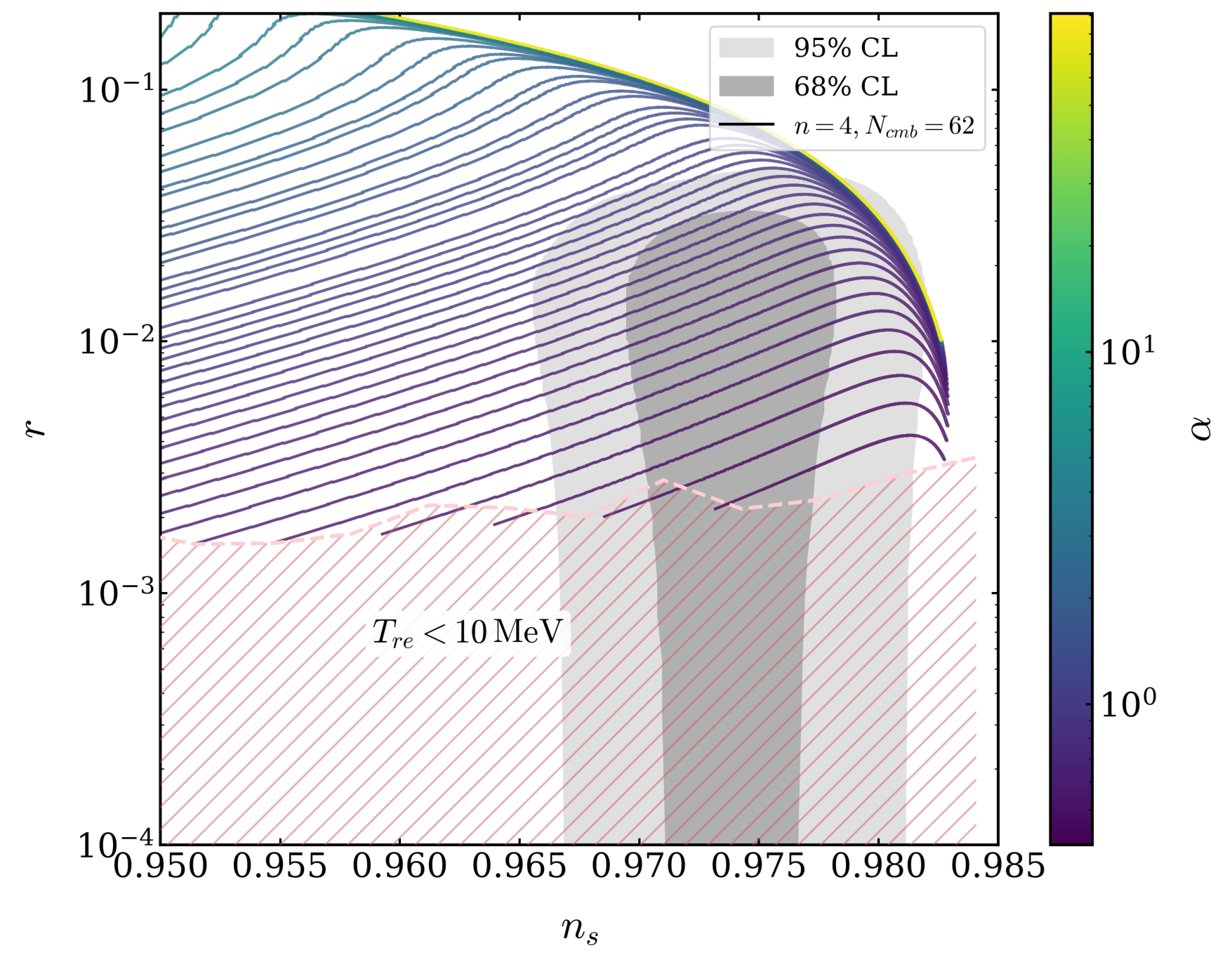}\hspace{0.25cm}
        \includegraphics[width=0.485\textwidth]{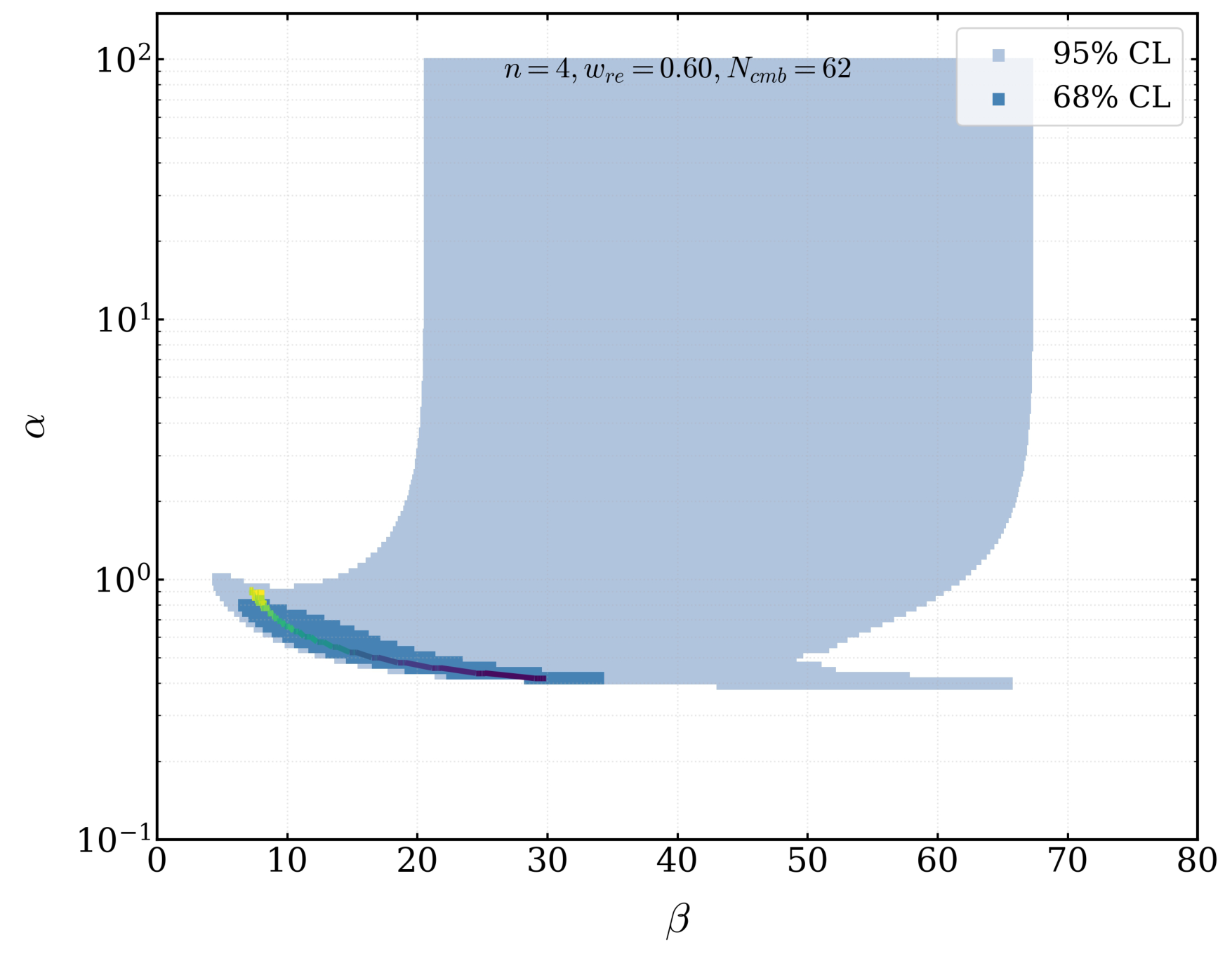}\\[-1.5em]
    \caption{
    Comparable to Fig.~\ref{fig-Ax-Back4}, but for the natural inflation potential with a negative cosine \eqref{eq:Natural-negative}. \textit{Left:} Each line represents the $(r,n_s)$ prediction for a fixed $\alpha$, compared against 68\% and 95\% C.L. posteriors from CMB observations (gray regions). 
        For each line, $\beta$ varies from 0 to 100. 
        The red hatched region is excluded by $T_{\rm reh} \lesssim 10 ~ {\rm MeV}$. \textit{Right:} The parameter spaces in the $(\alpha,\beta)$ plane, compatible with the $(n_s,r)$ posteriors, are shown in blue. The colored line corresponds to $\{\alpha,\beta\}$ that gives the best-fitted $n_s = 0.974$ \cite{Calabrese:2025} with varying reheating temperature $T_{\rm re}$.}
    \label{fig:nat-minus}
\end{figure}

In contrast to the positive cosine in the main text, the negative sign shifts the potential by a phase of $\pi$. I.e., the potential's maximum is located at $\phi \approx \pi \alpha M_{\rm Pl}$, while the minimum is at $\phi = 0$. Inflation therefore initiates at large field values near the hilltop and terminates as the inflaton rolls down toward zero. As the inflaton's evolution differs from the case of positive cosine, the field-dependent kinetic coupling $G(\phi) \propto \phi^\beta$ modulates the friction differently.
Hence, this case has the $\{\alpha,\beta\}$ parameter space compatible with CMB results in a regime complementary to that of the positive-cosine case.
In spite of the new behavior, the inflaton's behavior around the potential minimum remains the same, i.e., the potential can be approximated as $V_{\rm eff} \propto \phi^{2n}$, leading to the same equation of state parameter during reheating $w_{\rm re} = (n-1)/(n+1)$ as in the positive-sign case.
For $n=4$, the reheating bound in Sect.~\ref{sec:reheating} allows the inflation duration within $N_{\rm cmb} \in [56, 62.8]$, similar to the main case. 

In the following, we focus our analysis on two specific scenarios with $N=60$ and $N=62$ and compare them with the results from the positive case.
Fig.~\ref{fig:nat-minus} illustrates the predictions of $n_s$ and $r$ and compares them to the CMB results, in the same manner as Fig.~\ref{fig-Ax-Back4}.
Interestingly, the negative-cosine case achieves $r-n_s$ values in a region previously inaccessible to the positive-cosine case.
When scanning over the model parameter space ($\alpha$ and $\beta$), the right panel of Fig.~\ref{fig:nat-minus} suggests that $\alpha$ can be as low as 0.5 to explain the CMB results.
Furthermore, we observe that, as $\beta$ increases, the $r-n_s$ trajectories in the left panel exhibit a fixed-point toward a larger $n_s$ value, i.e., curves with different $\alpha$ starts overlapping when $\beta$ becomes sufficiently large.
While for smaller $N_{\rm cmb}$ this fixed point lies within the $68\%$ CL region, larger $\beta$ shifts it towards larger $n_s$ values, eventually exiting the $68\%$ and even the $95\%$ CL contours.
This behavior becomes clearer in the right panel of Fig. \ref{fig:nat-minus}.
For $N_{\rm cmb}=60$, the predictions remain within the $68\%$ CL region as $\beta$ increases. However, for $N_{\rm cmb}=62$, the model predictions fall outside the $68\%$ CL for $\beta \geq 35$ and cannot explain the CMB result when $\beta \geq 70$.

\FloatBarrier

\bibliographystyle{JHEP}
\bibliography{modified_grav.bib}

\end{document}